\documentclass[fleqn,usenatbib]{mnras}

\usepackage{newtxtext,newtxmath}

\usepackage[T1]{fontenc}

\DeclareRobustCommand{\VAN}[3]{#2}
\let\VANthebibliography\thebibliography
\def\thebibliography{\DeclareRobustCommand{\VAN}[3]{##3}\VANthebibliography}

\usepackage{graphicx}	
\usepackage{amsmath}	

\usepackage{pdflscape}
\DeclareUnicodeCharacter{2212}{-}
\DeclareUnicodeCharacter{2032}{'}
\usepackage{comment}
\usepackage{subcaption}
\usepackage{wrapfig}
\usepackage{float}
\usepackage{blindtext}
\usepackage{array}
\usepackage{makecell} 

\usepackage{amsmath}
\let\oldAA\AA
\renewcommand{\AA}{\text{\normalfont\oldAA}}
\usepackage{tablefootnote}
\usepackage[referable]{threeparttablex}
\usepackage{hyperref}
\usepackage{multirow}
\usepackage{orcidlink}

\newcommand{\gs}{\mathrel{\lower0.6ex\hbox{$\buildrel {\textstyle >}
 \over {\scriptstyle \sim}$}}}
\newcommand{\ls}{\mathrel{\lower0.6ex\hbox{$\buildrel {\textstyle <}
 \over {\scriptstyle \sim}$}}}

\newcommand{\lta}{\mathrel{\spose{\lower 3pt\hbox{$\mathchar"218$}}
     \raise 2.0pt\hbox{$\mathchar"13C$}}}
\newcommand{\gta}{\mathrel{\spose{\lower 3pt\hbox{$\mathchar"218$}}
     \raise 2.0pt\hbox{$\mathchar"13E$}}}

\newcommand{\oiiia}{\mbox{[O\,{\textsc{iii}}]}\,}
\newcommand{\oiii}{\mbox{[O\,{\sc iii]\sc{$\lambda$5007}}}\,}

\newcommand{\nii}{\mbox{[N\,{\sc ii]}}\,}
\newcommand{\niil}{\mbox{[N\,{\sc ii]\sc{$\lambda$6548}}}\,}
\newcommand{\niih}{\mbox{[N\,{\sc ii]\sc{$\lambda$6585}}}\,}

\defcitealias{2025MNRAS.541.1348P}{P25} 
\defcitealias{2025MNRAS.541.1329D}{D25}
\defcitealias{2025A&A...694A.178C}{CP25}
\defcitealias{2025ApJ...987..186F}{F25}

\title[Narrow-band H$\alpha$ LF and $\rho_{\rm{SFR}}$ into the EoR]{The \emph{JWST} Emission Line Survey (JELS): A narrow-band determination of the H$\alpha$ Luminosity Function and Cosmic Star Formation into the Epoch of Reionization}

\author[C. A. Pirie et al.]{C. A. Pirie\,\orcidlink{0009-0003-5303-6920}\,,$^{1}$\thanks{E-mail: corey.pirie@ed.ac.uk (CAP)}
K. J. Duncan\,\orcidlink{0000-0001-6889-8388}\,,$^{1}$
P. N. Best\,\orcidlink{0000-0001-5081-4801}\,,$^{1}$
D. J. McLeod\,\orcidlink{0000-0003-4368-3326}\,,$^{1}$
C. L. Hale\,\orcidlink{0000-0001-6279-4772}\,,$^{1,2}$
R. K. Cochrane\,\orcidlink{0000-0001-8855-6107}\,,$^{3,1}$\newauthor
S. R. Flury\,\orcidlink{0000-0002-0159-2613}\,,$^{1}$
H. M. O. Stephenson\,\orcidlink{0000-0002-0777-1591}\,,$^{4}$
M. I. Arnaudova\,\orcidlink{0000-0002-1128-0592}\,,$^{1}$
M. Brinch\,\orcidlink{0000-0002-0245-6365}\,,$^{5,6}$
J. S. Dunlop\,\orcidlink{0000-0002-1404-5950}\,,$^{1}$\newauthor
E. Ibar\,\orcidlink{0009-0008-9801-2224}\,,$^{5,6}$
Z Li\,\orcidlink{0000-0001-7373-3115}\,,$^{7}$
J. Matthee\,\orcidlink{0000-0003-2871-127X}\,,$^{8}$
R. J. McLure\,\orcidlink{0009-0005-9742-2318}\,,$^{1}$
L. Ossa-Fuentes\,\orcidlink{0009-0002-3124-1328}\,,$^{5,6}$
A. L. Patrick\,\orcidlink{0000-0003-0645-6853}\,,$^{1}$\newauthor
J. Selfridge\,\orcidlink{0000-0001-5575-7906}\,,$^{3}$
Ian Smail\,\orcidlink{0000-0003-3037-257X}\,,$^{7}$
D. Sobral\,\orcidlink{0000-0001-8823-4845}\,,$^{9,10}$
J. P. Stott\,\orcidlink{0000-0002-1679-9983}\,,$^{4}$ and 
A. M. Swinbank\,\orcidlink{0000-0003-1192-5837}\,$^{7}$
\vspace{0.2cm}\\
$^{1}$Institute for Astronomy, University of Edinburgh, Royal Observatory, Blackford Hill, Edinburgh, EH9 3HJ, UK\\
$^{2}$Astrophysics, Department of Physics, University of Oxford, Denys Wilkinson Building, Keble Road, Oxford, OX1 3RH, UK\\
$^{3}$Jodrell Bank Centre for Astrophysics, Alan Turing Building, University of Manchester, Oxford Road, Manchester M13 9PL, UK\\
$^{4}$School of Physics and Astronomy, Lancaster University, Lancaster LA1 4YB, UK\\
$^{5}$Instituto de F\'isica y Astronom\'ia, Universidad de Valpara\'iso, Avda. Gran Breta\~na 1111, Valpara\'iso, Chile\\
$^{6}$Millennium Nucleus for Galaxies (MINGAL)\\
$^{7}$Centre for Extragalactic Astronomy, Department of Physics, Durham University, South Road, Durham DH1 3LE, UK\\
$^{8}$Institute of Science and Technology Austria (ISTA), Am Campus 1, 3400 Klosterneuburg, Austria\\
$^{9}$Departamento de F\'isica, Faculdade de Ci\`encias, Universidade de Lisboa, Edif\'icio C8, Campo Grande, PT1749-016 Lisbon, Portugal\\
$^{10}$BNP Paribas Corporate \& Institutional Banking, Torre Ocidente Rua Galileu Galilei, 1500-392 Lisbon, Portugal\\
}

\date{Accepted XXX. Received YYY; in original form ZZZ}

\pubyear{2026}

\begin{document}
\label{firstpage}
\pagerange{\pageref{firstpage}--\pageref{lastpage}}
\maketitle

\begin{abstract}
The recent star-formation activity in galaxies can be optimally traced by the H$\alpha$ emission line, with the resulting H$\alpha$ luminosity function (LF) at a given epoch providing a reliable probe of cosmic star formation. We present the first narrow-band determination of the H$\alpha$ LF into the Epoch of Reionization (EoR) at $z \sim 6.1$, using 39 H$\alpha$ emitters selected from the \emph{JWST} Emission Line Survey (JELS). The observed and dust-corrected LFs are broadly consistent with recent slitless spectroscopic measurements but show notable discrepancies with predictions from cosmological hydrodynamical simulations, likely reflecting differences in emission-line and dust modelling. Fits combining multiple LF datasets help constrain the high-redshift faint--end slope of the H$\alpha$ LF ($-1.79 < \alpha_{\rm{H\alpha}} < -1.62$), although there remains uncertainty in the extent to which this evolves with redshift at these epochs. Integrating the JELS dust-corrected H$\alpha$ LFs yields a star-formation rate density of $\log_{10}(\rho_{\rm{SFR_{H\alpha}}} \,/\, \rm{M_{\odot}\,yr^{-1}\,Mpc^{-3}})=-1.93 \, ^{+ 0.14}_{- 0.12} \ \rm{or} \ -2.00 \, ^{+ 0.16}_{- 0.10}$, assuming a continuum-to-line extinction ratio of $\eta_{\rm{dust}}=A_{\rm{cont}}(\rm{6563 \, \AA})/A_{\rm{H\alpha}}=1 \ \rm{and} \ 0.44$, respectively. Both measurements are consistent within uncertainties with previous determinations under standard assumptions for the LF integration limit ($L_{\rm{H\alpha, \, lim}}$) and $\rm{SFR_{H\alpha}}$ calibration constant $\kappa_{\rm{H\alpha}}$, despite the uncertainties in the dust corrections. We explore the metallicity dependence of $\kappa_{\rm{H\alpha}}$, and we find $\rho_{\rm{SFR_{H\alpha}}}$ values that are 0.43 dex lower than the standard calibration and are no longer consistent with UV-based determinations of $\rho_{\rm{SFR}}$ at $z\sim6$. This work highlights the importance of narrow-band surveys in probing the faint H$\alpha$ population and providing new constraints on cosmic star-formation activity into the EoR.
\end{abstract}

\begin{keywords}
galaxies: evolution – galaxies: high-redshift – galaxies: emission lines – galaxies: star formation – surveys – reionization
\end{keywords}



\section{Introduction}
\label{sec:intro}

The cosmic star-formation history remains a key observational constraint on our understanding of galaxy formation and evolution from the early Universe to the present day. We have been able to measure the volume-averaged star-formation rate of galaxies, the star-formation rate density ($\rho_{\rm{SFR}}$), over a range of epochs from the local Universe to high redshift \citep[see a comprehensive review from][]{2014ARA&A..52..415M}.

Prior to the launch of the \emph{James Webb Space Telescope} \citep[\emph{JWST};][]{2023AAS...24110001R}, constraints on the SFR evolution of faint high-redshift galaxies were largely reliant on rest-frame UV emission \citep[e.g.][]{2013ApJ...763L...7E,2013MNRAS.432.2696M,2015ApJ...810...71F,2015MNRAS.450.3032M,2016MNRAS.459.3812M,2018ApJ...855..105O,2021AJ....162...47B,2022ApJ...940...55B}. The unprecedented capabilities of \emph{JWST} mean that rest-frame UV observations have been used to probe cosmic star-formation out to $z = 10 - 15$ \citep[e.g.][]{2022ApJ...940L..14N,2023ApJS..265....5H,2023MNRAS.523.1036B,2023MNRAS.518.6011D,2023MNRAS.520.4554D,2024MNRAS.533.3222D,2024MNRAS.527.5004M,2024ApJ...965..169A}. However, the UV-continuum is heavily impacted by dust attenuation, and samples arising from rest-frame UV-driven photometric redshift (photo-$z$) selections could be biased, with measurements depending on the prior assumptions about the UV-continuum slopes ($\beta$) and emission line properties \citep{2023Natur.622..707A,2023ApJ...958..141L}. Therefore, this selection technique carries a high risk that both the selection of the galaxy sample and the determination of their physical properties \citep[c.f.][]{2015MNRAS.452.2018O} can be systematically affected.

Several studies suggest that the Universe transitions from primarily obscured star-formation activity at $z$ $\lesssim$ 4 to primarily dust-unobscured at $z$ $\gtrsim$ 5 \citep[e.g.][]{2017MNRAS.466..861D,2020ApJ...902..112B,2021ApJ...909..165Z,2025arXiv250307774L}. This result may arise because dust obscuration increases strongly with stellar mass \citep[e.g.][]{2010MNRAS.409..421G,2010MNRAS.402.2017G,2013ApJ...763..145D,2013ApJ...777L...8K,2017ApJ...850..208W,2018MNRAS.476.3991M,2018MNRAS.476.3218C,2022ApJ...926..145S}, while the evolving galaxy stellar mass function (SMF) contains fewer massive galaxies at earlier cosmic times, reducing the relative contribution of massive, dusty galaxies to the total $\rho_{\rm{SFR}}$ \citep[e.g.][]{2021ApJ...909..165Z,2023A&A...677A..34G,2023A&A...677A.184W}. Conversely, other studies suggest no decline in the dust-obscured $\rho_{\rm{SFR}}$ out to the EoR \citep[e.g.][]{2020A&A...643A...8G,2021A&A...649A.152K,2024A&A...681A.118T,2024MNRAS.530..966G}, and significant dust reservoirs have been detected in galaxies out to $z \sim 8$  \citep[e.g.][]{2021Natur.597..489F,2023A&A...672A.108A,2024MNRAS.527.5808B,2025MNRAS.544.1502B}, thus the impact of dust on our view of cosmic star-formation could still be significant at high redshift.

Alternative SFR indicators, less affected by dust, have now been employed to measure $\rho_{\rm{SFR}}$ at $z > 2$ and out to the latter stages of the Epoch of Reionization (EoR; $5 \lesssim z \lesssim 7$). These include measurements from deep radio observations, tracing the non-thermal synchrotron emission from supernovae \citep[e.g.][]{2011ApJ...730...61K,2017A&A...602A...5N,2020ApJ...899...58L,2022ApJ...927..204E,2023MNRAS.523.6082C}, and from dust-reprocessed infrared (IR) emission, which is often calibrated or corrected by the rest-frame UV measurements to obtain the total $\rho_{\rm{SFR}}$. However, these observations were not sensitive enough in selecting faint galaxies and therefore relied on stacking, often on UV pre-selected samples \citep[e.g.][]{2009A&A...496...57M}. 

Another useful SFR indicator is the H$\alpha$ emission line -- a long-standing, well-calibrated and highly-sensitive SFR indicator \citep[e.g.][]{1976ApJ...203..587C,1983ApJ...272...54K}. This emission line results from the recombination of gas ionised by UV photons from star-forming regions, tracing stellar populations with lifetimes $\leq$ 10 Myr, unveiling the most recent star-formation activity within galaxies \citep[e.g.][and references therein]{1998ARA&A..36..189K,2012ARA&A..50..531K,2013seg..book..419C}. This indicator is also less impacted by dust obscuration compared to the UV-continuum \citep[e.g.][]{2001ApJ...548..681B}, and H$\alpha$ can be detected in highly dust-obscured galaxies \citep[e.g.][]{2013ApJ...767..151M,2016ApJ...827...57O,2020A&A...635A.119C,2021MNRAS.503.2622C,2025MNRAS.537..788H}. H$\alpha$ studies using ground-based telescopes were limited to $z < 2.6$ due to the atmospheric opacity limiting observations redward of the near-infrared (NIR) wavelength regime \citep[e.g. the High-$z$ Emission Line Survey, HiZELS;] []{2008MNRAS.388.1473G,2013MNRAS.428.1128S,2013ASSP...37..235B}. However, efforts were made to place constraints on $\rho_{\rm{SFR}}$ at $z > 2.5$ via observed \emph{Spitzer}/IRAC flux excesses (or upper limits) associated with H$\alpha$ emission, and were derived using spectral energy distribution (SED) fitting \citep[e.g.][]{2021ApJ...915...47A,2022ApJ...924...71A,2023ApJ...946..117B}, relying on underlying assumptions for the stellar continuum and emission line strengths which are hard to disentangle.

The launch of \emph{JWST} has expanded the redshift frontier for detecting the H$\alpha$ emission line, with detections up to $z \sim 6.5$ now possible with the NIR instruments and up to $z\sim9$ \citep{2024A&A...686A..85A} using the MIRI instrument \citep{2023PASP..135d8003W}. Pre-selected high-redshift galaxy candidates have been targeted for follow-up multi-object and IFU spectroscopy using NIRSpec \citep{2022A&A...661A..80J} to study their H$\alpha$ emission \citep[e.g.][]{2023A&A...677A.115C,2023ApJ...950L...1S,2025ApJ...980..242S,2023ApJ...955...54S,2024ApJ...962...24S,2024ApJ...976..193R}. However, \emph{JWST} has also enabled the untargeted search for H$\alpha$ emission line galaxies into the EoR, largely thanks to the NIRCam instrument \citep{2005SPIE.5904....1R,2023PASP..135b8001R}. These include slitless grism spectroscopic surveys \citep[][]{2023ApJ...950...66K,2023ApJ...950...67M,2023MNRAS.525.2864O,2023jwst.prop.3577E,2023jwst.prop.2883S} and medium-band imaging surveys \citep[e.g.][]{2023ApJS..268...64W,2023arXiv231012340E,2024ApJ...976..101S,2025arXiv250719706M}. It is through these surveys that it has become possible to start constraining the H$\alpha$ luminosity function (LF) at $z > 2.5$ for the first time \citep[e.g.][hereafter \citetalias{2025A&A...694A.178C} and \citetalias{2025ApJ...987..186F}, respectively]{2025A&A...694A.178C,2025ApJ...987..186F}.

The launch of \emph{JWST} has also enabled untargeted narrow-band observations to select emission line galaxies across multiple epochs of cosmic time, thanks to the \emph{JWST} Emission Line Survey \citep[JELS;][hereafter \citetalias{2025MNRAS.541.1329D}]{2025MNRAS.541.1329D}. JELS observations have led to the discovery of H$\alpha$ emitters at $z\sim6.1$ \citep[][hereafter \citetalias{2025MNRAS.541.1348P}]{2025MNRAS.541.1348P}, \oiiia emitters at $z\sim8.3$ \citepalias{2025MNRAS.541.1348P,2025MNRAS.541.1329D} and Paschen-$\alpha$ and Paschen-$\beta$ emitters at $z\sim1.5$ and $z\sim2.6$, respectively (see Ossa-Fuentes et al. \emph{submitted}; Brinch et al. \emph{submitted}). The advantage of narrow-band surveys is that they are highly sensitive \citepalias[2 -- 4$\times$ fainter line fluxes than slitless spectroscopic and medium-band selections;][]{2025MNRAS.541.1329D} and offer a clean selection of emission line galaxies based only on line strength. In \citetalias{2025MNRAS.541.1348P}, we analysed a robust sample of 35 H$\alpha$ emitters at $z \sim 6.1$, the highest-redshift sample of narrow-band selected H$\alpha$ emitters to date. These sources appear to be dust-poor (median $A_{V} \sim 0.23$), have established stellar components \citep{2025MNRAS.544.1412S} and exhibit recent upturns in their SFRs indicative of bursty star-formation histories (SFHs).

The full population statistics of this faint population have yet to be constrained and so we extend the narrow-band constraints on the H$\alpha$ LF \citep[c.f.][]{2013MNRAS.428.1128S} in this study and infer $\rho_{\rm{SFR}}$ into the EoR at $z>6$, exploiting the sensitivity afforded by JELS imaging. The paper is laid out as follows. In Section~\ref{sec:emission_line_selection}, we summarise the JELS narrow-band survey and the multi-wavelength observations, narrow-band detected catalogue creation steps and the H$\alpha$ emission line sample selection procedure. In Section~\ref{sec:halpha_lf_method}, we describe the methodology for computing the JELS observed and dust-corrected H$\alpha$ LFs. In Section~\ref{sec:halpha_lf}, we describe the fitting methodology for the H$\alpha$ LFs and compare the results to those from observations and simulations, including probing the evolution of the faint-end slope $\alpha_{\rm{H\alpha}}$ with redshift. We then infer the star-formation rate density $\rho_{\rm{SFR}}$ in Section \ref{sec:sfrd}, compare to previous studies and discuss the impact of different assumptions on our view of the cosmic star-formation history. Finally, we draw conclusions in Section~\ref{sec:conclusions}. We adopt the following cosmological parameters in all cosmological calculations: $H_{\rm{0}}$ = 70 km s$^{−1}$ Mpc$^{−1}$, $\Omega_{\rm{M}}$ = 0.3 and $\Omega_{\rm{\Lambda}}$ = 0.7. All magnitudes are in the AB system \citep{1974ApJS...27...21O,1983ApJ...266..713O}. All stellar mass and SFR values assume a \citet{2003PASP..115..763C} initial mass function (IMF).

\section{The Data and Emission line galaxy selection}
\label{sec:emission_line_selection}

\subsection{JELS and multi-wavelength ancillary observations}
\label{sec:observations}

The JELS narrow-band observations \citepalias[described in full in][]{2025MNRAS.541.1329D} utilised the \emph{JWST}/NIRCam long-wavelength filters F466N and F470N at $\sim$4.7$\mu$m. In parallel, we observed the same field in the short-wavelength channel using the F212N narrow-band (at $\sim$2.12 $\mu$m) and F200W broad-band filters, but the F212N imaging was not considered in this paper. This observation setup provided continuous coverage over an area of 63 arcmin$^{2}$ of the Cosmic Evolution Survey (COSMOS) field with central coordinates (RA, Dec) = (150.125, 2.333) deg. The JELS observations used a $3\times3$ mosaic strategy with 57 per cent overlap between columns, and adopted the `Medium8' observing strategy with 9 groups for the F466N filter and 10 groups for the F470N filter, which gave $\sim$1000s on-sky per observation. A 3-point intramodule dithering pattern, with two sub-pixel dithers at each location, was then used to account for bad pixels and cosmic rays. The on-sky integration time was $\sim$ 6 ks over the full mosaic with double-depth imaging ($\sim$ 12 ks) over the central $\sim$ 40 per cent of the mosaic. This totalled 43 hours of programme time. Data reduction of the JELS NIRCam imaging was performed using the PRIMER Enhanced NIRCam Image Processing Library (\textsc{PENCIL}; Magee et al. in \emph{preparation}) software. The reduced imaging was astrometrically aligned to GAIA DR3 \citep{2023A&A...674A...1G} and stacked to the same pixel scale of 0.03 arcsec.

Ancillary multi-wavelength imaging of the COSMOS field was folded into this analysis \citepalias[see Fig. 1 and Table 1 of ][]{2025MNRAS.541.1348P}. This includes data from the \emph{Hubble Space Telescope} (\emph{HST}), including the Cosmic Assembly Near-IR Deep Extragalactic Legacy Survey \citep[CANDELS;][]{2011ApJS..197...35G,2011ApJS..197...36K}, 3D-\emph{HST} \citep{2012ApJS..200...13B} and UVCANDELS \citep{2018hst..prop15647T}, observed using the WFC3 and ACS instruments and providing multi-wavelength coverage from $\sim$0.3 through to 1.6 $\mu$m. In addition, this same field contains the public \emph{JWST} Treasury Program, Public Release IMaging for Extragalactic Research \citep[PRIMER; GO 1837; PI:][]{2021JWST.prop.1837D} that provides \emph{JWST}/NIRCam imaging in 8 filters from $\sim$0.9 to 4.7$\mu$m with nearly 100 per cent overlapping coverage of the JELS field. All JELS and multi-wavelength imaging were then homogenised to a common point-spread function \citepalias[PSF; see Section 2 of][]{2025MNRAS.541.1348P}. 

Note, as discussed in Appendix A of \citetalias{2025MNRAS.541.1329D}, the \emph{JWST} imaging data utilised in \citetalias{2025MNRAS.541.1348P} were reduced using an earlier version of \textsc{PENCIL} (pipeline version 1.10.2 -- jwst$\_$1107.pmap) -- we refer to this imaging dataset and associated catalogues as v0.8. Some of the JELS v0.8 image frames, particularly in the F212N and F200W filters, were contaminated by scattered light, but have since been re-observed; these have been incorporated and all PRIMER and JELS images re-reduced with an updated version of the \textsc{PENCIL} pipeline (1.13.4 -- jwst$\_$1303.pmap), improving both their image quality and reaching 5$\sigma$ global depths (in 0.3 arcsec aperture diameters) of 26.51 and 26.60 mag for F466N and F470N images, respectively, compared to 26.28 and 26.24 mag in the v0.8 imaging. Lastly, we incorporate a depth weighted addition of both the JELS and PRIMER F200W imaging to obtain even deeper 2$\mu$m coverage with a 5$\sigma$ global depth of 29.06 mag in the JELS survey area. These are the JELS v1.0 images and associated catalogues used in this analysis; these additions improve the quality of the photo-$z$ and SED fitting analysis (see Section \ref{sec:halpha_sources} and \ref{sec:halpha_lum}).

\subsection{Narrow-band detection catalogues and photometric redshift analysis}
\label{sec:nb_cat}

In constructing our narrow-band (NB) detected catalogues, we follow the same steps described in Section 3 of \citetalias{2025MNRAS.541.1348P} but utilise the v1.0 imaging discussed in Section \ref{sec:observations}. Any deviations from the \citetalias{2025MNRAS.541.1348P} methodology will be explicitly highlighted in this section. In summary, we perform source detection using \textsc{SExtractor} \citep{1996A&AS..117..393B} in `dual mode', using the native-resolution (i.e. before PSF homogenisation) F466N and F470N images as the detection filters. Sources are required to meet a minimum signal-to-noise ratio (SNR) $>$ 5 in 0.3 arcsec diameter apertures. We then perform forced photometry in all the available PSF-homogenised imaging. Measurements are also performed in 0.6, 0.9 and 2.0 arcsec diameter apertures and in Kron apertures \citep{1980ApJS...43..305K}. The 0.3 arcsec diameter measurements are utilised for source detection as stated above, but also for excess source selection and photo-$z$ analysis, to maximise signal (see Section \ref{sec:halpha_sources}). The 0.6 arcsec diameter measurements are used for the line luminosity calculations and SED fitting analysis in \citetalias{2025MNRAS.541.1348P}, with the larger aperture better capturing the total fluxes for compact sources. A series of contamination cuts are then implemented \citepalias[also discussed in Section 3.3.1 of][]{2025MNRAS.541.1348P} to obtain our F466N and F470N narrow-band (NB) detected catalogues. Firstly, we require a minimum effective radius ($r_{e}$ > 1.5 pixels) since sources below this size are identified as hot pixels likely from cosmic rays. Secondly, we reject sources contained in bright star masked regions that are contaminated by stellar emission and diffraction spikes. 

For the v0.8 NB-detected catalogues, \citetalias{2025MNRAS.541.1348P} required sources to have high significance in the most sensitive PRIMER filter F356W (SNR(F356W) $>$ 5) as the final sample of emission line galaxies had high rates of contamination. The v1.0 dataset does not have the same contamination rates in our H$\alpha$ emission line candidate samples (see discussion in Section \ref{sec:halpha_sources}). Therefore, we decide not to implement the SNR(F356W) cut in this work to allow for a more complete emission line galaxy sample in the new v1.0 NB-detected catalogues. Photo-$z$'s for our NB-detected sources \citepalias[see Section 3.4 of][]{2025MNRAS.541.1348P} are derived by performing SED-fitting with \textsc{EAZY-Py} \citep{2008ApJ...686.1503B} utilising all filter coverage available for a given source. As with the v0.8 NB-detected catalogue, we evaluate the photo-$z$ performance by calculating bulk quality statistics for spectroscopically confirmed sources, using the normalised median absolute deviation, $\sigma_{\rm{NMAD}} = 1.48 \times \rm{median}(|\Delta z| / (1 + z_{\rm{spec}}))$, where $\Delta z = z_{\rm{phot}} - z_{\rm{spec}}$, and the absolute outlier fraction, $\rm{OLF} = |\Delta z| / (1 + z_{\rm{spec}}) > 0.15$ \citep[following common literature definitions; e.g.][]{2013ApJ...775...93D,2019A&A...622A...3D,2021A&A...648A...4D}. The results for the v1.0 NB-detected catalogue show $\sigma_{\rm{NMAD}} = 0.0302$ and OLF = 0.0790, which are slight improvements to the v0.8 NB-detected catalogues \citepalias{2025MNRAS.541.1348P}.

\subsection{Selecting excess sources to identify H$\alpha$ emitters}
\label{sec:halpha_sources}

A full description of the excess source criteria and selection of the H$\alpha$ emission line galaxies is described in Section 4 of \citetalias{2025MNRAS.541.1348P}. In summary, we utilise the JELS F466N, F470N and PRIMER F444W observations to select sources with colour excess in the detection filter compared to the ancillary narrow-band (NB $-$ NB) or ancillary broad-band (BB $-$ NB) filters. 

For both colour selections, we firstly account for the impact of colour across the different filters \citepalias[which for F466N or F470N compared to F444W is exacerbated by the relative wavelengths; see Fig. 2 of][]{2025MNRAS.541.1348P}. We then select genuine NB excess sources using the narrow-band excess parameter $\Sigma$ \citep[e.g.][]{1995MNRAS.273..513B,2013MNRAS.428.1128S}, thereby removing sources that are unlikely to have an intrinsic narrow-band colour excess, but are scattered to higher values due to local noise variations which become prominent for the faintest sources (see Fig. 8 of \citetalias{2025MNRAS.541.1348P}). For the BB $-$ NB selected sources, this is given by:

\begin{equation}
\label{eq:sig1}
\Sigma \ = \ \frac{1 - 10^{-0.4 (\rm{BB} \ - \ \rm{NB)}}}{10^{-0.4 (\rm{ZP} \ - \ \rm{NB})} \sqrt{\sigma_{\rm{NB}}^{2} \ + \ \sigma_{\rm{BB}}^{2}}}
\end{equation}

\noindent where ZP is the zero point magnitude of the NB filter, which is set to 23.9 mag, and $\sigma_{\rm{NB}}$ and $\sigma_{\rm{BB}}$ are the photometric flux density errors (in $\rm{\mu}$Jy) for the NB and BB filters, respectively, for each source. 

Towards brighter magnitudes, the BB $-$ NB colour corresponding to a given value of $\Sigma$ tends towards zero. However, additional systematic effects (e.g. colour continuum corrections) means there is still a degree of scatter at bright magnitudes. Therefore, we also apply an equivalent-width ($EW$) limit, which corresponds to a minimum BB $-$ NB colour for sources to be selected as an `excess source'. The excess criteria are met if $\Sigma > 3$ and F444W $-$ NB $>$ 0.3, corresponding to rest-frame H$\alpha$ equivalent-width $EW_{\rm{H\alpha}} > 24.2 \, \AA$ and $EW_{\rm{H\alpha}} > 25.7 \, \AA$ for the F466N and F470N-detected sources, respectively. In \citetalias{2025MNRAS.541.1348P}, we required F444W $-$ F470N $>$ 0.35 ($EW_{\rm{H\alpha}} > 29.0 \, \AA$) due to the higher rate of contamination resulting from the v0.8 F470N image scattered light \citepalias[see Section 2.2.2 of][]{2025MNRAS.541.1329D}, which caused a larger systematic scatter of F444W $-$ F470N colours around zero. Aligning the BB $-$ NB selection parameters for both F466N and F470N-detected sources shows the improvement in the F470N imaging, the subsequent reduction in contamination, and ensures homogeneity across the full JELS selection volume.

As discussed in \citetalias{2025MNRAS.541.1348P}, the excess source selection is also performed for the NB $-$ NB colour selection. However, \citetalias{2025MNRAS.541.1348P} showed that all of the H$\alpha$ emission line galaxy candidates were selected in the BB $-$ NB excess source selection, which yields more accurate measurements of properties such as the emission line flux \citepalias[see discussion in Section \ref{sec:halpha_lum} and Section 4.1.2 of ][]{2025MNRAS.541.1348P}. All H$\alpha$ emission line galaxy candidates in the v1.0 catalogue are BB -- NB selected, and so we utilise these colours for selecting and measuring the emission line properties. From the narrow-band detected excess sources, the first step in selecting H$\alpha$ emission line galaxy candidates is to impose two photo-$z$ conditions from the outputs discussed in Section \ref{sec:nb_cat}:

\begin{enumerate}

  \item $5.5 \leq z_{\rm{1,median}} \leq 6.5$
  
  \item $(z_{\rm{1,max}} - z_{\rm{1,min}}) / (1 + z_{\rm{1,median}}) < 0.4$
  
\end{enumerate} 

Here, $z_{\rm{1,min}}$, $z_{\rm{1,max}}$ and $z_{\rm{1,median}}$ correspond to the lower bound, upper bound and the median of the primary 90 per cent highest probability density (HPD) credible interval (CI) peak, respectively \citep[see e.g.][]{2019A&A...622A...3D}. These conditions require our sample to have narrow and well-defined photo-$z$ posteriors, as expected for line emission driving a strong narrow-band excess. We note that condition (i) allows for an explicitly broad redshift range, but the $z_{\rm{1,median}}$ values are all within the much smaller NB redshift windows for the H$\alpha$ emission line at $z\sim6.1$. We also acknowledge that there could be sources with secondary redshift peaks at $z\sim6.1$, and the impact this has on the LF analysis is explored in Section \ref{sec:p_z_analysis}. From the above steps applied to our v1.0 `excess source' catalogues, we obtain a preliminary sample of 44 H$\alpha$ emission line galaxy candidates. Visual inspection indicates that 39 of these are real and not artefacts; this represents a $\sim$89 per cent success rate. This is a considerable improvement over the v0.8 catalogue, where \citetalias{2025MNRAS.541.1348P} found that only 35 out of the original 71 H$\alpha$ candidates were considered real following visual inspection, representing only a $\sim$50 per cent success rate.

Most of the contaminants removed from the v1.0 catalogues are sources picked up at the edges of the JELS footprint, where artefacts are more likely due to the reduced number of exposures contributing to a given pixel \citepalias[see][]{2025MNRAS.541.1329D}. However, we report that there are two H$\alpha$ emission line galaxy candidates from the v0.8 catalogue which do not meet the v1.0 selection criteria. These sources were low-SNR detections in the v0.8 catalogue (SNR(NB) $\sim$ 5.4 and 5.0) and narrowly fail to meet the SNR(NB) $>$ 5.0 threshold in the v1.0 catalogues. These sources do appear to be robustly identified H$\alpha$ emitters and would contribute to the faint-end of the H$\alpha$ LF. However, lowering the required SNR threshold would come at the expense of purity from other spurious sources, so we choose to maintain the stricter threshold and account for the potentially missed sources through subsequent completeness corrections (see Section \ref{sec:completeness}).

\section{The H$\alpha$ luminosity function procedure}
\label{sec:halpha_lf_method}

\subsection{Corrected H$\alpha$ emission line luminosity}
\label{sec:halpha_lum}

Before computing the $z \sim 6.1$ H$\alpha$ luminosity function, we calculate the H$\alpha$ line fluxes for our galaxy candidates, $F_{\rm{H\alpha}}$, using the BB $-$ NB method \citep[e.g.][]{2013MNRAS.428.1128S}:

\begin{equation}
\label{eq:f_line_1}
F_{\rm{H\alpha}} \ = \ \Delta \lambda_{\rm{NB}} \ \frac{f_{\rm{NB}} \ - \ f_{\rm{BB}}}{1 \ - \ (\Delta \lambda_{\rm{NB}} / \Delta \lambda_{\rm{BB}})}
\end{equation}

\noindent where $\Delta \lambda_{\rm{NB}}$ and $\Delta \lambda_{\rm{BB}}$ are the NB and BB filter widths, respectively, and where $f_{\rm{NB}}$ and $f_{\rm{BB}}$ are the measured flux densities in the NB and BB filters, respectively. The H$\alpha$ luminosity ($L_{\rm{H\alpha}}$) is then calculated from the line flux: $L_{\rm{H\alpha}} = 4 \pi \ D^{2}_{L} \ F_{\rm{H\alpha}}$, where $D_{L}$ is the luminosity distance corresponding to the photometric redshift $z_{\rm{1,median}}$ of a given H$\alpha$ emission line galaxy. In \citetalias{2025MNRAS.541.1348P}, we utilise 0.6 arcsec diameter aperture measurements to capture a high fraction of the total H$\alpha$ flux, whereas in this work we utilise Kron-aperture \citep{1980ApJS...43..305K} photometry. While Kron-aperture photometry is often found to capture $\sim$90 per cent of the total source flux, potentially requiring a further $\sim$10 per cent aperture correction (see e.g. \citealt{2024MNRAS.527.5004M}; \citealt{2025A&A...704A.339S}; \citealt{2026MNRAS.545f1961L}; Ossa-Fuentes et al. \emph{submitted}; Brinch et al. \emph{submitted}), we test the suitability of the Kron-aperture measurements using the injected model sources in our completeness simulations (see Section \ref{sec:completeness}). We find that the recovered Kron-aperture flux densities are broadly consistent with the input total flux densities within the scatter, while applying an additional 10 per cent aperture correction introduces a systematic offset. We therefore adopt the Kron-aperture measurements to estimate the total H$\alpha$ luminosities. We note, however, that applying the additional 10 per cent aperture correction has a negligible impact on the final H$\alpha$ LF number densities (see Section \ref{sec:lf_corr_fit}).

Five H$\alpha$ emitters have spectroscopic follow-up observations from the CANDELS-Area Prism Epoch of Reionization Survey \citep[CAPERS; GO-6368, PI:][]{2024jwst.prop.6368D} and Director’s Discretionary program DD 6585 \citep[PI:][]{2024jwst.prop.6585C}, confirming their classifications. We retrieve this dataset from the \href{https://dawn-cph.github.io/dja/}{DAWN JWST Archive} \citep[DJA;][]{2025A&A...693A..60H,Brammer_Valentino_2025}, containing spectroscopic data from public JWST programs, uniformly extracted and reduced using the same pipeline based on \href{https://github.com/gbrammer/grizli}{grizli2} and \href{https://github.com/gbrammer/msaexp}{MSAexp3} \citep[see further details in][]{2025A&A...697A.189D,2025A&A...693A..60H,2025A&A...699A.358V,2026A&A...708A.203P}. The local continuum around the H$\beta$, \oiii, and H$\alpha$ emission lines are modelled with a linear function and subtracted. The continuum-subtracted spectra were then fitted simultaneously with Gaussian profiles, requiring all emission lines to share a common velocity offset and line width. The H$\alpha$ flux was measured from the best-fitting Gaussian profile. The photometric (see Eq. \ref{eq:f_line_1}) and spectroscopic measurements of the H$\alpha$ fluxes agree well, with the offsets ranging from 0.012 -- 0.205 dex (and a median offset of 0.026 dex), showing that the slit-loss corrections are mostly accurate. However, even the largest offsets are smaller than the H$\alpha$ LF bin widths utilised in this study (see Table \ref{table:halpha_lf_data}), and we account for the potential impact that the flux uncertainties have on the number densities in our LF analysis (see Section \ref{sec:prelim_lf}).

As discussed in \citetalias{2025MNRAS.541.1348P}, and in previous work with lower-$z$ samples of H$\alpha$ emitters \citep[e.g.][]{2012MNRAS.420.1926S}, the narrow-band filters will also have contamination due to the \niil and \niih emission lines for H$\alpha$ emission line galaxy candidates. \citet{2023ApJ...950L...1S} used spectral stacks of their $z$ = 5.0 $−$ 6.5 sample of star-forming galaxies to assess the average strength of the \nii emission lines in their sample. They showed negligible \niil emission and a stacked ratio of $\log_{10}(\rm{[NII]6585/H\alpha})$ = -−1.31 \citep[similar to results from][]{2023ApJ...955...54S} corresponding to a correction of 0.021 dex to $L_{\rm{H\alpha}}$. \citetalias{2025MNRAS.541.1329D} performed mock simulations of H$\alpha$ emission line recovery, assuming the same approximation for the \niih/H$\alpha$. They found that the 5$\sigma$ narrow-band detection limit resulted in all H$\alpha$ luminosities were accurate to $\pm0.02$ dex and precise to $\pm0.01$ dex. Therefore, we correct $L_{\rm{H\alpha}}$ measurements by 0.021 dex for all H$\alpha$ emitters, assuming that the average \nii/H$\alpha$ line ratio measured in \citet{2023ApJ...950L...1S} applies to the entire JELS H$\alpha$ sample.

Though the H$\alpha$ emission line is far less affected by dust compared to the UV-continuum \citep[e.g.][]{2001ApJ...548..681B}, it is not immune to dust extinction. Ideally, we would dust-correct $L_{\rm{H\alpha}}$ for each source, through measurements of the Balmer decrement or through comparison between the H$\alpha$ and far-infrared (FIR) determined SFRs \citep[e.g.][]{2009ApJ...703.1672K}. However, these sources do not have detections in current FIR datasets. In addition, the five H$\alpha$ emitters with spectroscopic datasets show low SNR for H$\beta$, compounded by the low spectral resolution of the NIRSpec PRISM observations \citep[see e.g.][]{2024A&A...691A.305S}, making measurements of the Balmer decrement highly uncertain. We instead utilise spectral energy distribution (SED) fitting of our H$\alpha$ emission line galaxy candidates using \textsc{Bagpipes} \citepalias[see Section 5.1 of][]{2025MNRAS.541.1348P} to obtain estimates of the $V$-band stellar continuum attenuation $A_{V}$, assuming the \citet{2018ApJ...859...11S} dust attenuation law. \citetalias{2025MNRAS.541.1348P} found no evidence for the dust attenuation law to be different from the \citet{2000ApJ...533..682C} attenuation law, on average. Therefore, we scale the $A_{V}$ values using \citet{2000ApJ...533..682C} dust law to obtain stellar continuum attenuation at 6563$\AA$, $A_{\rm{cont}}(\rm{6563 \, \AA}$), where H$\alpha$ emission originates. Any deviations from the assumed correction will be negligible and subdominant to measurement uncertainties for individual sources.

Another factor that needs to be considered is whether there is additional reddening of the H$\alpha$ emission line ($A_{\rm{H\alpha}}$) relative to $A_{\rm{cont}}(\rm{6563 \, \AA}$), given that H$\alpha$ emission originates within the dusty birth clouds surrounding the most massive, short-lived stars and the stellar continuum emission includes contributions from older stellar populations that have dispersed their natal clouds. This effect has been well characterised in the local Universe by \citet{2000ApJ...533..682C}, who found an empirical continuum-to-nebular colour excess ratio of $E(B-V)_{\rm{cont}}/E(B-V)_{\rm{line}} = 0.44$. However, other work has questioned whether this relation is valid for typical high-redshift star-forming galaxies (e.g. \citealt{2015ApJ...806..259R,2020ApJ...899..117S,2020ApJ...902..123R}; see also \citealt{2026MNRAS.546ag049F} for more massive high-redshift galaxies), likely reflecting the diversity in dust laws affecting both the stellar continuum and nebular emission across the galaxy population \citep{2020ARA&A..58..529S,2026ApJ...999...15R,2025ApJ...989..209S}. 

Our SED fitting procedure differs slightly to that described in \citetalias{2025MNRAS.541.1348P}, where the relative dust reddening of the stellar continuum and nebular emission, parameterised as $\eta_{\rm{dust}} = A_{\rm{cont}}(\rm{6563 \, \AA})/A_{\rm{H\alpha}}$, was originally allowed to vary. We then calculate $A_{\rm{H\alpha}}$ on a source-by-source basis by applying the relevant $\eta_{\rm{dust}}$ factor to the $A_{\rm{cont}}(\rm{6563 \, \AA})$ values, derived from the SED-derived $A_{V}$ values. In this work, we consider the following values for $\eta_{\rm{dust}}$: i) $\eta_{\rm{dust}}=1$ (i.e. no additional reddening of the H$\alpha$ emission line relative to the underlying continuum), and ii) $\eta_{\rm{dust}}=0.44$, corresponding to the empirical ratio $E(B-V)_{\rm{cont}}/E(B-V)_{\rm{line}} = 0.44$ measured by \citet{2000ApJ...533..682C}, under the simplifying assumption that the same dust law applies to both the stellar continuum and nebular emission\footnote{The empirical relation $E(B-V)_{\rm{cont}}/E(B-V)_{\rm{line}} = 0.44$ was derived by \citet{2000ApJ...533..682C} assuming a \citet{1989ApJ...345..245C} extinction curve for the nebular emission, which results in $A_{\rm{H\alpha}}/A_{\rm{cont}}(\rm{6563 \, \AA}) \sim 1.72$. We adopt the \citet{2000ApJ...533..682C} dust law for both the reddening of the stellar continuum and nebular emission, resulting in $A_{\rm{H\alpha}}/A_{\rm{cont}}(\rm{6563 \, \AA}) = 2.27$, leading to a H$\alpha$-to-continuum dust correction that is $\sim$32 per cent larger than that implied by the original \citet{2000ApJ...533..682C} prescription (see also \citealt{2014ApJ...795..165S}).}{}.

The median $A_{\rm{H\alpha}}$ values for our sample are $\sim$0.20 and 0.24 mag assuming $\eta_{\rm{dust}}=1$ and 0.44, respectively. These results show that both assumptions produce remarkably similar H$\alpha$ dust corrections, with the standard deviation in the $A_{\rm{H\alpha}}$ differences being only $\sim$0.06 mag. This is because the narrow-band flux is dominated by H$\alpha$ emission, and so the SED fit provides a stronger constraint on the reddening of the emission line compared to the underlying stellar continuum. Consequently, different assumptions for $\eta_{\rm{dust}}$ can alter the SED-derived $A_{V}$ values while still producing similar values of $A_{\rm{H\alpha}}$. The largest differences occur for the lowest-luminosity ($\log_{10}(L_{\rm{H\alpha}} \,/\, \rm{erg \, s^{-1}}) < 42$) H$\alpha$ emitters, which are typically dust poor and have inferred $A_{V}$ values that are increasingly dominated by the photometric uncertainties; for these sources, adopting $\eta_{\rm{dust}}=0.44$ produces systematically larger dust corrections than assuming $\eta_{\rm{dust}}=1$. Given these results, we incorporate both assumptions for $\eta_{\rm{dust}}$ to produce two versions of the dust-corrected H$\alpha$ LF, reflecting the expectation that the population-average $\eta_{\rm{dust}}$ likely lies between these limiting cases. This approach also follows recent studies of the H$\alpha$ LF at similar epochs, which adopt comparable dust-correction prescriptions either through fixed assumptions or Bayesian inference \citepalias{2025A&A...694A.178C,2025ApJ...987..186F}.

\subsection{Preliminary H$\alpha$ luminosity function}
\label{sec:prelim_lf}

After calculating the luminosity for each H$\alpha$ emitter, the luminosity function (LF) is calculated by computing the source count in each luminosity bin, centred on $\log(L_{c} \,/\, \rm{erg \, s^{-1}})$, divided by the survey volume $V_{\rm{filter}}$ covered by the filter profiles and luminosity bin width $\Delta(\log_{10}L)$ \citep[e.g. ][]{2009MNRAS.398...75S,2012MNRAS.420.1926S,2013MNRAS.428.1128S}:

\begin{equation}
\label{eq:lf_uncorr}
\phi(\log_{10} (L_{\rm{c}})) = \frac{1}{\Delta(\log_{10} L)} \sum_{|\log_{10} \frac{L_{\rm{i}}}{L_{\rm{c}}}| < \frac{\Delta(\log_{10} L)}{2}} \frac{1}{V_{\rm{filter}}}
\end{equation}

\noindent for a given source $i$ and luminosity bin centre $c$. The bin widths are guided by the Freedman–Diaconis rule \citep[][see also Ossa-Fuentes et al. in \emph{submitted}]{freedman81,birge06}, which accounts for both the sample size and scatter in the individual $L_{\rm{H\alpha}}$ measurements. The FWHM for each filter (520.36$\AA$ and 494.59$\AA$ for the F466N and F470N filters, respectively) is then utilised to determine the minimum and maximum redshifts probed, assuming a Tophat transmission function; we apply corrections to account for the true shape of the filter transmission function in Section \ref{sec:filter_corr}. The total JELS observed area is $\sim$63 arcmin$^{2}$, with 1.5 per cent masked out due to stellar contamination \citepalias[see Section 3.3.2 of][]{2025MNRAS.541.1348P}. The unmasked survey area and redshift ranges probed give co-moving volumes for the F466N and F470N-detected H$\alpha$ samples of $\sim$11910 and $\sim$11350 Mpc$^{3}$, respectively.

The uncertainties in the number densities are calculated by taking the Poissonian uncertainty \citep[see Eq. 9 and 12 of][]{1986ApJ...303..336G}. In addition, the scatter of sources between adjacent bins due to luminosity measurement uncertainties is estimated using Monte Carlo sampling, and this is added in quadrature to the Poissonian uncertainty for each bin. At this stage, we compute two separate LFs for each narrow-band filter, ready to then correct for completeness (see Section \ref{sec:completeness}) and the filter transmission profile (see Section \ref{sec:filter_corr}).

\begin{figure}
    \centering
    \includegraphics[width=\linewidth]{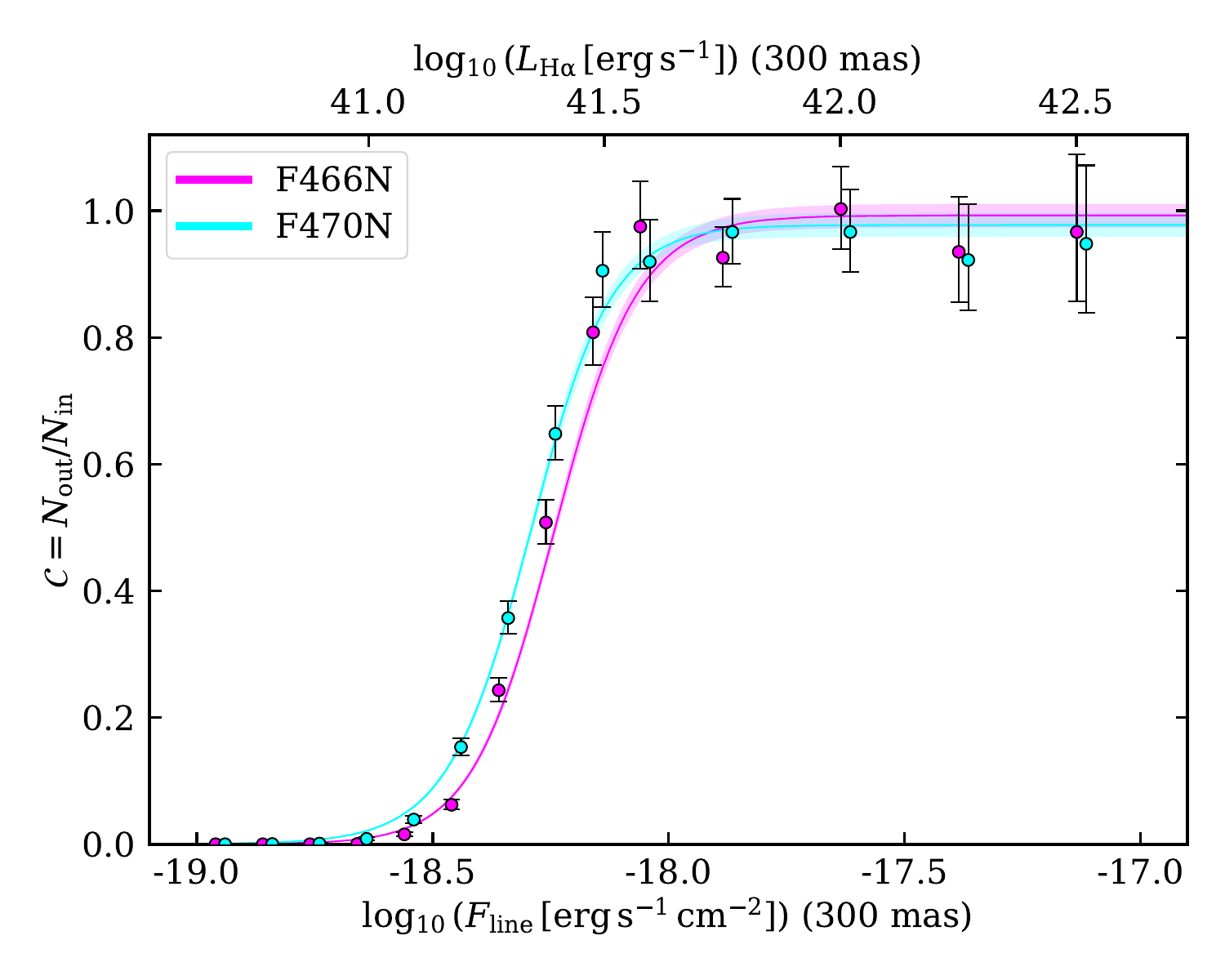}

    \caption{The completeness correction factor ($\mathcal{C}$), derived from the completeness simulations of the F466N and F470N imaging (see Section \ref{sec:completeness}). We calculate $\mathcal{C}$ from the ratio of histogram number counts for the injected sources recovered ($N_{\rm{out}}$) over to those inputted ($N_{\rm{in}}$), binned as a function of emission line flux (lower axis) and propagated luminosity (upper axis), measured in 0.3 arcsec diameter apertures. The uncertainties in $N_{\rm{in}}$ are Poissonian \citep{1986ApJ...303..336G}. The uncertainties in $N_{\rm{out}}$ follow a Poisson-binomial distribution, treating each recovered source as contributing probabilistically to each flux bin according to its recovered line flux uncertainty, following the methodology of \citet{2022ApJS..260....1F}. The fractional uncertainties for $N_{\rm{in}}$ and $N_{\rm{out}}$ are then added in quadrature to give the fractional uncertainty for $\mathcal{C}$. The F466N and F470N-derived $\mathcal{C}$ values are offset by 0.01 dex to the left and right, respectively, for visualisation purposes. Note, $\mathcal{C}>1$ is possible due to bin migration between the input and recovered line flux values. The simulations account for effective completeness due to the detection and selection of the `excess source' sample (see Section \ref{sec:nb_cat}), as well as any flux biases. In addition, the lines show the fitted 1D generalised logistic function (GLF) to the F466N and F470N completeness correction factors.}

    \label{fig:nb_comp_plot}

\end{figure}

\subsection{Completeness corrections}
\label{sec:completeness}

Further corrections to the LFs are required due to the incompleteness of the faintest sources as a result of the survey depth, due to changes in emitter fluxes from processes like Eddington bias, and due to selection functions applied to obtain our sample of H$\alpha$ emitters. Specifically, we assess the impact that applying the source detection (SNR and $r_{e}$ cuts) and excess source selection (colour and $\Sigma$ cuts) criteria of our sample affects the sources we can recover, and their measured fluxes. To do this, we run a series of completeness simulations where we inject model sources into both the F466N and F470N images, assuming the sources are detected at the centre of the narrow-band transmission profiles; we correct for the effect of the filter shape on the LF in Section \ref{sec:filter_corr}. The input parameters for each source include their axis ratio ($0.2 \leq b/a\leq 1$), half-light (effective) radius ($0.5 \leq r_{e} \, [\rm{pixels}] \leq 14.5$) and position angle, that are sampled on uniformly spaced discrete values. In addition, the total narrow-band magnitude is drawn from a uniform distribution between 20 and 28 mag for each source. These magnitudes are rescaled (given the input model parameters) to magnitudes measured in 0.3 arcsec diameter apertures, which is the choice of aperture size used for source detection and selection (see Section \ref{sec:nb_cat}). The input line fluxes are then calculated using both the scaled narrow-band magnitudes and randomly drawing from an $EW_{\rm{H\alpha}}$ distribution to separate the line and continuum components and thus also to obtain information on the F444W flux density/magnitude. We use the log-normal $EW_{\rm{H\alpha}}$ distribution derived by \citet{2024MNRAS.533.1111E}, with mean $\mu_{EW} = 600 \, \AA$ and standard deviation $\sigma_{EW} = 0.27$ dex. This distribution broadly matches the distribution for the JELS sample \citepalias{2025MNRAS.541.1348P}, and we test whether using a broader distribution would influence the results. We find that the choice of $EW_{\rm{H\alpha}}$ distribution made a negligible difference to the completeness function since this is dominated by the detection criteria (primarily the SNR cut). 

\textsc{SExtractor} is then re-run with the same parameters used in obtaining the F466N and F470N detected catalogues \citepalias[see Table 2 of][]{2025MNRAS.541.1348P}. To obtain a sample representative of our H$\alpha$ sample, sources from the input catalogue are drawn from a magnitude distribution that uses a first pass of the JELS H$\alpha$ LF and the same $EW_{\rm{H\alpha}}$ distribution discussed above to convert from luminosity to magnitude space. In addition, we sample from the JELS half-light radius distributions \citep[$r_{e}$; measured using GALFIT from][]{2025MNRAS.544.1412S}. The input simulated sources (selected from the magnitude and $r_{e}$ distribution) recovered in the output \textsc{SExtractor} catalogue are then cross-matched to the input simulated catalogue. From the selected output catalogue of recovered sources, the narrow-band magnitudes are re-measured in 0.3 arcsec diameter apertures and converted into line flux values using the same $EW_{\rm{H\alpha}}$ assigned to a source in the input catalogue. We then apply the detection and narrow-band excess selection criteria (described in Section \ref{sec:nb_cat}) to determine which sources would be recovered as excess sources.

We generate two histograms by binning the number of input simulated sources ($N_{\rm{in}}$) and recovered sources ($N_{\rm{out}}$) as a function of line flux, and define the completeness correction factor $\mathcal{C} = N_{\rm{out}}/N_{\rm{in}}$. The median ratio of recovered-to-input line flux, binned across line flux space, is approximately unity, but with scatter that increases towards fainter line fluxes. Consequently, sources in the input histogram migrate to different line flux bins in the recovered histogram, which our method accounts for when calculating the uncertainties in $\mathcal{C}$ from the recovered line flux uncertainties (see Fig. \ref{fig:nb_comp_plot}). This migration may also arise when faint mock sources are injected near real neighbouring sources or imaging artefacts (including noise spikes), making them more likely to be selected and recovered with higher flux densities than their input values. Therefore, some completeness correction factors can exceed unity.

We show the completeness correction factor $\mathcal{C}$ in Fig. \ref{fig:nb_comp_plot} for the F466N and F470N imaging as a function of line flux; the upper axis indicates the propagated H$\alpha$ luminosity (assuming all of the narrow-band emission arises from the emission line and sources are at $z \sim 6.1$). We fit the $\mathcal{C}$ values with a 1D Generalised Logistic Function (GLF), adopting a Bayesian approach using Markov Chain Monte Carlo (MCMC) sampling with the \textsc{emcee} package \citep{2013PASP..125..306F}. We show in Fig. \ref{fig:nb_comp_plot} a good fit to the data points, and this function is applied to the H$\alpha$ LF on a bin-by-bin basis, accounting for the median aperture correction in each bin (and the median dust correction for the dust-corrected LFs; see Section \ref{sec:halpha_lf_dust_corr}). For the F466N and F470N images, average completeness reaches $\sim$50 per cent at $\log_{10}(F_{\rm{H\alpha}} \,/\, \rm{erg \, s^{-1} \, cm^{-2}}) \sim -18.24$ and $-18.29$, respectively, which corresponds to $\log_{10}(L_{\rm{H\alpha}} \,/\, \rm{erg \, s^{-1}}) \sim 41.39$ and 41.34. There is an increased depth reached in the v1.0 imaging, particularly in the F470N filter. The impact of the completeness corrections on the observed H$\alpha$ LF can be seen in Fig. \ref{fig:obs_lf_data_points}, where we only include luminosity bins with a combined source count-weighted completeness across the F466N and F470N filters (see Section \ref{sec:lf_corr_fit}) of $\langle \mathcal{C} \rangle \gtrsim 0.5$ in the LF analysis. The uncertainty in the completeness function is difficult to quantify due to systematics from our modelling assumptions (e.g. morphology and size distributions), and so we apply an additional uncertainty, corresponding to 20 per cent of the size of the completeness correction applied, in quadrature to the number density uncertainties in all LF calculations \citep[see e.g.][]{2013MNRAS.428.1128S}.

\subsection{Filter profile corrections}
\label{sec:filter_corr}

The filter transmission functions for the F466N and F470N filters are not perfect top hats, which was assumed earlier when calculating the survey volumes. In fact, the volume probed varies with intrinsic luminosity, where luminous H$\alpha$ emitters are detected over a larger volume compared to faint emitters. This is because luminous sources are more likely to be detected in the wings of the transmission function (though they are detected as fainter emitters), whereas low-luminosity emitters will only be detected near the peak of the transmission function and so will occupy a lower effective volume.

\begin{figure}
    \centering
    \includegraphics[width=\linewidth]{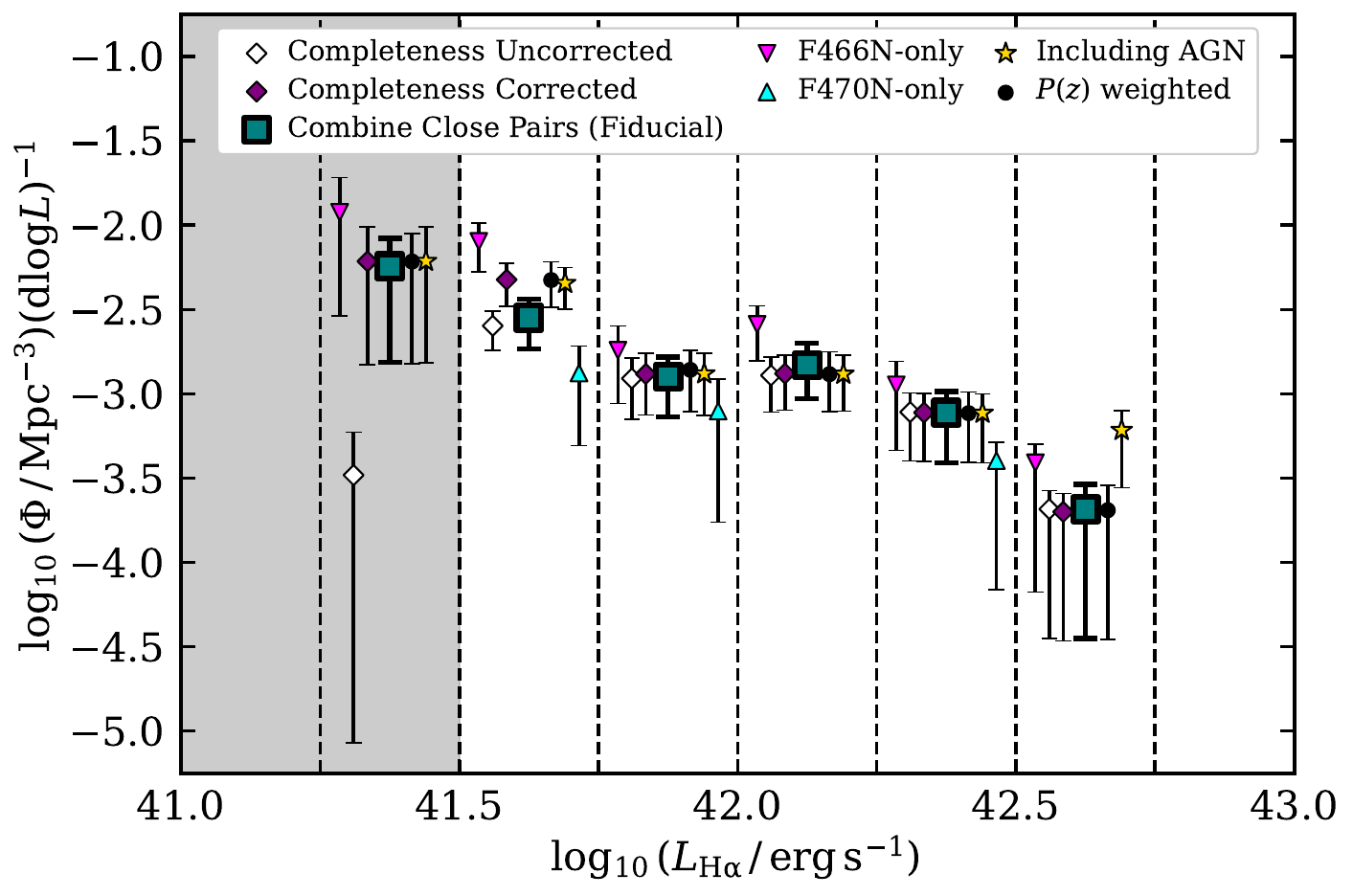}

    \caption{The observed H$\alpha$ LF data points for the JELS H$\alpha$ sample at $z \sim 6.1$. Here, we show number densities uncorrected for source completeness (empty diamond markers), corrected for source completeness
    (purple diamond markers; see Section \ref{sec:completeness}), the F466N and F470N-only number densities for their respective volumes (downward and upwards pointing markers, respectively), the inclusion of the AGN candidates (star markers; see Section \ref{sec:agn}), combining close galaxy pairs (square markers; see Section \ref{sec:mergers}) and applying $P(z)$ weighting on a source-by-source basis (circular markers; see Section \ref{sec:p_z_analysis}). All number densities are filter profile corrected, which results in $\sim$5 -- 15 per cent differences across the observed LF (see Section \ref{sec:filter_corr}). The grey shaded region shows luminosity bins with below 50 per cent completeness. Combining the close pairs results in the most substantial changes in the H$\alpha$ LF number densities (though consistent within uncertainties), particularly at the faint end, and so we adopt this completeness corrected version as our fiducial LF (see Table \ref{table:halpha_lf_data}). We do not apply $P(z)$ weighting since there is a negligible difference in the number densities. These results highlight that the statistical uncertainties and cosmic variance (highlighted by the difference in F466N and F470N-only LFs) are the dominant sources of uncertainty in the H$\alpha$ LF binning and subsequent analysis. Note, small horizontal offsets are made between the different H$\alpha$ LFs for each bin for visualisation purposes: the vertical dotted lines indicate the luminosity bin limits and all data points correspond to the central luminosity of the bin.}

    \label{fig:obs_lf_data_points}

\end{figure}

A series of further simulations are run following the methodology of previous narrow-band surveys \citep[e.g.][]{2009MNRAS.398...75S,2012MNRAS.420.1926S,2013MNRAS.428.1128S,2018MNRAS.476.4725S} where we utilise the preliminary H$\alpha$ LF (which assumes a Tophat volume) accounting for completeness and draw randomly from the best fitting Schechter function (see methodology described in Section \ref{sec:halpha_lf} for the final H$\alpha$ LF). From the randomly drawn luminosities, we assign H$\alpha$ redshifts spanning the full wavelength coverage of the relevant narrow-band in a uniform manner. We then calculate the luminosity recovered by applying the true transmission function and then recompute the LF. The factor between the recovered LF (accounting for the filter transmission function) and the first pass LF (assuming a Tophat transmission function) is then applied as a correction factor to the relevant luminosity bins for the LFs derived in the F466N and F470N filters. Here, we confirm that the number of brighter sources is underestimated compared to the fainter sources. For the observed H$\alpha$ LF (see Fig. \ref{fig:obs_lf_data_points}), the number density decreases by $\sim$5 per cent for faintest bin ($\log_{10}(L_{\rm{H\alpha}} \,/\, \rm{erg \ s^{-1}}) = 41.50 - 41.75$) and increases by $\sim$15 per cent for the brightest bin ($\log_{10}(L_{\rm{H\alpha}} \,/\, \rm{erg \ s^{-1}}) = 42.50 - 42.75$) when applying the filter profile corrections. The simulation is run again on the new filter profile corrected number densities, drawing luminosities from the updated Schechter fit to the new LF, which yielded a nearly identical LF, and showing that the procedure provides a robust correction to the survey volume.

\subsection{AGN contamination}
\label{sec:agn}

As discussed in \citetalias{2025MNRAS.541.1348P}, we identified two H$\alpha$ candidates exhibiting high stellar masses inferred from SED fitting that have compact, PSF-like morphologies \citep[see also Section 2.1 of][]{2025MNRAS.544.1412S}, and were identified as ID 2768 and 7810 in Fig. A1 in \citetalias{2025MNRAS.541.1348P}. With the updated photometry in the v1.0 catalogues and applying the relevant aperture corrections (see Section \ref{sec:halpha_lum}), the inferred stellar masses for these two sources are 10$^{10.8}$ and 10$^{10.0}$ $\rm{M_{\odot}}$, respectively. Integrating the \citet{2024MNRAS.533.1808W} galaxy SMF at $z\sim6$ beyond $\log_{10}(M_{\star}/\rm{M_{\odot}}) = 10$, to obtain the expected number density of star-forming galaxies beyond this mass range, and multiplying by the JELS survey volume results in $\sim$20 per cent chance of detecting one source. This reduces to $\sim$0.006 per cent chance of detecting one source beyond $\log_{10}(M_{\star}/\rm{M_{\odot}}) = 10.8$, implying that there is a low probability that these two sources would be found in the JELS survey, assuming they are star-forming galaxies. In addition, both sources meet several photometric and compactness selection criteria \citep[with blue UV and red optical SEDs; e.g.][]{2024ApJ...964...39G,2025ApJ...986..126K,2026ApJ...997...48B} for `little red dots' \citep[LRDs;][]{2024ApJ...963..129M}, thought to be a newly discovered population of AGN by \emph{JWST} \citep[e.g.][]{2025A&A...697A.189D,2025arXiv250316596N,2025ApJ...980L..27I,2026ApJ..1004..153K}. 

We show the impact that including these AGN candidates have on the observed H$\alpha$ LF in Fig. \ref{fig:obs_lf_data_points}, where there is an $\sim$200 per cent increase in the number density for the brightest bin ($\log_{10}(L_{\rm{H\alpha}} \,/\, \rm{erg \ s^{-1}}) = 42.50 - 42.75$). We note that the dust corrections, assuming no AGN emission in the SED fitting, results in the $L_{\rm{H\alpha}}$ values beyond the dynamic range of the rest of the sample used to construct the dust-corrected H$\alpha$ LFs ($\log_{10}(L_{\rm{H\alpha}} \,/\, \rm{erg \, s^{-1}})>43$). Given the discussion points above, it is likely that significant AGN contribution is unaccounted for in these galaxies. Therefore, we remove these sources from some LF analyses so that we are capturing only star-formation activity, to avoid biasing measurements of $\rho_{\rm{SFR}}$ (see Section \ref{sec:sfrd}). The rest of the sample show more extended morphologies, but we cannot rule out that they contain some AGN activity. \citetalias{2025A&A...694A.178C} found that the percentage of broad H$\alpha$ emitters (with FWHM $>$ 800 km s$^{-1}$) in their sample compared to the number of sources utilised in their LF analysis was 1.56 per cent at $z \sim 4.45$ and 2.72 per cent at $z \sim 5.3$. They found no broad H$\alpha$ emitters in their $z \sim 6.15$ sample. Given that these two AGN candidates make up $\sim$5 per cent of the JELS H$\alpha$ sample, the percentage of AGN removal in this work is reasonable when compared to the statistics from previous studies.

\subsection{Combining galaxy close pairs}
\label{sec:mergers}

Other studies of emission line galaxies have found evidence of clustering on small scales through autocorrelation function analysis, including an excess on scales of $<$1 -- 2 arcsec \citep[e.g.][]{2023ApJ...950...67M,2024ApJ...974..275E,2026ApJ...997..207L}. These scales have typically defined the maximum cross-matching distances for galaxy pairs/groups to be combined into single systems. However, the limited number statistics makes it difficult to quantify the small-scale clustering of the JELS H$\alpha$ sample (see further discussion in Section \ref{sec:cv}). Despite this, we find three pairs of H$\alpha$ emitters with cross-matched separation distances $<$1 arcsec (or $<5.7$ kpc), then find a random distribution for cross-matches at distances $>10$ arcsec \citep[see e.g. Fig. 4 of][]{2023ApJ...950...67M}, but no cross-matches on scales between 1 and 10 arcsec. This could indicate that these close pairs of H$\alpha$ emitters in our JELS sample are examples of small-scale clustering.

How these close pairs are treated, as individual sources or as components of larger galaxies, will impact the corresponding H$\alpha$ LF. We quantify this impact by combining these close pairs into galaxy systems and re-computing the observed H$\alpha$ LF. Due to the decrease in overall sample size (37 to 34) from combining the close pairs and adding their individual H$\alpha$ luminosities, we find a flattening of the observed H$\alpha$ LF (see Fig. \ref{fig:obs_lf_data_points}). The most significant changes in the number densities are the $\sim$40 per cent reduction for the $\log_{10}(L_{\rm{H\alpha}} \,/\, \rm{erg \ s^{-1}}) = 41.50 - 41.75$ bin and the $\sim$12 per cent increase for the $\log_{10}(L_{\rm{H\alpha}} \,/\, \rm{erg \ s^{-1}}) = 42.00 - 42.25$ bin, but these differences are still consistent with the uncertainty in the number densities when not combining the close pairs. Following the methodology of other H$\alpha$ LF studies (e.g. \citetalias{2025A&A...694A.178C}; \citetalias{2025ApJ...987..186F}; \citealt{2026ApJ...997..207L}), we choose to merge our three H$\alpha$ emitter pairs in the LF analysis by summing the total $L_{\rm{H\alpha}}$ measurements for a given system.

\subsection{Cosmic variance}
\label{sec:cv}

Due to the narrow wavelength (and hence redshift) range probed by narrow-band imaging, we are probing smaller cosmic volumes than equivalent slitless spectroscopic surveys \citepalias[e.g.][]{2025A&A...694A.178C} or medium-band selections. Therefore, our analysis will be affected by cosmic variance (CV). The impact of CV depends on both the survey dimensions and the clustering (bias) of the galaxy sample, which in turn is largely driven by galaxy stellar mass. We use the work from \citet{2011ApJ...731..113M} to estimate the CV, yielding a CV range of $\sim$92 -- 186 per cent for $\log_{10}(M_{\star}/\rm{M_{\odot}}) \sim 7.5$ -- 9.5 (the approximate stellar mass range of our JELS H$\alpha$ sample). This corresponds to an increase in the LF number density uncertainties by $\sim$0.30 dex.

We also estimate CV directly from the difference in both sample size and LF results for the F466N and F470N filters, individually. We have identified 31 and 6 H$\alpha$ emitters, detected in the F466N and F470N filters, respectively. Given that both filters select H$\alpha$ emitters to the same approximate sensitivity and over a similar cosmic volume, we can approximate the survey as two different sight-lines: the difference in number count between these could be evidence of significant clustering for the JELS sample; c.f. also Ossa-Fuentes et al. (\emph{submitted}). Firstly, we calculate CV by taking the factor difference in H$\alpha$ number count between filters and dividing by two, since we combine the F466N and F470N LFs (see Section \ref{sec:lf_corr_fit}). This results in an increase in the uncertainties of global properties by $\sim$0.41 dex. Alternatively, we take the F466N and F470N-only LFs (see Fig. \ref{fig:obs_lf_data_points}) and infer the characteristic number densities $\Phi_{\rm{H\alpha}}^{\star}$ from the best-fitting Schechter functions to the H$\alpha$ LFs (see description in Section \ref{sec:halpha_lf}) and then estimate the CV assuming $\Delta \Phi_{\rm{H\alpha}}^{\star}/2$ and find an uncertainty increase of $\sim$0.30 dex, comparable to predictions from \citet{2011ApJ...731..113M}.

Our sample is not sufficiently large to perform clustering analysis, such as the two-point correlation function, and our above estimates do not account for the dependence of CV on the H$\alpha$ luminosity \citep[e.g.][]{2010MNRAS.404.1551S,2017MNRAS.469.2913C} and/or stellar mass \citep[e.g.][]{2018MNRAS.475.3730C}. Given that clustering has been observed in the COSMOS field at this epoch \citep[e.g.][\citetalias{2025A&A...694A.178C}, Ossa-Fuentes et al. \emph{submitted}]{2024MNRAS.527.6591B,2024ApJ...974...41H} and that we likely require larger survey volumes to overcome CV (see \citealt{2015MNRAS.451.2303S}; \citealt{2024ApJ...961..102M}), we acknowledge that the true impact of CV is still unknown for this sample and so do not incorporate the above estimates into inferences of global properties calculated from the LF analysis. However, we assume the above estimates are upper limits on CV given that the results in Section \ref{sec:halpha_lf} show good agreement with previous LF analyses at similar redshifts \citepalias[e.g.][]{2025A&A...694A.178C,2025ApJ...987..186F}.

\subsection{Redshift posterior analysis}
\label{sec:p_z_analysis}

As discussed in Section \ref{sec:halpha_sources}, we apply photometric redshift cuts to the narrow-band excess source sample to obtain different populations of emission line galaxies, including our $z \sim 6.1$ H$\alpha$ emitter sample. The uncertainties on the inferred redshifts are smaller than the redshift range probed by the narrow-band transmission profiles \citepalias[as discussed in Section 4.2 of][]{2025MNRAS.541.1348P}, but could be prone to degeneracies with other emission lines, particularly at low SNR, which is common for the highest redshift galaxy samples. Therefore, we first assess the robustness of the H$\alpha$ classifications using the redshift posterior distributions, $P(z)$, and test the impact on sample selections. The results of these tests can be found in Appendix \ref{app:pz_fractions}. In summary, there is little degeneracy between the redshift solutions for the high-redshift ($z_{\rm{1,median}} > 5.5$) emission line galaxy samples, where the integrated $P(z) > 0.95$ around the primary redshift solutions for the vast majority of the sample. We find a negligible probability that the high-redshift sample is at low-redshift ($z < 5.5$). Lastly, we find that the low-redshift ($z_{\rm{1,median}} < 5.5$) sample has $<0.1$ per cent probability of being at higher redshift ($z > 5.5$). Therefore, we are confident that our redshift selection criteria do not significantly affect the completeness of our H$\alpha$ sample utilised for the LF analysis, with the five H$\alpha$ emitters with spectroscopic data (see Section \ref{sec:halpha_lum}) providing further validation of the photometric redshifts.

To verify the above results, we repeated the LF analysis by binning the fraction of $P(z)$ present in the H$\alpha$ redshift window for each source rather than on individual sources meeting the photo-$z$ criteria discussed in Section \ref{sec:halpha_sources}. We find negligible differences in $\log_{10}(\Phi)$ results (see Fig. \ref{fig:obs_lf_data_points}) with the largest correction being to the $\log_{10}(L_{\rm{H\alpha}} \,/\, \rm{erg \, s^{-1}}) = 41.75 \, - 42.00$ LF bin, resulting in $\sim5$ per cent increase, well within the uncertainty on $\log_{10}(\Phi)$. This bin contains two candidate \oiiia emitters (see sources ID 4212 and 9241 in Table \ref{tab:high_z_pz}) with non-negligible probability of being H$\alpha$ emitters. These results show that the uncertainties are driven mainly by the statistical uncertainties, cosmic variance, and the dust corrections employed, and not by the redshift selection criteria. Therefore, we do not apply $P(z)$ weighting for the H$\alpha$ LF analysis.

\begin{table}
\centering
  \caption{Data points for the narrow-band determined (observed and dust-corrected) H$\alpha$ luminosity functions at $z \sim 6.1$, for the fiducial assumption outlined in Section \ref{sec:lf_corr_fit}. For each luminosity bin $\log (L_{\rm{H\alpha}})$, we show the total number of sources detected $N_{\rm{H\alpha}}$, the average completeness correction factor $\langle \mathcal{C} \rangle$ and the number density $\log_{10}(\Phi \,/\, \rm{Mpc^{-3}})$ along with the associated uncertainties. Note, lower uncertainty values denoted as NaN correspond to bins where the lower uncertainty is greater than the number density value, giving an undefined logarithmic lower uncertainty. These bins are assigned sufficiently large lower uncertainties in log-space, carrying negligible weight in the Schechter function fitting described in Section \ref{sec:halpha_lf}.}
  \label{table:halpha_lf_data}

  \renewcommand{\arraystretch}{1.5} 
  \begin{tabular}{c c c c}
    \hline
    $\log_{10}(L_{\rm{H\alpha}} \,/\, \rm{erg \, s^{-1}})$ & $N_{\rm{H\alpha}}$ & $\langle \mathcal{C} \rangle$ & $\log_{10}(\Phi \,/\, \rm{Mpc^{-3}})$ \\
    \hline
    \multicolumn{4}{c}{Observed} \\
    \hline
    41.25 -- 41.50 & 3 & 0.08 & --2.24 $^{+ 0.17}_{- 0.57}$\\
    41.50 -- 41.75 & 11 & 0.65 & --2.55 $^{+ 0.11}_{- 0.18}$\\
    41.75 -- 42.00 & 7 & 0.95 & --2.90 $^{+ 0.12}_{- 0.24}$\\
    42.00 -- 42.25 & 8 & 0.96 & --2.83 $^{+ 0.13}_{- 0.20}$\\
    42.25 -- 42.50 & 4 & 0.98 & --3.11 $^{+ 0.13}_{- 0.30}$\\
    42.50 -- 42.75 & 1 & 0.98 & --3.68 $^{+ 0.15}_{- 0.77}$\\
    \hline
    \multicolumn{4}{c}{Dust-corrected: $\eta_{\rm{dust}} = 1$} \\
    \hline
    41.25 -- 41.50 & 2 & 0.04 & --2.08 $^{+ 0.17}_{- 0.93}$\\
    41.50 -- 41.75 & 9 & 0.48 & --2.51 $^{+ 0.12}_{- 0.21}$\\
    41.75 -- 42.00 & 7 & 0.78 & --2.82 $^{+ 0.12}_{- 0.26}$\\
    42.00 -- 42.25 & 6 & 0.97 & --2.97 $^{+ 0.12}_{- 0.29}$\\
    42.25 -- 42.50 & 8 & 0.98 & --2.82 $^{+ 0.12}_{- 0.20}$\\
    42.50 -- 42.75 & 1 & 0.98 & --3.70 $^{+ 0.15}_{- 1.06}$\\
    42.75 -- 43.00 & 1 & 0.97 & --3.67 $^{+ 0.15}_{- 1.62}$\\
    \hline
    \multicolumn{4}{c}{Dust-corrected: $\eta_{\rm{dust}} = 0.44$} \\
    \hline
    41.25 -- 41.50 & 2 & 0.04 & --2.10 $^{+ 0.17}_{- 0.87}$\\
    41.50 -- 41.75 & 7 & 0.46 & --2.64 $^{+ 0.13}_{- 0.26}$\\
    41.75 -- 42.00 & 7 & 0.88 & --2.89 $^{+ 0.12}_{- 0.27}$\\
    42.00 -- 42.25 & 8 & 0.97 & --2.85 $^{+ 0.12}_{- 0.23}$\\
    42.25 -- 42.50 & 8 & 0.98 & --2.82 $^{+ 0.12}_{- 0.21}$\\
    42.50 -- 42.75 & 1 & 0.98 & --3.68 $^{+ 0.15}_{- \rm{NaN}}$\\
    42.75 -- 43.00 & 1 & 0.97 & --3.63 $^{+ 0.15}_{- \rm{NaN}}$\\
    \hline
   \end{tabular}
   
\centering
\end{table}

\subsection{The corrected H$\alpha$ luminosity function}
\label{sec:lf_corr_fit}

After implementing aperture, \nii and dust corrections to the $L_{\rm{H\alpha}}$ values, accounting for source completeness, correcting the volume for the filter profiles, removing AGN candidates and combining galaxy close pairs, we re-calculate a corrected H$\alpha$ LF for each filter:

\begin{equation}
\label{eq:lf_corr}
\phi(\log_{10}(L_{c})) = \frac{1}{\Delta(\log_{10} L)} \sum_{|\log_{10} \frac{L_{\rm{i}}}{L_{\rm{c}}}| < \frac{\Delta(\log_{10} L)}{2}} \frac{1}{\mathcal{C}_{\rm{c}} \ V_{\rm{filter},\,c}} 
\end{equation}

\noindent where the same definitions as in Eq. \ref{eq:lf_uncorr} apply, $\mathcal{C}_{\rm{c}}$ is the completeness correction factor and $V_{\rm{filter}, \, c}$ is the corrected survey volume, respectively, for a given luminosity bin centre. We calculate the combined number density as the volume-weighted mean of the F466N and F470N LFs. The corresponding uncertainties are propagated independently in quadrature using the same volume weights. The final observed and dust-corrected H$\alpha$ LF number densities at $z \sim 6.1$ are shown in Table \ref{table:halpha_lf_data}.

\section{Fitting the H$\alpha$ luminosity function at the Epoch of Reionization}
\label{sec:halpha_lf}

We fit the H$\alpha$ LF using the Schechter function \citep{1976ApJ...203..297S}\footnote{There is insufficient evidence to suggest that H$\alpha$ LF at $z\lesssim6.5$ deviates from an exponential decline beyond $L_{\star}$ (e.g. \citealt{2013MNRAS.428.1128S}; \citetalias{2025A&A...694A.178C}, \citetalias{2025ApJ...987..186F}), with only studies of the UV LF finding a preference for a double-power law functional form at $z\gtrsim7$ \citep[e.g.][]{2017MNRAS.466.3612B,2020MNRAS.493.2059B,2023MNRAS.518.6011D,2023MNRAS.524.4586V,2026A&A...707A.239V}}:

\begin{equation}
\label{eq:log_schechter}
\Phi(L) = \frac{\rm{d} n }{\rm{d}(\log_{10} L )} = \ln(10) \ \Phi_{\rm{H\alpha}}^{\star} \ \left( \frac{L}{L_{\rm{H\alpha}}^{\star}} \right) ^{\alpha_{\rm{H\alpha}} + 1} \ e^{-(L/L_{\rm{H\alpha}}^{\star})}
\end{equation}

\noindent where $\Phi_{\rm{H\alpha}}^{\star}$ is the characteristic number density, $L_{\rm{H\alpha}}^{\star}$ is the characteristic luminosity and $\alpha_{\rm{H\alpha}}$ is the faint-end slope. For the fitting procedure, we adopt a Bayesian approach using MCMC sampling with the \textsc{emcee} package (like in Section \ref{sec:completeness}) to fit a Schechter function and infer the best-fitting parameters. Initially, we set broad uniform priors: $-6.0 < \log_{10}(\Phi_{\rm{H\alpha}}^{\star} \,/\, \rm{Mpc^{-3}}) < -2.0$, $41.0 < \log_{10}(L_{\rm{H\alpha}}^{\star} \,/\, \rm{erg \, s^{-1}}) < 45.0$ and $-2.5 < \alpha_{\rm{H\alpha}} < 0$. We perform MCMC sampling with 10000 iterations across 100 chains to obtain posterior distributions for the parameters, and derive the 16th, 50th (median) and 84th percentiles. 

\begin{figure}
    \centering
    \includegraphics[width=\linewidth]{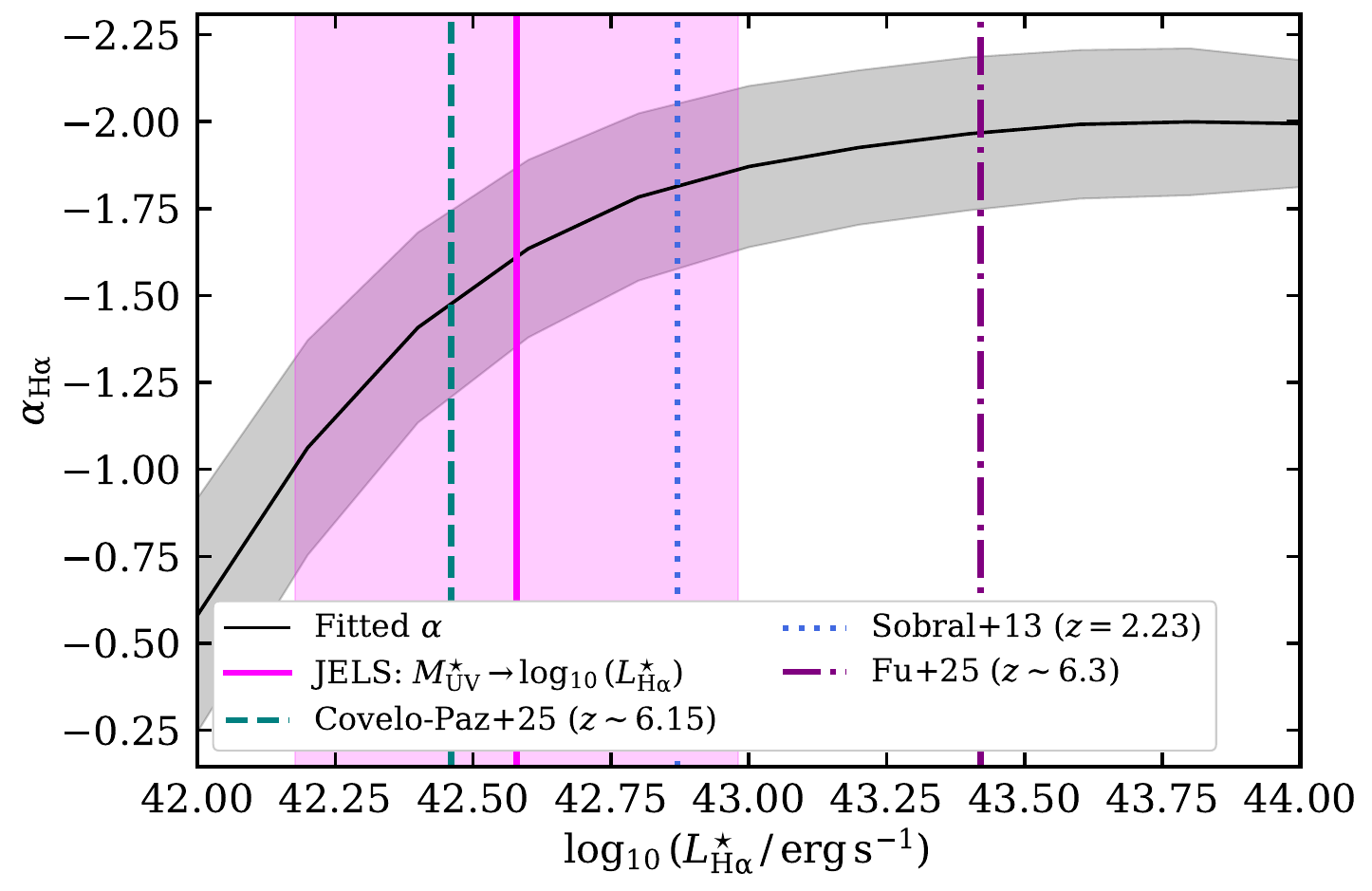}

    \caption{Results of the median faint-end slope ($\alpha_{\rm{H\alpha}}$) as a function of fixed $L_{\rm{H\alpha}}^{\star}$ in the range: $42.0 < \log_{10}(L_{\rm{H\alpha}}^{\star} \,/\, \rm{erg \, s^{-1}}) < 44.0$. The dark grey shaded region shows the 16th and 84th percentiles on the inferred $\alpha_{\rm{H\alpha}}$, for a fixed $L_{\rm{H\alpha}}^{\star}$ value. The vertical lines show the following $L_{\rm{H\alpha}}^{\star}$ choices: i) Translating the measured $M_{\rm{UV}}^{\star}$ from the $z \sim 6$ UV luminosity function \citep{2021AJ....162...47B} into a characteristic H$\alpha$ luminosity $L_{\rm{H\alpha}}^{\star}$ using the median $L_{\rm{UV}}/L_{\rm{H\alpha}}$ for the JELS H$\alpha$ sample (solid line). ii) From the FRESCO $z \sim 6.15$ H$\alpha$ LF \citepalias[dashed line;][]{2025A&A...694A.178C}, which utilise a 3-parameter Schechter fit with weakly informative Gaussian priors centred on results from \citet{2023ApJ...946..117B}. iii) From the HiZELS $z \sim 2.23$ H$\alpha$ LF \citep[dotted line;][]{2013MNRAS.428.1128S}, assuming this value does not evolve across redshift. iv) From the MAGNIF $z \sim 6.3$ H$\alpha$ LF \citepalias[dot-dashed line;][]{2025ApJ...987..186F}, which incorporates the data points from \citetalias{2025A&A...694A.178C} but predicts a significantly larger $L_{\rm{H\alpha}}^{\star}$ than the above studies at $z>6$. These results show broadly that if the break in the LF occurs at luminosities greater than the range our LF probes ($\log_{10}(L_{\rm{H\alpha}}^{\star} \,/\, \rm{erg \, s^{-1}}) > 43.0$), then the faint-end slope must be steep and well-constrained: $-2.00 \lesssim \alpha_{\rm{H\alpha}} \lesssim -1.87$. However, if the break occurs at lower luminosities and at more physically motivated values ($42.2 \lesssim \log_{10}(L_{\rm{H\alpha}}^{\star}) \lesssim 43.0$), then the faint-end slopes are strongly degenerate with $L_{\rm{H\alpha}}^{\star}$, and thus less well-constrained: $-1.87 \lesssim \alpha_{\rm{H\alpha}} \lesssim -1.06$. For the observed LF fitting procedure, we adopt the first inferred $L_{\rm{H\alpha}}^{\star}$ value as the centre of the Gaussian prior with the prior width (1$\sigma$; see definition in the text) shown as the vertical and lighter shaded region showing there is still flexibility allowed when fitting $L_{\rm{H\alpha}}^{\star}$.}

    \label{fig:2_param_schechter_test}

\end{figure}

Due to the degeneracies between Schechter function parameters and the limited dynamic range probed by the JELS sample, we find that $L_{\rm{H\alpha}}^{\star}$ is poorly constrained when all parameters are fitted independently \citepalias[this is also reported in the H$\alpha$ LF results from MAGNIF at $z \sim 6.3$;][]{2025ApJ...987..186F}. To address this, we assume a more informative prior on $L_{\rm{H\alpha}}^{\star}$, while allowing the same freedom in parameter space for fitting $\Phi_{\rm{H\alpha}}^{\star}$ and $\alpha_{\rm{H\alpha}}$. To investigate the importance of the $L_{\rm{H\alpha}}^{\star}$ prior, we first perform 2-parameter fits to $\Phi(L | \Phi_{\rm{H\alpha}}^{\star}, \alpha_{\rm{H\alpha}})$, while holding $L_{\rm{H\alpha}}^{\star}$ fixed at a range of different luminosities, $42.0 < \log_{10}(L_{\rm{H\alpha}}^{\star} \,/\, \rm{erg s^{-1}}) < 44.0$. Given that one goal of the JELS survey is to probe the faintest H$\alpha$ emission line galaxies, we investigate how our choice of $L_{\rm{H\alpha}}^{\star}$ impacts the posterior distributions on the fitted faint-end slope $\alpha_{\rm{H\alpha}}$. The results, shown in Fig. \ref{fig:2_param_schechter_test}, indicate that if $L_{\rm{H\alpha}}^{\star}$ is fixed to a luminosity beyond the dynamic range of our data points ($\log_{10}(L_{\rm{H\alpha}} \,/\, \rm{erg \, s^{-1}}) > 43$), then $\alpha_{\rm{H\alpha}}$ is well-constrained and largely independently of $L_{\rm{H\alpha}}^{\star}$. However, for lower values of $L_{\rm{H\alpha}}^{\star}$, there is a significant degeneracy between $\alpha_{\rm{H\alpha}}$ and $L_{\rm{H\alpha}}^{\star}$.

Studies utilising \emph{HST} have shown evidence that the characteristic SFR ($\rm{SFR_{\star}}$) does not evolve strongly from $z \sim 2$ to $z \sim 6$ \citep[e.g.][]{2015ApJ...803...34B,2016MNRAS.456.3194P,2016ApJ...833..254S} which suggests that $L_{\rm{H\alpha}}^{\star}$ should be comparable to the value determined at lower redshift, $\log_{10}(L_{\rm{H\alpha}}^{\star} \,/\, \rm{erg \ s^{-1}}) = 42.87$ at $z=2.23$ \citep{2013MNRAS.428.1128S}. With the advent of \emph{JWST}, there are also now constraints on H$\alpha$ LF at $z>6$ through untargeted spectroscopic surveys, including from the First Reionization Epoch Spectroscopically Complete Observations \citepalias[FRESCO;][]{2025A&A...694A.178C} and the Medium-band Astrophysics with the Grism of NIRCam in Frontier Fields \citepalias[MAGNIF;][]{2025ApJ...987..186F}. However, there is significant disagreement on the inference of $L_{\rm{H\alpha}}^{\star}$, with \citetalias{2025A&A...694A.178C} finding $\log_{10}(L_{\rm{H\alpha}}^{\star} \,/\, \rm{erg \ s^{-1}}) = 42.46$ and \citetalias{2025ApJ...987..186F} finding that $\log_{10}(L_{\rm{H\alpha}}^{\star} \,/\, \rm{erg \ s^{-1}}) = 43.42$, for the observed H$\alpha$ LF. 

An alternative determination of $L_{\rm{H\alpha}}^{\star}$ can be made using our JELS H$\alpha$ sample, where the sample median H$\alpha$-to-UV luminosity ratio ($L_{\rm{H\alpha}}$/$L_{\rm{UV}}$) can be utilised to convert the characteristic UV luminosity at this epoch \citep[$M_{\rm{UV}}^{\star}$;][]{2021AJ....162...47B} into a characteristic H$\alpha$ luminosity, $L_{\rm{H\alpha}}^{\star}$. Specifically, we run 10$^{6}$ Monte Carlo iterations drawing from Gaussian distributions centred on each source's measured $L_{\rm{H\alpha}}$ and $L_{\rm{UV}}$ values, with standard deviations set to the uncertainties in the luminosities. From the Monte Carlo iterations, we obtain a result where $\log_{10}(L_{\rm{H\alpha}}^{\star} \,/\, \rm{erg \, s^{-1}})$ = 42.58 $\pm$ 0.40, for the dust uncorrected H$\alpha$ luminosities. This range incorporates previously determined characteristic luminosity values from UV-based $M_{\rm{UV}}^{\star}$ and directly measured H$\alpha$ determinations of $L_{\rm{H\alpha}}^{\star}$ (e.g. \citealt{2013MNRAS.428.1128S}; \citealt{2021AJ....162...47B}; \citetalias{2025A&A...694A.178C}) as shown in Fig. \ref{fig:2_param_schechter_test}. Given these results, we fit a Schechter function to the observed H$\alpha$ LF (see the dust-correction methodology in Section \ref{sec:halpha_lf_dust_corr}), adopting the physically motivated Gaussian prior centred on $\log_{10}(L_{\rm{H\alpha}}^{\star} \,/\, \rm{erg \ s^{-1}})=42.58$ and with a standard deviation of 0.40, while maintaining the uniform priors for $\alpha_{\rm{H\alpha}}$ and $\Phi_{\rm{H\alpha}}^{\star}$, as discussed in Section \ref{sec:lf_corr_fit}.

The best-fitting Schechter parameters for each H$\alpha$ LF can be found in Table \ref{table:halpha_lf_schechter_params}. Note, we also separately fit the LF including data from \citetalias{2025A&A...694A.178C} to better constrain the fitting on $L_{\rm{H\alpha}}^{\star}$ and hence the faint-end slope $\alpha_{\rm{H\alpha}}$, given the degeneracy between the two parameters (see discussion in Section \ref{sec:alpha_evolution}). These results can also be found in Table \ref{table:halpha_lf_schechter_params}. The inferred H$\alpha$ LFs (both observed and dust-corrected versions) and their fitted Schechter functions are also shown in Fig. \ref{fig:halpha_lfs}. 

\begin{table}
\centering
\caption{Best fitting Schechter function parameters for the observed and dust-corrected H$\alpha$ luminosity functions. The quoted values are the median of the posterior distributions, and the lower and upper uncertainties correspond to the difference between the median and 16th and 84th percentiles, respectively. The FRESCO H$\alpha$ LF data points are added to better constrain the bright end of the JELS observed and dust-corrected LFs and hence constrain the faint-end slope $\alpha_{\rm{H\alpha}}$. However, we note that they assume $\eta_{\rm{dust}}=1$ for their inferences on $A_{\rm{H\alpha}}$, which could bias results when incorporated into the dust-corrected LF assuming $\eta_{\rm{dust}}=0.44$.}
\label{table:halpha_lf_schechter_params}

\renewcommand{\arraystretch}{1.7} 
\begin{threeparttable}
\setlength{\tabcolsep}{2pt} 
\newcolumntype{L}[1]{>{\centering\arraybackslash}m{#1}}

\begin{tabular}{L{2.0cm} c c c}
\hline
Survey & $\log_{10}(L_{\rm{H\alpha}}^{\star} \,/\, \rm{erg\,s^{-1}})$ & $\log_{10}(\Phi_{\rm{H\alpha}}^{\star} \,/\, \rm{Mpc^{-3}})$ & $\alpha_{\rm{H\alpha}}$\\
\hline
\multicolumn{4}{c}{Observed} \\
\hline
JELS & 42.56 $^{+0.68}_{-0.43}$ & --3.50 $^{+0.61}_{-0.96}$ & --1.58 $^{+0.68}_{-0.40}$\\
JELS+FRESCO & 42.77 $^{+0.60}_{-0.38}$ & --3.81 $^{+0.57}_{-0.84}$ & --1.72 $^{+0.45}_{-0.28}$\\
\hline
\multicolumn{4}{c}{Dust-corrected: $\eta_{\rm{dust}}=1$} \\
\hline
JELS & 42.62 $^{+0.82}_{-0.44}$ & --3.49 $^{+0.61}_{-1.11}$ & --1.56 $^{+0.69}_{-0.40}$\\
JELS+FRESCO & 43.02 $^{+0.68}_{-0.40}$ & --4.04 $^{+0.60}_{-0.87}$ & --1.79 $^{+0.34}_{-0.22}$\\
\hline
\multicolumn{4}{c}{Dust-corrected: $\eta_{\rm{dust}}=0.44$} \\
\hline
JELS & 42.33 $^{+0.84}_{-0.27}$ & --3.07 $^{+0.30}_{-1.07}$ & --1.06 $^{+0.69}_{-0.67}$\\
JELS+FRESCO & 42.85 $^{+0.77}_{-0.35}$ & --3.76 $^{+0.51}_{-0.94}$ & --1.62 $^{+0.41}_{-0.27}$\\
\hline
\end{tabular}

\end{threeparttable} 
\end{table}

\begin{figure*}
\centering
    \includegraphics[width=\linewidth]{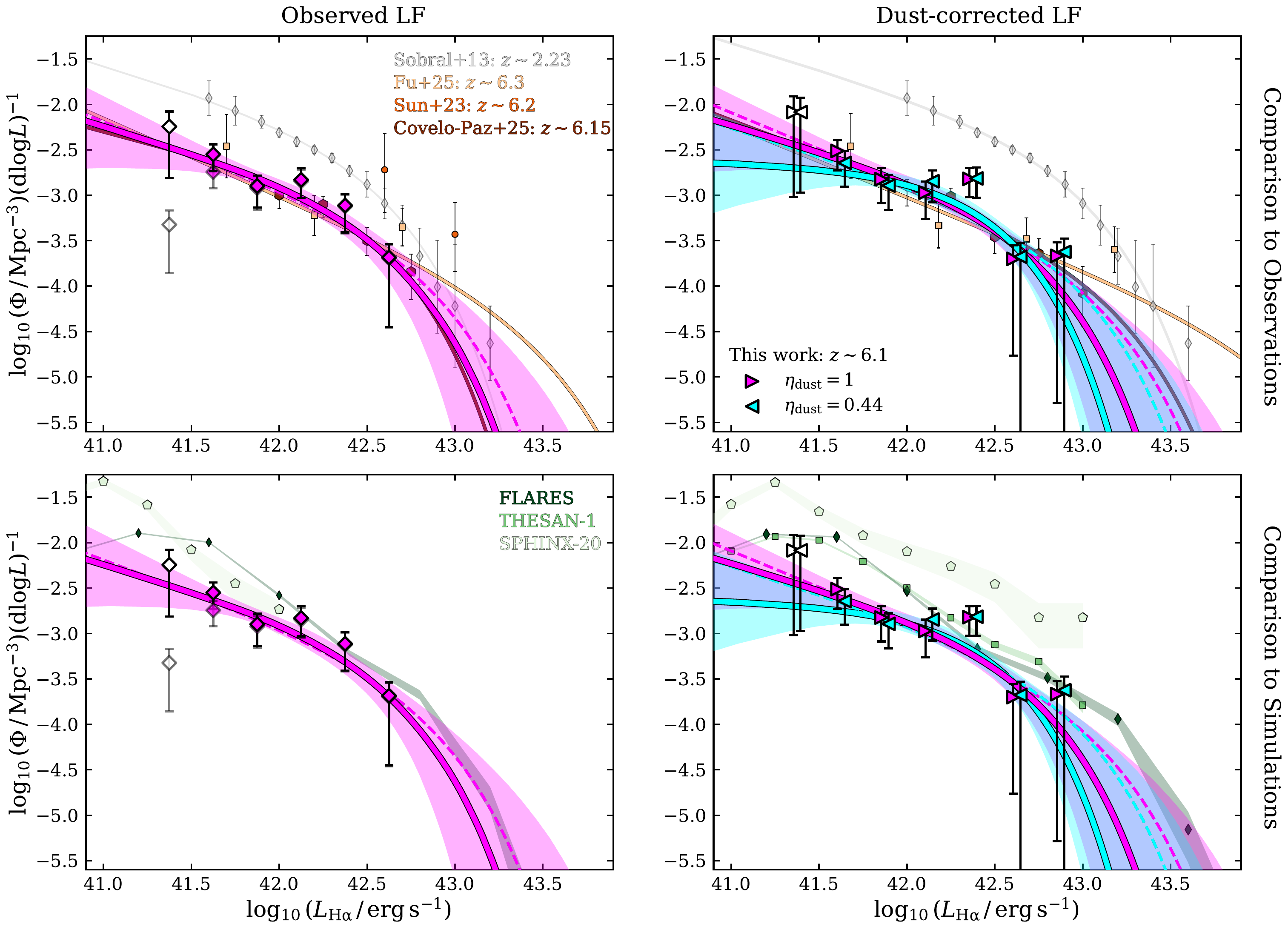}

\caption{Left panels: The observed H$\alpha$ luminosity function as determined by the JELS H$\alpha$ $z \sim 6.1$ sample, with the number densities shown in the pink diamond points. The partially faded points correspond to the number densities not corrected for completeness. Right panels: The dust-corrected H$\alpha$ luminosity functions where the pink right-pointing markers show the dust correction assuming equivalent reddening of the stellar continuum and nebular emission ($\eta_{\rm{dust}}=1$) and the blue left-pointing markers show the dust correction assuming $\eta_{\rm{dust}}=0.44$. The white markers show data points with $\lesssim$50 per cent completeness. These dust-corrected points are offset by 0.02 dex to the left and right, respectively, for visualisation purposes. Note, the attenuation on the stellar continuum at the H$\alpha$ wavelength, $A_{\rm{cont}}(6563 \, \AA)$, is determined from the inferred $A_{V}$ from SED fitting scaled by the \citet{2000ApJ...533..682C} attenuation slope (see Section \ref{sec:halpha_lum}). The fitted Schechter functions (using data points with $\gtrsim$50 per cent completeness) are shown in the same colours with the solid lines, and the shaded regions correspond to the $1\sigma$ uncertainty region (16th to 84th percentiles). The dashed lines with the same colours show the fitted Schechter functions to the combined JELS and FRESCO number densities. The top panels show comparison to observational studies at similar epochs utilising \emph{JWST}/NIRCam slitless spectroscopic surveys (\citealt{2023ApJ...953...53S}; \citetalias{2025A&A...694A.178C}; \citetalias{2025ApJ...987..186F}), and for comparison, also show the lower-redshift narrow-band survey results \citep[HiZELS;][]{2013MNRAS.428.1128S}. The bottom panels show the comparison to cosmological simulations at $z \sim 6$: SPHINX--20 \citep{2023OJAp....6E..44K}, FLARES \citep{2021MNRAS.500.2127L,2024MNRAS.527.7337V} and THESAN--1 \citep{2022MNRAS.514.3857K}. We note that only the intrinsic H$\alpha$ LF results are available for THESAN--1.}

\label{fig:halpha_lfs}

\end{figure*}

\subsection{Fitting the dust-corrected H$\alpha$ luminosity function}
\label{sec:halpha_lf_dust_corr}

As described in Section \ref{sec:halpha_lum}, we correct the $L_{\rm{H\alpha}}$ values for dust extinction by scaling our inferred SED-fitted $A_{V}$ values by the \citet{2000ApJ...533..682C} attenuation slope to obtain $A_{\rm{cont}}(6563 \, \AA)$ and applying the relevant $\eta_{\rm{dust}}=1$ and 0.44 factors to obtain $A_{\rm{H\alpha}}$. We also fold the uncertainties in SED-derived $A_{V}$ values into the dust-corrected $L_{\rm{H\alpha}}$, meaning the uncertainties in LF number densities also account for the uncertainties in the dust corrections and are incorporated into the Monte Carlo sampling method described in Section \ref{sec:halpha_lum}. A Schechter function is then fit to each version of the dust-corrected LF using the same method described in Section \ref{sec:halpha_lf} but with the Gaussian prior on $L_{\rm{H\alpha}}^{\star}$, driven by the dust-corrected $L_{\rm{H\alpha}}$ values rather than the observed values. Specifically, using the conversion from $M_{\rm{UV}}^{\star}$ at $z \sim 6$ \citep{2021AJ....162...47B}, we obtain the following $L_{\rm{H\alpha}}^{\star}$ priors: i) $\eta_{\rm{dust}} = 1$: $\log_{10}(L_{\rm{H\alpha}}^{\star} \,/\, \rm{erg \, s^{-1}})$ = 42.67 $\pm$ 0.52 and ii) $\eta_{\rm{dust}} = 0.44$: $\log_{10}(L_{\rm{H\alpha}}^{\star} \,/\, \rm{erg \, s^{-1}})$ = 42.69 $\pm$ 0.61. Note, the larger standard deviation values on the Gaussian priors originate from the propagated uncertainties in the inferred $A_{V}$ values. The dust-corrected H$\alpha$ LFs and the best-fitting Schechter function parameters can be found in Tables \ref{table:halpha_lf_data} and \ref{table:halpha_lf_schechter_params}, respectively, and these LFs are visualised in the right panels of Fig. \ref{fig:halpha_lfs}. We also include the \citetalias{2025A&A...694A.178C} dust-corrected LF data points into our fits to further constrain the faint-end slope $\alpha_{\rm{H\alpha}}$ (see discussion in Section \ref{sec:alpha_evolution}). We note that both the JELS-only and the combined JELS and FRESCO LFs show $L_{\rm{H\alpha}}^{\star}$ values that are largely prior driven, given that both the values and uncertainties are similar to the adopted priors.

\subsection{Comparison to observations}
\label{sec:halpha_lf_obs_comparison}

In Fig. \ref{fig:halpha_lfs}, we present the observed H$\alpha$ LF results at $z \sim 6.1$. We compare to a series of H$\alpha$ LF results at $z>6$ from \emph{JWST} surveys utilising slitless spectroscopy: i) FRESCO \citepalias{2025A&A...694A.178C}, ii) Commissioning NIRCam grism data \citep{2023ApJ...953...53S}, and iii) MAGNIF \citepalias{2025ApJ...987..186F}. In addition, we show the narrow-band determined H$\alpha$ LF at $z \sim 2.23$ \citep{2013MNRAS.428.1128S}. Our LF determination shows generally good agreement within uncertainties and in overlapping luminosity dynamic range with other studies, mainly from \citetalias{2025A&A...694A.178C}. Disagreement in the number densities occurs at the bright end ($\log_{10}(L_{\rm{H\alpha}} \,/\, \rm{erg \, s^{-1}}) = 42.50 - 42.75$) when comparing results from \citetalias{2025ApJ...987..186F} to this study and \citetalias{2025A&A...694A.178C}. In addition, results from \citet{2023ApJ...953...53S} predict higher number densities in their observed H$\alpha$ LF compared to our results and others at this epoch. However, their uncertainties are substantial ($\sim$0.5 dex for each bin) due to small-number statistics. We gain $\sim 0.5$ dex luminosity coverage at the faint end of the LF compared to \citetalias{2025A&A...694A.178C}, reaching a depth comparable to \citetalias{2025ApJ...987..186F}, which draws on the power of gravitational lensing to reach such faint luminosities. This demonstrates the power of JELS and the need for narrow-band surveys in obtaining faint populations of emission line galaxies and constraining the faint end of emission line LFs (see further discussion about the faint-end slope in Section \ref{sec:alpha_evolution}).

As discussed in Section \ref{sec:halpha_lum}, assuming $\eta_{\rm{dust}}=1$ and 0.44 in the SED fitting results in very similar sample median $A_{\rm{H\alpha}}$ values, indicating that the $\eta_{\rm{dust}}$ assumption has little impact on the average dust correction. However, the differences in $A_{\rm{H\alpha}}$ become significant for sources with $\log_{10}(L_{\rm{H\alpha}} \,/\, \rm{erg \, s^{-1}}) \lesssim 42.0$ (as discussed in Section \ref{sec:halpha_lum}). This can be seen in the dust-corrected H$\alpha$ LFs shown in the right-hand panels of Fig. \ref{fig:halpha_lfs}, where assuming $\eta_{\rm{dust}}=0.44$ results in a shallower faint-end slope ($\alpha_{\rm{H\alpha}}$) than for the LF assuming $\eta_{\rm{dust}}=1$. However, we note the number densities remain consistent within uncertainties for both assumptions.

In addition to the number densities for each JELS dust-corrected H$\alpha$ LF being consistent within uncertainties, they are also consistent with results from both \citetalias{2025A&A...694A.178C} and \citetalias{2025ApJ...987..186F}, within the overlapping dynamic range ($41.5 < \log_{10}(L_{\rm{H\alpha}} \,/\, \rm{erg \, s^{-1}}) < 43.0$). The inferred Schechter functions for the JELS dust-corrected H$\alpha$ LFs also agree with results from \citetalias{2025A&A...694A.178C} within the 1$\sigma$ regions across the full luminosity dynamic range as shown in Fig. \ref{fig:halpha_lfs}. However, \citetalias{2025ApJ...987..186F} predict a large number density in their brightest bin ($\log_{10}(L_{\rm{H\alpha}} \,/\, \rm{erg \, s^{-1}}) = 42.93 - 43.43$), and infer $\log_{10}(L_{\rm{H\alpha}}^{\star} \,/\, \rm{erg \, s^{-1}}) = 44.11$ \citepalias[also incorporating data from][]{2025A&A...694A.178C} which is 1.49 and 1.78 dex larger than derived for the JELS H$\alpha$ LFs, assuming $\eta_{\rm{dust}}=1$ and 0.44 respectively, and 1.03 dex larger than results from \citetalias{2025A&A...694A.178C}. Therefore, their Schechter function is inconsistent within the 1$\sigma$ regions of the JELS H$\alpha$ LF at $\log_{10}(L_{\rm{H\alpha}} \,/\, \rm{erg \, s^{-1}}) \gtrsim 43.0$.

These results demonstrate the importance of obtaining accurate and precise dust properties of H$\alpha$ emitters (e.g. from follow-up spectroscopy of the Balmer decrements) across the luminosity dynamic range to constrain the shape of the H$\alpha$ LF at this epoch. This has already been highlighted in previous studies utilising follow-up spectroscopy of high-redshift sources, finding a range of $\eta_{\rm{dust}}$ values, a range of nebular attenuation curves and non-unity dust covering fractions to explain the difference in Balmer and Paschen decrements \citep{2020ApJ...902..123R,2026ApJ...999...15R}. In addition, wider area surveys \citep[e.g. COSMOS-3D wide area slitless spectroscopic (grism) survey;][]{2024jwst.prop.5893K} will better sample the bright end of the H$\alpha$ LF at this epoch, placing tighter constraints on $L_{\rm{H\alpha}}^{\star}$, which currently limits the precision with which we can measure the faint-end slope $\alpha_{\rm{H\alpha}}$ due to degeneracies, even with current faint-end constraints from JELS (see further discussion in Section \ref{sec:alpha_evolution}).

\subsection{Comparison to simulations}
\label{sec:halpha_lf_sim_comparison}

In Fig. \ref{fig:halpha_lfs} (lower panels), we compare these observations to publicly available predictions from hydrodynamical galaxy simulations of the early Universe, specifically: SPHINX--20 \citep{2023OJAp....6E..44K}, FLARES \citep{2021MNRAS.500.2127L,2024MNRAS.527.7337V} and THESAN--1 \citep{2022MNRAS.514.3857K}. Given the large systematic uncertainties in correcting the observational data points (i.e. dust corrections), the limitations in simulation volumes and the difficulties of realistic forward modelling to match observables, performing a detailed quantitative comparison with simulations is beyond the scope of this analysis. However, we highlight a few notable discrepancies. Firstly, when comparing the dust-corrected LFs with the direct simulation output (lower-right panel), regardless of the assumed $\eta_{\rm{dust}}$ for dust-correcting these observations, most simulations over-predict the number density with varying severity. In particular, the SPHINX--20 number densities are consistently $\gtrsim0.5$ dex larger than in the JELS dust-corrected LFs (see further discussions in Sections \ref{sec:sfrd_v1}). The number densities from both FLARES and THESAN--1 show better agreement with the JELS LFs for intermediate luminosities ($42.0 \lesssim \log_{10}(L_{\rm{H\alpha}} \,/\, \rm{erg \, s^{-1}}) \lesssim 43.0$) within uncertainties, but these simulations over-predict the number densities both at lower and higher luminosities.

If we instead compare the observed LF against the equivalent simulation prediction (lower left panel, which adopts the dust attenuation properties of each simulation), it is a more varied picture. FLARES continues to over-predict the H$\alpha$ LF number densities at $\log_{10}(L_{\rm{H\alpha}} \,/\, \rm{erg \, s^{-1}}) \lesssim 42.0$, while SPHINX--20 shows number densities consistent with the JELS LFs within uncertainties. The assumed treatment of dust and emission line modelling within the respective simulations has a significant impact on the corresponding observables. These results (and others studying the H$\alpha$ LF) give further observational constraints for simulations to test galaxy formation models and, in particular, how dust and emission lines are modelled for high-redshift galaxies.

\subsection{Constraining the faint-end slope $\alpha_{\rm{H\alpha}}$ and its evolution with redshift}
\label{sec:alpha_evolution}

As discussed in Section \ref{sec:halpha_lf_obs_comparison}, the JELS H$\alpha$ sample expands the faint-end coverage of the LF by $\sim0.5$ dex compared to results from FRESCO \citepalias{2025A&A...694A.178C}, improving the statistical uncertainties at the faint end. However, the limited cosmic volume probed by our sample means that we obtain poor constraints on $L_{\rm{H\alpha}}^{\star}$, resulting in large degeneracies between the inferred Schechter function parameters and lower precision on the faint-end slope $\alpha_{\rm{H\alpha}}$ (see Table \ref{table:halpha_lf_schechter_params}).

We take the approach of \citetalias{2025ApJ...987..186F} and incorporate both the observed and dust-corrected LF number densities from \citetalias{2025A&A...694A.178C} to better constrain $L_{\rm{H\alpha}}^{\star}$, and so improve the constraints on $\alpha_{\rm{H\alpha}}$. This was justified given the good agreement shown in the number densities in the overlapping luminosity dynamic range, though we note cosmic variance could still be an important factor due to the different observing fields probed, both with reported galaxy overdensities at $z\sim6$ (see Section \ref{sec:cv}). We report the fitted Schechter parameters in Table \ref{table:halpha_lf_schechter_params} and show the Schechter function fits in Fig. \ref{fig:halpha_lfs} as the dashed lines. The combined JELS and FRESCO dust-corrected LFs now show similar shapes, with better constrained 1$\sigma$ regions (see definition in Fig \ref{fig:halpha_lfs}) compared to the JELS-only fits.

For the combined JELS and FRESCO observed H$\alpha$ LF, we infer $\alpha_{\rm{H\alpha}} = -1.72 \, ^{+ 0.45}_{- 0.28}$, which is consistent within uncertainties with the JELS-only results reported by \citetalias{2025A&A...694A.178C}, with fractional uncertainties lower by a factor of $\sim$1.5. Interestingly, we find that the fractional uncertainties on $\alpha_{\rm{H\alpha}}$ are comparable to the FRESCO-only results, initially suggesting no improvement in the precision on $\alpha_{\rm{H\alpha}}$. Since the FRESCO-only results predict a lower $L_{\rm{H\alpha}}^{\star}$ value, the increased precision on their derived $\alpha_{\rm{H\alpha}}$ is not driven by the $\alpha_{\rm{H\alpha}} - L_{\rm{H\alpha}}^{\star}$ degeneracy. We instead test the impact that the Schechter parameter prior choices have on these results and fit the FRESCO-only number densities using the prior choices discussed in Section \ref{sec:halpha_lf}. We obtain $\alpha_{\rm{H\alpha}} = -1.55 \, ^{+ 0.72}_{- 0.45}$, with fractional uncertainties $\sim$2 times larger; this indicates that prior choices could significantly influence the derived uncertainties, but that in a like-for-like comparison, including the JELS number densities increases the precision on $\alpha_{\rm{H\alpha}}$. \citetalias{2025ApJ...987..186F} also combine their observed H$\alpha$ LF results with FRESCO and achieve fractional uncertainties in $\alpha_{\rm{H\alpha}}$ that are a factor of $\sim$1.5 lower than the combined JELS and FRESCO results. However, they infer $L_{\rm{H\alpha}}^{\star}$ that is 0.65 dex larger than for our results, meaning that there is less degeneracy between $\alpha_{\rm{H\alpha}}$ and $L_{\rm{H\alpha}}^{\star}$ and hence lower uncertainties on $\alpha_{\rm{H\alpha}}$ itself. As discussed in Section \ref{sec:halpha_lf_obs_comparison}, this discrepancy in $L_{\rm{H\alpha}}^{\star}$ values is likely driven by the increased number density of the brightest LF bin for MAGNIF ($\log_{10}(L_{\rm{H\alpha}} \,/\, \rm{erg \, s^{-1}}) = 42.93 - 43.43$), which is inconsistent with our results and could be indicative of cosmic variance between surveys.

For the $\eta_{\rm{dust}} = 1$ and 0.44 dust-corrected LFs, we include the \citetalias{2025A&A...694A.178C} dust-corrected number densities as for the observed LF. The addition of the FRESCO data points in the $\eta_{\rm{dust}} = 1$ dust-corrected LF is consistent with their $\eta_{\rm{dust}}$ assumption, but is then inconsistent with the $\eta_{\rm{dust}} = 0.44$ dust-corrected LF. Given the uncertainties involved in the individual dust corrections and the likely non-uniform impact the $\eta_{\rm{dust}}$ assumption has on the LF number densities (seen to cause bigger discrepancies at the faint-end based on the JELS-only results), we apply no correction factors to the \citetalias{2025A&A...694A.178C} data points when included in the $\eta_{\rm{dust}} = 0.44$ dust-corrected LF. Therefore, the above caveats should be considered when interpreting these results. When combining the FRESCO dataset with the $\eta_{\rm{dust}} = 1$ and 0.44 dust-corrected LFs, we obtain $\alpha_{\rm{H\alpha}} = -1.79 \, ^{+ 0.34}_{- 0.22}$ and $\alpha_{\rm{H\alpha}} = -1.62 \, ^{+ 0.41}_{- 0.27}$, respectively. Both $\alpha_{\rm{H\alpha}}$ values are consistent with each other within uncertainties, and with the FRESCO-only and MAGNIF results, highlighting that the larger dust corrections at the faint end have less impact on the inferred $\alpha_{\rm{H\alpha}}$ when the bright end is better constrained.

\begin{figure}
    \centering
    \includegraphics[width=\linewidth]{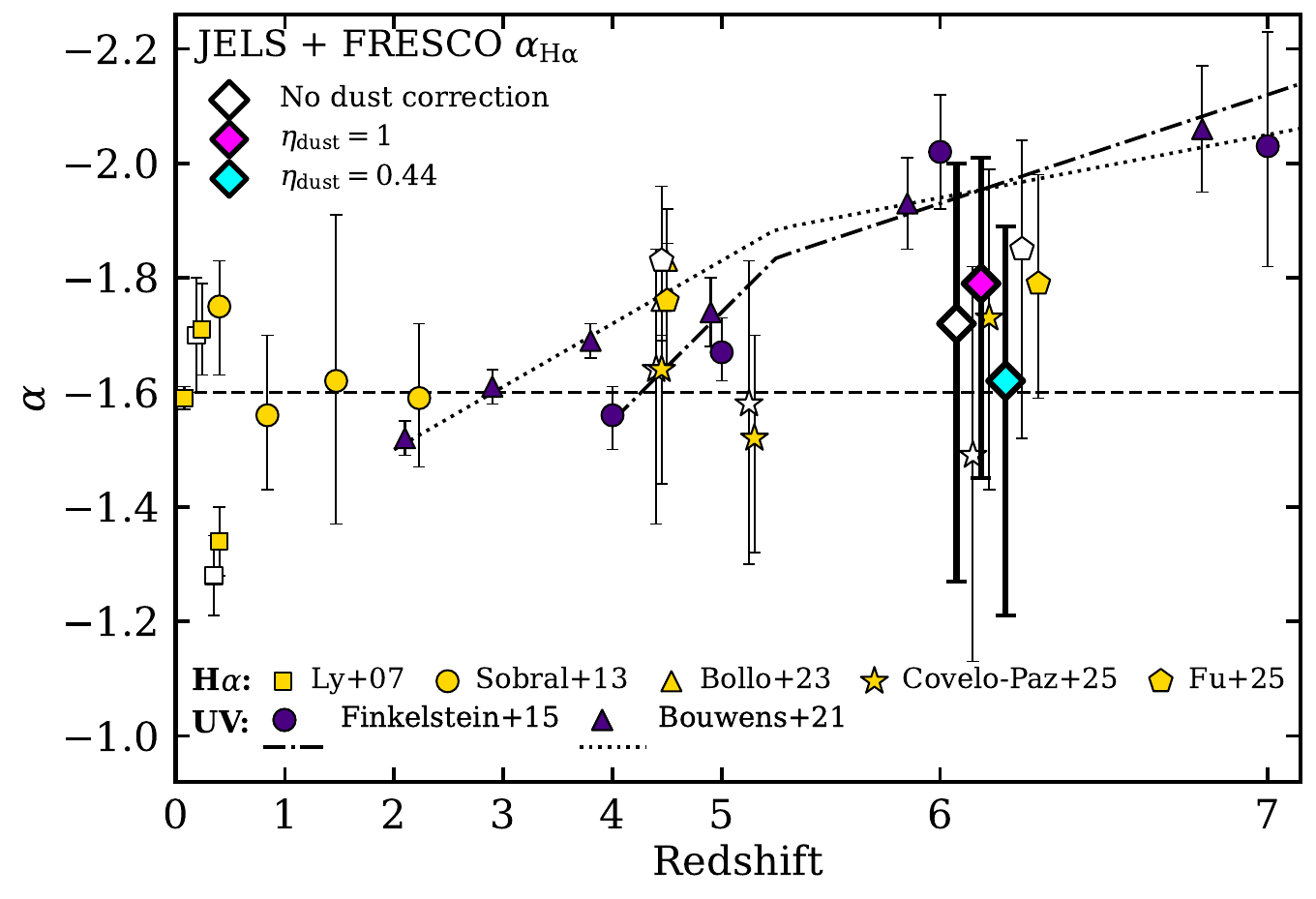}

    \caption{The evolution of the faint-end slope $\alpha_{\rm{H\alpha}}$ for the H$\alpha$ LF from $z \sim 0.24$ to $z \sim 6.3$ (\citealt{2007ApJ...657..738L}; \citealt{2013MNRAS.428.1128S}; \citealt{2023ApJ...946..117B}; \citetalias{2025A&A...694A.178C}; \citetalias{2025ApJ...987..186F}) along with the faint-end slope $\alpha_{\rm{UV}}$ for the UV LF from $z \sim 2.1$ to $z \sim 6.8$ \citep{2021AJ....162...47B}. The empty and filled data points show the inferred faint-end slopes for the observed and dust-corrected LFs, respectively. The fitted $\alpha_{\rm{H\alpha}}$ values utilising the JELS and FRESCO LF data points are shown as the diamond markers. Note, we apply small redshift shifts to results where $\alpha_{\rm{H\alpha}}$ is inferred for both the observed and dust-corrected LFs to better visualise the difference in values and uncertainties. The dashed line shows $\alpha_{\rm{H\alpha}}$ = --1.6, which is the predicted faint-end slope value from previous studies showing no evolution with redshift for $z < 2.5$ \citep[e.g.][]{2013MNRAS.428.1128S} and for $z > 2.5$ \citepalias{2025A&A...694A.178C,2025ApJ...987..186F}. The dotted and dot-dashed lines show fits to the evolution of $\alpha_{\rm{UV}}$ for the relevant UF LF studies (see legend). Note, the redshift axis scaling is increased by a factor of 3 at $z>5.5$ to better highlight the differences in both H$\alpha$ and UV results in this redshift regime.}
        
    \label{fig:alpha_evolution}

\end{figure}

We plot the combined JELS and FRESCO inferred $\alpha_{\rm{H\alpha}}$ for the observed and dust-corrected LF Schechter fits in Fig. \ref{fig:alpha_evolution} to show comparison with literature across redshift. Given the uncertainties in $\alpha_{\rm{H\alpha}}$, the evolution of the faint-end slope with redshift is not clear. The results are consistent with there being no redshift evolution with $\alpha_{\rm{H\alpha}} \sim -1.6$ from low-redshift to the EoR (e.g. \citealt{2007ApJ...657..738L}, \citealt{2013MNRAS.428.1128S}; \citetalias{2025A&A...694A.178C}; \citetalias{2025ApJ...987..186F}), but equally are not inconsistent with the observed evolution in the UV LF faint-end slope $\alpha_{\rm{UV}}$ \citep[c.f.][]{2021AJ....162...47B,2015ApJ...810...71F}. As discussed in Section \ref{sec:halpha_lf_obs_comparison}, these results highlight the need to further constrain the H$\alpha$ LF both above and significantly below $L_{\rm{H\alpha}}^{\star}$ to accurately constrain $\alpha_{\rm{H\alpha}}$ and its evolution from $z\sim2$ to the EoR at $z>6$.

\section{The star formation rate density into the Epoch of Reionization}
\label{sec:sfrd}

\subsection{The luminosity density $\rho_{L_{\rm{H\alpha}}}$}
\label{sec:rho_halpha}

We can now integrate the dust-corrected H$\alpha$ LFs discussed in Section \ref{sec:halpha_lf_dust_corr}, assuming a lower integration limit $L_{\rm{H\alpha, \, lim}}$, to compute the H$\alpha$ luminosity density $\rho_{L_{\rm{H\alpha}}}$:

\begin{equation}
\label{eq:rho_halpha}
\rho_{L_{\rm{H\alpha}}} = \int_{L_{\rm{H\alpha, \, lim}}}^{\infty} L \, \Phi(L) \, \rm{d} L = \Phi_{\rm{H\alpha}}^{\star} \, L_{\rm{H\alpha}}^{\star} \, \Gamma(\alpha_{\rm{H\alpha}}+2, \, L_{\rm{H\alpha, \, lim}}/L_{\rm{H\alpha}}^{\star}),
\end{equation}

\noindent where the Schechter function integral can be solved analytically and $\Gamma$ is the incomplete gamma function:

\begin{equation}
\label{eq:gamma_func}
\Gamma(a,x) = \int_{x}^{\infty} t^{a-1} e^{-t} dt,
\end{equation}

\noindent In this case, $a = \alpha_{\rm{H\alpha}}+2$ and $x = L_{\rm{H\alpha, \, lim}}/L_{\rm{H\alpha}}^{\star}$. For direct comparison to previous H$\alpha$ LF studies \citepalias{2025A&A...694A.178C,2025ApJ...987..186F}, we take $\log_{10}(L_{\rm{H\alpha, \, lim}} \,/\, \rm{erg\,s^{-1}})$ = 40.65 as our standard assumption. This $L_{\rm{H\alpha, \, lim}}$ value follows from the \citet{1998ARA&A..36..189K} UV and H$\alpha$ SFR--luminosity calibrations (see further discussion in Section \ref{sec:sfrd_v1}), where the $M_{\rm{UV}}$ lower integration limit from previous UV LF studies \citep[$M_{\rm{UV, \, lim}} = -17$ or $L_{\rm{UV, \, lim}}$ = 0.03 $L^{\star}_{z=3}$; e.g.][]{2015ApJ...803...34B,2024ApJ...965..169A,2024MNRAS.533.3222D} is converted into the lower limit in the star-formation rate function (SFRF), and then translated to $L_{\rm{H\alpha, \, lim}}$. Our standard $L_{\rm{H\alpha, \, lim}}$ probes $\sim$1 dex fainter than our faintest H$\alpha$ LF bin and minimises the impact of the uncertainty in faint-end slope $\alpha_{\rm{H\alpha}}$ on the inferred $\rho_{L_{\rm{H\alpha}}}$. For additional comparison with literature values \citep[e.g.][]{2013MNRAS.428.1128S}, we also compute $\rho_{L_{\rm{H\alpha}}}$ with $L_{\rm{H\alpha, \, lim}}$ = 0, 0.01 $L_{\rm{H\alpha}}^{\star}$ and 0.03 $L_{\rm{H\alpha}}^{\star}$, where we justify reporting results adopting $L_{\rm{H\alpha, \, lim}}$ = 0, given that lower-redshift ($z<1$) results constrain the faint-end of the H$\alpha$ LF to much lower $L_{\rm{H\alpha}}$ \citep[e.g.][]{2016A&A...591A.151G,2019A&A...631A..10R} while showing no faint-end turn-over \citep[see also discussions for the UV LF;][]{2017ApJ...835..113L,2017ApJ...843..129B}. The assumptions, and corresponding derived values for $\rho_{L_{\rm{H\alpha}}}$, are provided in Table \ref{table:sfrd}. 

For our $\eta_{\rm{dust}} = 1$ dust-corrected H$\alpha$ LF, we find the following range of $\rho_{L_{\rm{H\alpha}}}$ corresponding to $L_{\rm{H\alpha, \, lim}}$ = 0 -- 0.03 $L_{\rm{H\alpha}}^{\star}$: $\log_{10}(\rho_{L_{\rm{H\alpha}}} \,/\, \rm{erg \, s^{-1} \, Mpc^{-3}}) = 39.33 - 39.43$ (0.10 dex). Integrating to zero also increases $\rho_{L_{\rm{H\alpha}}}$ by 0.06 dex compared to the standard assumption, but as discussed above, the uncertainty in $\alpha_{\rm{H\alpha}}$ can have a significant impact on the inferred $\rho_{L_{\rm{H\alpha}}}$. For our $\eta_{\rm{dust}} = 0.44$ dust-corrected H$\alpha$ LF, we find the following range in luminosity density: $\log_{10}(\rho_{L_{\rm{H\alpha}}} \,/\, \rm{erg \, s^{-1} \, Mpc^{-3}}) = 39.29 - 39.32$ (0.03 dex). We note that the variations in $\rho_{L_{\rm{H\alpha}}}$ due to the different $L_{\rm{H\alpha, \, lim}}$ are smaller than the uncertainties in the derived values, which are largely driven by the uncertainty in $\alpha_{\rm{H\alpha}}$. The $\eta_{\rm{dust}}=0.44$ derived $\rho_{L_{\rm{H\alpha}}}$ results are also systematically lower than those for $\eta_{\rm{dust}}=1$, mostly driven by the shallower $\alpha_{\rm{H\alpha}}$. This effect of $\alpha_{\rm{H\alpha}}$ is also seen if we calculate $\rho_{L_{\rm{H\alpha}}}$ using the LFs including the FRESCO dataset: these are 0.07 and 0.10 dex higher than the equivalent for the JELS-only dataset (see Table \ref{table:sfrd}), again due largely to the steeper $\alpha_{\rm{H\alpha}}$.

\begin{table*}
\centering
  \caption{Estimates of the H$\alpha$ luminosity density ($\rho_{L_{\rm{H\alpha}}}$) and the cosmic star-formation rate density ($\rho_{\rm{SFR_{H\alpha}}}$) calculated from both dust-corrected H$\alpha$ LFs (see Section \ref{sec:halpha_lf_dust_corr}) assuming different luminosity integration limits ($L_{\rm{H\alpha, \, lim}}$) and different SFR calibration constants ($\kappa_{\rm{H\alpha}}$). The first row (for each dust-corrected calculation) shows the standard assumption and the value we compare to previous literature, and the subsequent rows show deviations from the standard assumption.}
  \label{table:sfrd}

  \renewcommand{\arraystretch}{1.5} 
  \begin{threeparttable}
  \begin{tabular}{c c c}
  \hline
  Integral Assumptions & $\log_{10}$($\rho_{L_{\rm{H\alpha}}} \,/\, \rm{erg \, s^{-1} \, Mpc^{-3}})$ & $\log_{10}(\rho_{\rm{SFR_{H\alpha}}} \,/\, \rm{M_{\odot}} \, yr^{-1} \, Mpc^{-3})$\\
  \hline
  \multicolumn{3}{c}{$\eta_{\rm{dust}} = 1$} \\
  \hline
  
  Standard assumption $^{\ref{tn:1}}$ & 39.37 $^{+ 0.14}_{- 0.12}$ & --1.93 $^{+ 0.14}_{- 0.12}$\\
  
  Standard assumption + FRESCO $^{\ref{tn:2}}$ & 39.44 $^{+ 0.11}_{- 0.10}$ & --1.85 $^{+ 0.11}_{- 0.10}$\\

  $L_{\rm{H\alpha, \, lim}}$ = 0 & 39.43 $^{+ 0.35}_{- 0.16}$ & --1.87 $^{+ 0.35}_{- 0.16}$\\
  
  $L_{\rm{H\alpha, \, lim}}$ = 0.01 $L_{\rm{H\alpha}}^{\star}$ & 39.37 $^{+ 0.14}_{- 0.12}$ & --1.93 $^{+ 0.14}_{- 0.12}$\\
   
  $L_{\rm{H\alpha, \, lim}}$ = 0.03 $L_{\rm{H\alpha}}^{\star}$ & 39.33 $^{+ 0.13}_{- 0.10}$ & --1.97 $^{+ 0.13}_{- 0.10}$\\

  $\log_{10}(\kappa_{\rm{H\alpha}} \, [\rm{(M_{\odot} \, yr^{-1})/(erg \, s^{-1})}])$ = --41.73 $^{\ref{tn:3}}$ & -- & --2.36 $^{+ 0.14}_{- 0.12}$\\
  
  \hline
  \multicolumn{3}{c}{$\eta_{\rm{dust}} = 0.44$} \\
  \hline
  
  Standard assumption $^{\ref{tn:1}}$ & 39.30 $^{+ 0.16}_{- 0.10}$ & --2.00 $^{+ 0.16}_{- 0.10}$\\
  
  Standard assumption + FRESCO $^{\ref{tn:2}}$ & 39.40 $^{+ 0.14}_{- 0.10}$ & --1.89 $^{+ 0.14}_{- 0.10}$\\

  $L_{\rm{H\alpha, \, lim}}$ = 0 & 39.32 $^{+ 0.26}_{- 0.12}$ & --1.98 $^{+ 0.26}_{- 0.12}$\\
  
  $L_{\rm{H\alpha, \, lim}}$ = 0.01 $L_{\rm{H\alpha}}^{\star}$ & 39.31 $^{+ 0.17}_{- 0.11}$ & --1.99 $^{+ 0.17}_{- 0.11}$\\
   
  $L_{\rm{H\alpha, \, lim}}$ = 0.03 $L_{\rm{H\alpha}}^{\star}$ & 39.29 $^{+ 0.15}_{- 0.10}$ & --2.00 $^{+ 0.15}_{- 0.10}$\\

  $\log_{10}(\kappa_{\rm{H\alpha}} \, [\rm{(M_{\odot} \, yr^{-1})/(erg \, s^{-1})}])$ = --41.73 $^{\ref{tn:3}}$ & -- & --2.43 $^{+ 0.16}_{- 0.10}$\\
   
  \hline
  \end{tabular}

  \begin{tablenotes}
    \item[1] \label{tn:1} Standard assumptions: i) Integrate the LF down to $\log_{10}(L_{\rm{H\alpha, \, lim}} \,/\, \rm{erg \, s^{-1}})$ = 40.65 and ii) Applying the \citet{1998ARA&A..36..189K} SFR calibration ($\log_{10}(\kappa_{\rm{H\alpha}} \, [\rm{(M_{\odot} \, yr^{-1})/(erg \, s^{-1})}])$ = --41.30; see Section \ref{sec:sfrd_v1}) to calculate $\rho_{\rm{SFR_{H\alpha}}}$.
    \item[2] \label{tn:2} Utilising the standard assumptions but also fitting the LF and calculating $\rho_{\rm{SFR_{H\alpha}}}$ by incorporating the data points from the FRESCO results \citepalias{2025A&A...694A.178C}.
    \item[3] \label{tn:3} $\kappa_{\rm{H\alpha}}$ derived instead from \textsc{BPASS} stellar population synthesis (SPS) models, which include binary star populations with stellar metallicity $Z$ = 0.001 (see Section \ref{sec:sfrd_v1}).
  \end{tablenotes}

  \end{threeparttable}  
   
\centering
\end{table*}

\begin{figure*}
    \centering

    \includegraphics[width=\linewidth]{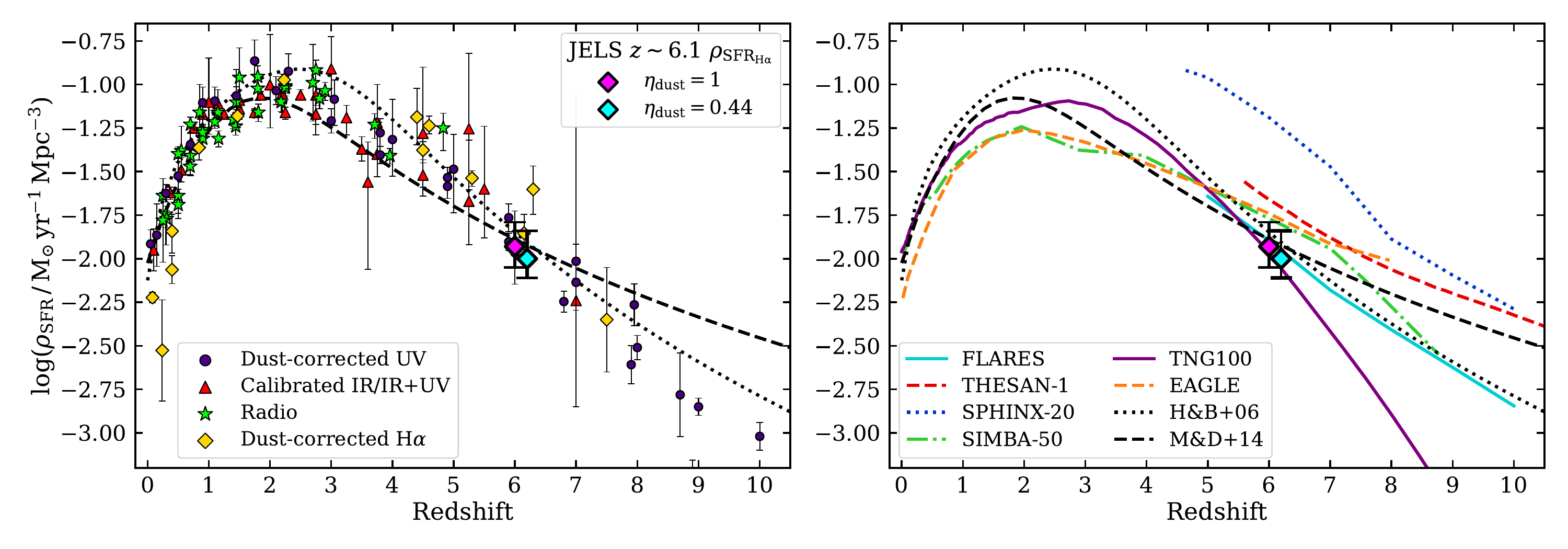}

    \caption{The H$\alpha$-determined star-formation rate density $\rho_{\rm{SFR_{H\alpha}}}$ at $z \sim 6.1$ as constrained by the JELS sample of H$\alpha$ emission line galaxies assuming the relevant dust corrections to the H$\alpha$ LFs (see section \ref{sec:halpha_lf_dust_corr} and Fig. \ref{fig:halpha_lfs}). The left panel shows the compilation of $\rho_{\rm{SFR}}$ measurements from multi-wavelength surveys from $z = 0 - 10$: i) dust-corrected rest-frame UV \citep[circle markers;][]{2014ARA&A..52..415M,2022ApJ...940...55B,2023MNRAS.523.1036B,2023MNRAS.518.6011D,2024MNRAS.533.3222D}, ii) dust-corrected H$\alpha$ (square markers; \citealt{2007ApJ...657..738L}; \citealt{2013MNRAS.428.1128S}; \citealt{2023ApJ...946..117B}; \citealt{2023ApJ...952..143R}; \citetalias{2025A&A...694A.178C}; \citetalias{2025ApJ...987..186F}), iii) Calibrated IR/IR+UV \citep[triangle markers;][]{2014ARA&A..52..415M,2017MNRAS.466..861D,2020A&A...643A...8G,2023MNRAS.518.6142A,2024A&A...681A.118T} and iv) radio \citep[star markers;][]{2011ApJ...730...61K,2017A&A...602A...5N,2022ApJ...927..204E,2023MNRAS.523.6082C}. Note, all measurements from observations are corrected to the \citet{2003PASP..115..763C} IMF and to the same luminosity/SFR integration limit where applicable. The right panel shows the compilation of $\rho_{\rm{SFR}}$ measurements from simulations for $z = 0 - 10$: FLARES \citep{2021MNRAS.501.3289V}, THESAN--1 \citep{2022MNRAS.511.4005K}, SPHINX--20 \citep{2023OJAp....6E..44K} SIMBA--50 \citep{2024MNRAS.534..361S}, Illustris--TNG100 \citep{2018MNRAS.473.4077P} and EAGLE \citep{2017MNRAS.472..919K}. In addition, we plot the evolution of $\rho_{\rm{SFR}}$ with redshift as determined by \citet{2006ApJ...651..142H} and \citet{2014ARA&A..52..415M} with the black dotted and dashed lines, respectively.}

    \label{fig:sfrd_multiwav}

\end{figure*}

\subsection{The star-formation rate density: assuming a constant $\kappa_{\rm{H\alpha}}$}
\label{sec:sfrd_v1}

The star-formation rate density ($\rho_{\rm{SFR_{H\alpha}}}$) can be calculated from the H$\alpha$ luminosity density $\rho_{L_{\rm{H\alpha}}}$, assuming a H$\alpha$ SFR calibration constant, $\kappa_{\rm{H\alpha}}$ in units of ($\rm{M_{\odot}\,yr^{-1}}$)/(erg s$^{-1}$), to give us:

\begin{equation}
\label{eq:sfrd_v1}
\rho_{\rm{SFR_{H\alpha}}} = \kappa_{\rm{H\alpha}} \times \rho_{L_{\rm{H\alpha}}}
\end{equation}

\noindent in units of $\rm{M_{\odot}\,yr^{-1}}\,Mpc^{-3}$. $\kappa_{\rm{H\alpha}}$ is dependent on the properties of the stellar populations of high-redshift galaxies, which are not well understood (see further discussion in Section \ref{sec:sfrd_v2}). Therefore, it is standard in the literature to compare $\rho_{\rm{SFR_{H\alpha}}}$ assuming a constant $\kappa_{\rm{H\alpha}}$ value. We compute $\rho_{\rm{SFR_{H\alpha}}}$ using the commonly adopted calibration constant, $\kappa_{\rm{H\alpha}}$ = 10$^{-41.30}$ ($\rm{M_{\odot}\,yr^{-1}}$)/(erg s$^{-1}$) from \citet{1998ARA&A..36..189K}, which uses the single stellar population synthesis models that have solar metallicity ($Z_{\star} = 0.02$), with an upper mass limit of 100 $\rm{M_{\odot}}$ and a constant SFR over 100 Myr.

Fig. \ref{fig:sfrd_multiwav} shows the measured $\rho_{\rm{SFR_{H\alpha}}}$ for the standard assumption in comparison to previous studies, with literature values corrected to the same IMF and integration limit where appropriate. The plot shows that regardless of the adopted $\eta_{\rm{dust}}$, the $\rho_{\rm{SFR_{H\alpha}}}$ values only differ by 0.07 dex, well within uncertainties, indicating that the difference in LF shape and inferred Schechter function parameters have little impact on global properties derived from the LF. We see excellent agreement with previous UV-determined $\rho_{\rm{SFR}}$ at $z \sim 6$ \citep[e.g.][see zoomed in version around $z\sim6$ in Fig. \ref{fig:sfrd_uv_halpha_zoom}]{2014ARA&A..52..415M,2022ApJ...940...55B}. In addition, we find that the $\rho_{\rm{SFR_{H\alpha}}}$ inference assuming $\eta_{\rm{dust}}=1$ and 0.44 agrees well with \citetalias{2025A&A...694A.178C}. We find less agreement with results from \citetalias{2025ApJ...987..186F}, potentially due to cosmic variance, though we note that the results are only marginally inconsistent within uncertainties (as shown in Fig. \ref{fig:sfrd_multiwav}).

We also compare the $\rho_{\rm{SFR_{H\alpha}}}$, assuming both $\eta_{\rm{dust}}=1$ and 0.44, to results from simulations and find agreement within uncertainties with FLARES \citep{2021MNRAS.501.3289V} and Illustris--TNG100 \citep{2018MNRAS.473.4077P}. Results from THESAN--1 \citep{2022MNRAS.511.4005K}, SIMBA--50 \citep{2024MNRAS.534..361S} and EAGLE \citep{2017MNRAS.472..919K} predict slightly higher $\rho_{\rm{SFR}}$ at $z\sim6$, marginally outwith the uncertainties of the JELS-determined $\rho_{\rm{SFR_{H\alpha}}}$. The differences between these simulation results probably stem from differences in the definitions of SFR (instantaneous or timescale averaged), as well as from some simulations imposing lower limits on galaxy SFRs when computing $\rho_{\rm{SFR}}$. For example, FLARES sets a lower limit on the 100 Myr averaged SFR ($\rm{SFR_{100}} > 0.1 \, \rm{M_{\odot}} \, yr^{-1}$) when computing $\rho_{\rm{SFR}}$, explaining the lower values. In contrast, SIMBA--50, EAGLE and THESAN--1 include the total SFR from all star-forming particles, with no lower limit imposed in their calculations of $\rho_{\rm{SFR}}$, explaining the higher values. Results from SPHINX--20 \citep{2023OJAp....6E..44K} predict substantially higher $\rho_{\rm{SFR}}$ compared to JELS and the other simulations. This may partly reflect the SFR definition and imposed lower limit (10 Myr averaged SFR: $\rm{SFR_{10}}>0.3 \, \rm{M_{\odot} \, yr^{-1}}$), but may also arise from the simulated galaxies having larger stellar masses and hence higher SFRs and dust content, for a given $L_{\rm{H\alpha}}$, than typical galaxies observed at this redshift \citep[see discussion in][]{2026A&A...707A.184K}.

\subsection{The star-formation rate density: $\kappa_{\rm{H\alpha}}$ as a function of $L_{\rm{H\alpha}}$}
\label{sec:sfrd_v2}

As discussed in Section \ref{sec:sfrd_v1}, $\kappa_{\rm{H\alpha}}$ is dependent on the properties of the stellar populations within galaxies at a given epoch. Such properties include, but are not limited to, the IMF shape and the upper mass limit, metallicity ($Z_{\star}$) and the abundance of single and binary stellar populations. The IMF could depend on galaxy environment and could evolve with redshift as the conditions of the Universe change, but this is difficult to measure \citep[see reviews;][]{2018PASA...35...39H,2024ARA&A..62...63H}, and so we do not investigate this property in this work. However, metallicity and the abundance of binary stellar populations are important to consider for galaxy populations into the EoR. Our H$\alpha$ sample primarily probes the low stellar mass regime \citepalias[$\log_{10}(M_{\star}/\rm{M_{\odot}}) < 9.0$;][]{2025MNRAS.541.1348P} and hence we expect lower metallicities for our sample given the evolution of the mass-metallicity relation to higher redshifts \citep[e.g.][]{2021ApJ...914...19S,2023ApJS..269...33N,2024A&A...684A..75C}. In addition, the ionization properties of high-redshift galaxies indicate that they contain populations of massive stars \citep[e.g.][]{2016ApJ...826..159S,2019MNRAS.487.2038C,2020MNRAS.495.4430T}, that are predicted to be predominantly in binary systems \citep[or higher multiplicity systems; e.g.][]{2012Sci...337..444S}. As a result, alternative binary stellar population and spectral synthesis models, including \textsc{BPASS} \citep{2017PASA...34...58E,2018MNRAS.479...75S}, have been developed to model binary and metal-poor stellar populations. 

Given the discussion above, the \citet{1998ARA&A..36..189K} assumption for $\kappa_{\rm{H\alpha}}$ may not be valid for high-redshift galaxy samples, including our H$\alpha$ sample at $z\sim6.1$. \textsc{BPASS} models have been employed to infer $\rm{SFR_{H\alpha}}$ for high-redshift galaxies, usually adopting stellar metallicities $Z_{\star} \sim 0.001 - 0.004$ \citep[e.g.][]{2019ApJ...871..128T,2022ApJ...926...31R,2023ApJ...950L...1S,2024ApJ...977..133C,2025ApJ...984..188K,2025arXiv250810099S}. These lower-metallicity calibrations result in lower $\kappa_{\rm{H\alpha}}$, meaning galaxies are expected to produce higher $L_{\rm{H\alpha}}$ per unit SFR, reflecting the greater ionizing photon production efficiencies of binary, lower-metallicity massive stars. However, not all H$\alpha$ emitters are expected to have low $Z_{\star}$ values; more luminous sources are likely more massive and hence have higher $Z_{\star}$ values. 

\begin{figure}
    \centering
    \includegraphics[width=\linewidth]{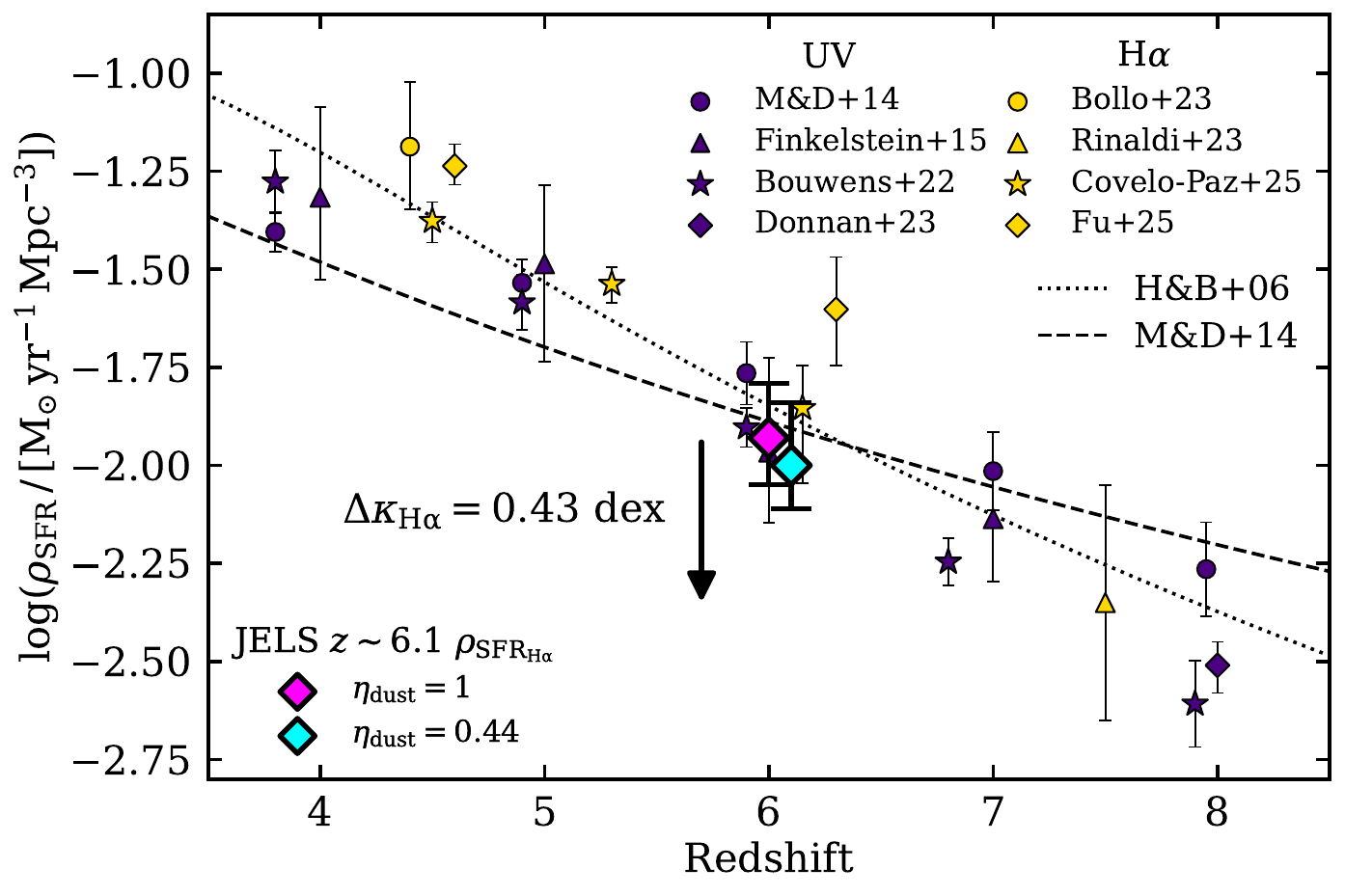}

    \caption{The JELS measured star-formation rate density value $\rho_{\rm{SFR_{H\alpha}}}$ as shown in Fig. \ref{fig:sfrd_multiwav} but only including literature from the H$\alpha$ and UV derived $\rho_{\rm{SFR}}$ in the redshift range: $3.5 < z < 8.5$. The points show $\rho_{\rm{SFR_{H\alpha}}}$ measurements with the standard assumptions on $\kappa_{\rm{H\alpha}}$ \citep{1998ARA&A..36..189K} and $\kappa_{\rm{UV}}$ \citep{2014ARA&A..52..415M} as in Fig. \ref{fig:sfrd_multiwav}. The downward arrows show the decrease in the H$\alpha$ derived $\rho_{\rm{SFR}}$ (0.43 dex) if we assume a BPASSv2.2 SPS model with metallicity $Z = 0.001$ (see Section \ref{sec:sfrd_v2}) instead of the standard relations. Note, the UV-derived $\rho_{\rm{SFR}}$ using the same BPASS SPS model decreases by 0.02 dex, well within the error bars. Therefore, the uncertainty on $\rho_{\rm{SFR}}$ can be compared to the impact of the assumption on $\kappa_{\rm{H\alpha}}$ and $\kappa_{\rm{UV}}$.}

    \label{fig:sfrd_uv_halpha_zoom}

\end{figure}

We investigate the luminosity dependence of $\kappa_{\rm{H\alpha}}$ using galaxy scaling relations in Appendix \ref{app:kappa_halpha_l_dependence}. The results show that the lowest luminosity bin probed in this analysis ($\log_{10}(L_{\rm{H\alpha}} \,/\, \rm{erg \ s^{-1}}) = 41.50 - 41.75$) corresponds to $Z_{\star}\sim0.0007$, while the highest luminosity bin, for the dust-corrected H$\alpha$ LFs ($\log_{10}(L_{\rm{H\alpha}} \,/\, \rm{erg \ s^{-1}}) = 42.75 - 43.0$), corresponds to $Z_{\star}\sim0.0021$, consistent with previous studies utilising low-$Z$ SPS models to infer $\rm{SFR_{H\alpha}}$. We then use the BPASS models to translate this $Z_{\star}$ range into $\kappa_{\rm{H\alpha}} = 10^{-41.74} - 10^{-41.70} \ (\rm{M_{\odot} \ yr^{-1}})/(\rm{erg \ s^{-1}})$. Incorporating the full $\kappa_{\rm{H\alpha}}$ grid from Appendix \ref{app:kappa_halpha_l_dependence} results in a 0.43 dex decrease in $\rho_{\rm{SFR_{H\alpha}}}$ when using BPASS SPS models compared to our fiducial result using the \citet{1998ARA&A..36..189K} standard assumption (see reported values in Table \ref{table:sfrd}). From the BPASS-derived $\rho_{\rm{SFR_{H\alpha}}}$ and $\rho_{L_{\rm{H\alpha}}}$, we calculate a luminosity-weighted $\kappa_{\rm{H\alpha}} = 10^{-41.73} \ (\rm{M_{\odot} \ yr^{-1}})/(\rm{erg \ s^{-1}})$, corresponding to $Z_{\star} \sim 0.001$. Adopting the $Z_{\star} \sim 0.001$ BPASS model gives the UV SFR calibration constant $\kappa_{\rm{UV}} = 10^{-43.46}$ ($\rm{M_{\odot}\,yr^{-1}})/(erg\, s^{-1})$, which is very similar to the UV calibrations commonly used in the literature \citep[e.g. 0.02 dex offset from][]{2014ARA&A..52..415M}. 

We visualise the difference in chosen SFR calibration values for the H$\alpha$ determinations of $\rho_{\rm{SFR}}$ in Fig. \ref{fig:sfrd_uv_halpha_zoom}, where the corrections to the H$\alpha$ measurements (denoted by the downward arrow) are significantly larger than the quoted uncertainties. In contrast, the shift in the UV $\rho_{\rm{SFR}}$ measurements are still well within their uncertainties. Correcting the previous $\rho_{\rm{SFR_{H\alpha}}}$ results using the \textsc{BPASS} inferred $\kappa_{\rm{H\alpha}}$ suggests that $\rho_{\rm{SFR}}$ is under-predicted at $z\sim6$ by $\sim$0.4 -- 0.6 dex, compared to what is suggested by UV studies. This could be due to uncertain dust corrections for the UV indicator, biases in sample selection and subsequent determination of physical properties. However, the BPASS-adjusted H$\alpha$ and UV determinations of $\rho_{\rm{SFR}}$ are in better agreement at $z<6$, which could be indicative that larger area surveys and sample sizes are required to overcome cosmic variance (as discussed in Section \ref{sec:halpha_lf_obs_comparison}). 

\citet{2026A&A...707A.184K} have attempted to constrain the $L_{\rm{H\alpha}}$ -- SFR relation using the SPHINX--20 simulations, building on BPASS SPS models by including more realistic galaxy-scale physics (e.g. realistic star-formation and metal-enrichment histories). They find that applying their derived $L_{\rm{H\alpha}}$ -- SFR relation to the \citetalias{2025ApJ...987..186F} H$\alpha$ LF results in a $\sim$0.05 dex reduction to the fiducial $\rho_{\rm{SFR_{H\alpha}}}$ results. However, we again note the potential caveats from the simulated galaxy properties discussed in Section \ref{sec:sfrd_v1} which will influence their results. Furthermore, the \citetalias{2025ApJ...987..186F} H$\alpha$ LF predicts a large $L_{\rm{H\alpha}}^{\star}$ value (see discussion in Section \ref{sec:halpha_lf_obs_comparison}), meaning the bright end of the H$\alpha$ LF (and hence larger SFR sources) contribute more significantly to $\rho_{\rm{SFR_{H\alpha}}}$, resulting in a smaller reduction compared to our results. Nonetheless, these comparisons motivate future work to better understand the properties of stellar populations in high-redshift galaxies, given the sensitivity of $\rho_{\rm{SFR_{H\alpha}}}$ to the underlying properties (such as the stellar metallicity and multiplicity), and provide observational constraints for galaxy formation simulations.

\section{Conclusions}
\label{sec:conclusions}

In this paper, we use deep narrow-band imaging from the JELS survey \citepalias{2025MNRAS.541.1329D,2025MNRAS.541.1348P} to make new constraints on the H$\alpha$ LF (observed and dust-corrected) and infer the star-formation rate density $\rho_{\rm{SFR_{H\alpha}}}$ into the EoR. We summarise the key results below:

\begin{enumerate}

  \item We update the narrow-band selection procedure with deeper imaging from the JELS and PRIMER surveys to identify 39 robust H$\alpha$ emission line galaxies, including three pairs of mergers and two AGN candidates, at $z \sim 6.1$. We combine the three pairs of H$\alpha$ sources into single systems to remain consistent previous studies of emission-line galaxies and remove the AGN candidates so that our H$\alpha$ LF analysis focuses on the star-formation activity in our sample.
  
  \item The narrow-band determined H$\alpha$ LF at $z \sim 6.1$ has demonstrated increased sensitivity for selecting H$\alpha$ emission line galaxies compared to \emph{JWST}/NIRCam grism selections \citepalias[e.g.][]{2025A&A...694A.178C}, expanding the faintest H$\alpha$ LF bin measurements by $\sim$0.5 dex, and reaching a depth comparable to grism surveys utilising gravitational lensing to increase sensitivity in selecting faint H$\alpha$ emitters \citepalias{2025ApJ...987..186F}. We note, however, that the small cosmic volume probed by the JELS H$\alpha$ sample means that the H$\alpha$ LFs could be significantly impacted by cosmic variance, as is indicated by the difference in the number of emitters detected in two independent volumes (31 and 6 sources in the F466N and F470N filters, respectively), which indicates a high level of clustering of these sources. We estimate that the cosmic variance uncertainty is $\sim$0.3 dex, which may impact global properties inferred from the LF analysis. More work is required to accurately quantify this (e.g. two-point correlation function in clustering analysis).

  \item We fitted the H$\alpha$ LF using a Schechter function and place informative priors on $L_{\rm{H\alpha}}^{\star}$ using constraints from the UV LF at $z\sim6$ \citep{2021AJ....162...47B} and the JELS average $L_{\rm{H\alpha}}$/$L_{\rm{UV}}$ ratio to improve estimates and decrease degeneracies in the inferred Schechter parameters. Both the observed and dust-corrected H$\alpha$ LF number densities and the fitted Schechter functions show excellent agreement with other studies at $z\sim6$, within the overlapping luminosity dynamic range of the datasets ($41.5 \lesssim \log_{10}(L_{\rm{H\alpha}} \,/\, \rm{erg \, s^{-1}}) \lesssim 43.0$) and within uncertainties. 
  
  \item The biggest disagreements between LF analyses occur when extrapolating the Schechter functions to the bright end of the LFs ($\log_{10}(L_{\rm{H\alpha}} \,/\, \rm{erg \ s^{-1}})\gtrsim43.0$) compared to \citetalias{2025ApJ...987..186F}, though the 1$\sigma$ regions for the observed and dust-corrected LFs are consistent with results from \citetalias{2025A&A...694A.178C}. While both dust-corrected JELS H$\alpha$ LFs are consistent within uncertainties, the assumption on $\eta_{\rm{dust}}$ has a significant impact on the inferred LF parameters and overall LF shape. 
  
  \item We also find general disagreement between the JELS H$\alpha$ LFs and those from simulations, particularly between the JELS dust-corrected LF and the SPHINX--20 intrinsic H$\alpha$ LF, which has number densities $>$0.5 dex larger across the luminosity dynamic range. We speculate that this results from differences in the treatment of dust and emission line modelling between different simulations. These results present an excellent opportunity to challenge galaxy formation models in cosmological simulations at high-redshift.

  \item We incorporate the data from \citetalias{2025A&A...694A.178C} to further constrain $L_{\rm{H\alpha}}^{\star}$ for the observed and dust-corrected H$\alpha$ LFs, and constrain the faint-end slope $\alpha_{\rm{H\alpha}}$ to a precision of $\sim$0.22 -- 0.45. We show that the evolution of $\alpha_{\rm{H\alpha}}$ is still uncertain, with uncertainties consistent with there being no evolution with redshift, but also with the increasing steepness of the faint-end slope of the UV LF to higher redshift \citep[e.g.][]{2015ApJ...810...71F,2021AJ....162...47B}. Despite these results, we show that the JELS data points improve the statistical uncertainties at the faint end of the H$\alpha$ LF and hence the precision on $\alpha_{\rm{H\alpha}}$, under the same assumptions for the Schechter function priors.

  \item From the JELS-determined dust-corrected H$\alpha$ LFs, we find that the inferred star-formation rate density $\rho_{\rm{SFR_{H\alpha}}}$ agrees well with previous EoR studies, assuming the \citet{1998ARA&A..36..189K} SFR calibration constant $\kappa_{\rm{H\alpha}}$. For $\eta_{\rm{dust}}=1$ and $\eta_{\rm{dust}}=0.44$, we obtain $\log_{10}(\rho_{\rm{SFR_{H\alpha}}}\,/\,\rm{M_{\odot}\,yr^{-1}\,Mpc^{-3}})=-1.93 \ \rm{and} \ -2.00$, respectively. These results show that the $\eta_{\rm{dust}}$ assumption has negligible impact on the inferred $\rho_{\rm{SFR_{H\alpha}}}$, although this is dependent on the adopted lower integration limit ($L_{\rm{H\alpha, , lim}}$). 

  \item Using the \textsc{BPASS} v2.2 SPS models, we find the luminosity-weighted $\kappa_{\rm{H\alpha}} = 10^{-41.73} \, \rm{(M_{\odot} \, yr^{-1})/(erg \, s^{-1})}$, consistent for each dust-corrected LF, giving an average stellar metallicity $Z_{\star} = 0.001$. Utilising the $Z_{\star} = 0.001$ BPASS model to determine $\kappa_{\rm{H\alpha}}$ results in a 0.43 dex decrease in $\rho_{\rm{SFR_{H\alpha}}}$, compared to standard H$\alpha$ calibrations. However, using the $Z_{\star} = 0.001$ BPASS model to determine the UV SFR calibration $\kappa_{\rm{UV}}$ has a negligible impact on the inferred $\rho_{\rm{SFR}}$ (0.02 dex decrease compared to standard calibrations). The BPASS-derived results for $\rho_{\rm{SFR_{H\alpha}}}$ are $\sim$0.4 -- 0.6 dex lower than typical UV determinations at $z\sim6$, which could be due to uncertain dust corrections and biased sample selections. However, we note that this also could be due to current H$\alpha$ surveys having lower sample sizes and probing lower cosmic volumes, meaning cosmic variance may have a significant impact. In addition, the true nature of stellar populations in the high-redshift Universe is still not well understood and could therefore impact our inference of $\rho_{\rm{SFR}}$.
  
\end{enumerate}

\noindent JELS has increased the statistical precision of observational constraints of the faint end of the H$\alpha$ LF. However, improved constraints on $\alpha_{\rm{H\alpha}}$ require better constraints on the bright-end of the H$\alpha$ LF to determine the location of $L_{\rm{H\alpha}}^{\star}$, which will also be possible with future, larger area slitless spectroscopic surveys \citep[e.g. COSMOS-3D;][]{2024jwst.prop.5893K} undertaken with \emph{JWST}/NIRCam. In addition, targeted spectroscopic follow-up of current H$\alpha$ samples are required to better constrain their physical properties, including their dust properties. These additional datasets will improve both the statistical and systematic uncertainties in the current H$\alpha$ LF results and hence the inferred $\rho_{\rm{SFR}}$ at this epoch.

\section*{ACKNOWLEDGMENTS}
Several authors acknowledge the support of the UK Science and Technology Facilities Council (STFC) via grants ST/W507441/1 (CAP), ST/Y000951/1 (CAP, PNB, CLH and MIA), ST/X001075/1 (ZL, AMS and IS), and through an Ernest Rutherford Fellowship (KJD; grant number ST/W003120/1). DJM and JSD acknowledge the support of the Royal Society through the award of a Royal Society University Research Professorship to JSD. CLH acknowledges support from STFC through grant ST/Y000951/1. RKC is grateful for support from the Leverhulme Trust via a Leverhulme Early Career Fellowship. MB gratefully acknowledges financial support from ANID -- MILENIO - NCN2024$\_$112 and funding by project ALMA-ANID N$^\circ$31230049. EI gratefully acknowledges financial support from ANID - MILENIO - NCN2024$\_$112 and ANID FONDECYT Regular 1221846. JM acknowledges funding by the European Union (ERC, AGENTS, 101076224). LO was supported by the National Doctoral Degree Scholarship given by the National Research and Development Agency of Chile (ANID grant number 21220499). For the purpose of open access, the author has applied a Creative Commons Attribution (CC BY) licence to any Author Accepted Manuscript version arising from this submission.

\section*{Data Availability}

The data underlying this article are available in the Mikulski Archives for Space Telescopes (MAST) at \url{https://doi.org/10.179
09/8v6n-ad45}. Higher level data products, including reduced mosaics in the JELS filters, as well as associated catalogues are publicly available through the University of Edinburgh \href{https://datashare.ed.ac.uk}{DataShare}. Any other data produced for the article will be shared on reasonable request to the corresponding author.



\bibliographystyle{mnras}
\bibliography{ref} 

@ARTICLE{2017ApJ...850..208W,
       author = {{Whitaker}, Katherine E. and {Pope}, Alexandra and {Cybulski}, Ryan and {Casey}, Caitlin M. and {Popping}, Gerg{\"o} and {Yun}, Min S.},
        title = "{The Constant Average Relationship between Dust-obscured Star Formation and Stellar Mass from z = 0 to z = 2.5}",
      journal = {\apj},
         year = 2017,
        month = dec,
       volume = {850},
       number = {2},
          eid = {208},
        pages = {208},
          doi = {10.3847/1538-4357/aa94ce},
archivePrefix = {arXiv},
       eprint = {1710.06872},
 primaryClass = {astro-ph.GA},
       adsurl = {https://ui.adsabs.harvard.edu/abs/2017ApJ...850..208W}
}

@ARTICLE{1996A&AS..117..393B,
       author = {{Bertin}, E. and {Arnouts}, S.},
        title = "{SExtractor: Software for source extraction.}",
      journal = {\aaps},
         year = 1996,
        month = jun,
       volume = {117},
        pages = {393-404},
          doi = {10.1051/aas:1996164},
       adsurl = {https://ui.adsabs.harvard.edu/abs/1996A&AS..117..393B}
}

@ARTICLE{1995MNRAS.273..513B,
       author = {{Bunker}, A.~J. and {Warren}, S.~J. and {Hewett}, P.~C. and {Clements}, D.~L.},
        title = "{On near-infrared H\&alpha searches for high-redshift galaxies}",
      journal = {\mnras},
         year = 1995,
        month = mar,
       volume = {273},
       number = {2},
        pages = {513-516},
          doi = {10.1093/mnras/273.2.513},
archivePrefix = {arXiv},
       eprint = {astro-ph/9501026},
 primaryClass = {astro-ph},
       adsurl = {https://ui.adsabs.harvard.edu/abs/1995MNRAS.273..513B}
}

@ARTICLE{2009MNRAS.398...75S,
       author = {{Sobral}, D. and {Best}, P.~N. and {Geach}, J.~E. and {Smail}, Ian and {Kurk}, J. and {Cirasuolo}, M. and {Casali}, M. and {Ivison}, R.~J. and {Coppin}, K. and {Dalton}, G.~B.},
        title = "{HiZELS: a high-redshift survey of H{\ensuremath{\alpha}} emitters - II. The nature of star-forming galaxies at z = 0.84}",
      journal = {\mnras},
         year = 2009,
        month = sep,
       volume = {398},
       number = {1},
        pages = {75-90},
          doi = {10.1111/j.1365-2966.2009.15129.x},
archivePrefix = {arXiv},
       eprint = {0901.4114},
 primaryClass = {astro-ph.CO},
       adsurl = {https://ui.adsabs.harvard.edu/abs/2009MNRAS.398...75S}
}

@ARTICLE{2012MNRAS.420.1926S,
       author = {{Sobral}, David and {Best}, Philip N. and {Matsuda}, Yuichi and {Smail}, Ian and {Geach}, James E. and {Cirasuolo}, Michele},
        title = "{Star formation at z=1.47 from HiZELS: an H>{\ensuremath{\alpha}}+[O II] double-blind study}",
      journal = {\mnras},
         year = 2012,
        month = mar,
       volume = {420},
       number = {3},
        pages = {1926-1945},
          doi = {10.1111/j.1365-2966.2011.19977.x},
archivePrefix = {arXiv},
       eprint = {1109.1830},
 primaryClass = {astro-ph.CO},
       adsurl = {https://ui.adsabs.harvard.edu/abs/2012MNRAS.420.1926S}
}

@ARTICLE{2013MNRAS.428.1128S,
       author = {{Sobral}, David and {Smail}, Ian and {Best}, Philip N. and {Geach}, James E. and {Matsuda}, Yuichi and {Stott}, John P. and {Cirasuolo}, Michele and {Kurk}, Jaron},
        title = "{A large H{\ensuremath{\alpha}} survey at z = 2.23, 1.47, 0.84 and 0.40: the 11 Gyr evolution of star-forming galaxies from HiZELS★}",
      journal = {\mnras},
         year = 2013,
        month = jan,
       volume = {428},
       number = {2},
        pages = {1128-1146},
          doi = {10.1093/mnras/sts096},
archivePrefix = {arXiv},
       eprint = {1202.3436},
 primaryClass = {astro-ph.CO},
       adsurl = {https://ui.adsabs.harvard.edu/abs/2013MNRAS.428.1128S}
}

@ARTICLE{1998ARA&A..36..189K,
       author = {{Kennicutt}, Robert C., Jr.},
        title = "{Star Formation in Galaxies Along the Hubble Sequence}",
      journal = {\araa},
         year = 1998,
        month = jan,
       volume = {36},
        pages = {189-232},
          doi = {10.1146/annurev.astro.36.1.189},
archivePrefix = {arXiv},
       eprint = {astro-ph/9807187},
 primaryClass = {astro-ph},
       adsurl = {https://ui.adsabs.harvard.edu/abs/1998ARA&A..36..189K}
}

@ARTICLE{2012ARA&A..50..531K,
       author = {{Kennicutt}, Robert C. and {Evans}, Neal J.},
        title = "{Star Formation in the Milky Way and Nearby Galaxies}",
      journal = {\araa},
         year = 2012,
        month = sep,
       volume = {50},
        pages = {531-608},
          doi = {10.1146/annurev-astro-081811-125610},
archivePrefix = {arXiv},
       eprint = {1204.3552},
 primaryClass = {astro-ph.GA},
       adsurl = {https://ui.adsabs.harvard.edu/abs/2012ARA&A..50..531K}
}

@ARTICLE{2010MNRAS.409..421G,
       author = {{Garn}, Timothy and {Best}, Philip N.},
        title = "{Predicting dust extinction from the stellar mass of a galaxy}",
      journal = {\mnras},
         year = 2010,
        month = nov,
       volume = {409},
       number = {1},
        pages = {421-432},
          doi = {10.1111/j.1365-2966.2010.17321.x},
archivePrefix = {arXiv},
       eprint = {1007.1145},
 primaryClass = {astro-ph.GA},
       adsurl = {https://ui.adsabs.harvard.edu/abs/2010MNRAS.409..421G}
}

@ARTICLE{2013ApJ...763..145D,
       author = {{Dom{\'\i}nguez}, A. and {Siana}, B. and {Henry}, A.~L. and {Scarlata}, C. and {Bedregal}, A.~G. and {Malkan}, M. and {Atek}, H. and {Ross}, N.~R. and {Colbert}, J.~W. and {Teplitz}, H.~I. and {Rafelski}, M. and {McCarthy}, P. and {Bunker}, A. and {Hathi}, N.~P. and {Dressler}, A. and {Martin}, C.~L. and {Masters}, D.},
        title = "{Dust Extinction from Balmer Decrements of Star-forming Galaxies at 0.75 <= z <= 1.5 with Hubble Space Telescope/Wide-Field-Camera 3 Spectroscopy from the WFC3 Infrared Spectroscopic Parallel Survey}",
      journal = {\apj},
         year = 2013,
        month = feb,
       volume = {763},
       number = {2},
          eid = {145},
        pages = {145},
          doi = {10.1088/0004-637X/763/2/145},
archivePrefix = {arXiv},
       eprint = {1206.1867},
 primaryClass = {astro-ph.CO},
       adsurl = {https://ui.adsabs.harvard.edu/abs/2013ApJ...763..145D}
}

@ARTICLE{2014ARA&A..52..415M,
       author = {{Madau}, Piero and {Dickinson}, Mark},
        title = "{Cosmic Star-Formation History}",
      journal = {\araa},
         year = 2014,
        month = aug,
       volume = {52},
        pages = {415-486},
          doi = {10.1146/annurev-astro-081811-125615},
archivePrefix = {arXiv},
       eprint = {1403.0007},
 primaryClass = {astro-ph.CO},
       adsurl = {https://ui.adsabs.harvard.edu/abs/2014ARA&A..52..415M}
}

@ARTICLE{2006ApJ...651..142H,
       author = {{Hopkins}, Andrew M. and {Beacom}, John F.},
        title = "{On the Normalization of the Cosmic Star Formation History}",
      journal = {\apj},
         year = 2006,
        month = nov,
       volume = {651},
       number = {1},
        pages = {142-154},
          doi = {10.1086/506610},
archivePrefix = {arXiv},
       eprint = {astro-ph/0601463},
 primaryClass = {astro-ph},
       adsurl = {https://ui.adsabs.harvard.edu/abs/2006ApJ...651..142H}
}

@ARTICLE{2017MNRAS.466..861D,
       author = {{Dunlop}, J.~S. and {McLure}, R.~J. and {Biggs}, A.~D. and {Geach}, J.~E. and {Micha{\l}owski}, M.~J. and {Ivison}, R.~J. and {Rujopakarn}, W. and {van Kampen}, E. and {Kirkpatrick}, A. and {Pope}, A. and {Scott}, D. and {Swinbank}, A.~M. and {Targett}, T.~A. and {Aretxaga}, I. and {Austermann}, J.~E. and {Best}, P.~N. and {Bruce}, V.~A. and {Chapin}, E.~L. and {Charlot}, S. and {Cirasuolo}, M. and {Coppin}, K. and {Ellis}, R.~S. and {Finkelstein}, S.~L. and {Hayward}, C.~C. and {Hughes}, D.~H. and {Ibar}, E. and {Jagannathan}, P. and {Khochfar}, S. and {Koprowski}, M.~P. and {Narayanan}, D. and {Nyland}, K. and {Papovich}, C. and {Peacock}, J.~A. and {Rieke}, G.~H. and {Robertson}, B. and {Vernstrom}, T. and {Werf}, P.~P. van der and {Wilson}, G.~W. and {Yun}, M.},
        title = "{A deep ALMA image of the Hubble Ultra Deep Field}",
      journal = {\mnras},
         year = 2017,
        month = apr,
       volume = {466},
       number = {1},
        pages = {861-883},
          doi = {10.1093/mnras/stw3088},
archivePrefix = {arXiv},
       eprint = {1606.00227},
 primaryClass = {astro-ph.GA},
       adsurl = {https://ui.adsabs.harvard.edu/abs/2017MNRAS.466..861D}
}

@ARTICLE{2020ApJ...902..112B,
       author = {{Bouwens}, Rychard and {Gonz{\'a}lez-L{\'o}pez}, Jorge and {Aravena}, Manuel and {Decarli}, Roberto and {Novak}, Mladen and {Stefanon}, Mauro and {Walter}, Fabian and {Boogaard}, Leindert and {Carilli}, Chris and {Dudzevi{\v{c}}i{\={u}}t{\.{e}}}, Ugn{\.{e}} and {Smail}, Ian and {Daddi}, Emanuele and {da Cunha}, Elisabete and {Ivison}, Rob and {Nanayakkara}, Themiya and {Cortes}, Paulo and {Cox}, Pierre and {Inami}, Hanae and {Oesch}, Pascal and {Popping}, Gerg{\"o} and {Riechers}, Dominik and {van der Werf}, Paul and {Weiss}, Axel and {Fudamoto}, Yoshi and {Wagg}, Jeff},
        title = "{The ALMA Spectroscopic Survey Large Program: The Infrared Excess of z = 1.5-10 UV-selected Galaxies and the Implied High-redshift Star Formation History}",
      journal = {\apj},
         year = 2020,
        month = oct,
       volume = {902},
       number = {2},
          eid = {112},
        pages = {112},
          doi = {10.3847/1538-4357/abb830},
archivePrefix = {arXiv},
       eprint = {2009.10727},
 primaryClass = {astro-ph.GA},
       adsurl = {https://ui.adsabs.harvard.edu/abs/2020ApJ...902..112B}
}

@ARTICLE{2020A&A...643A...8G,
       author = {{Gruppioni}, C. and {B{\'e}thermin}, M. and {Loiacono}, F. and {Le F{\`e}vre}, O. and {Capak}, P. and {Cassata}, P. and {Faisst}, A.~L. and {Schaerer}, D. and {Silverman}, J. and {Yan}, L. and {Bardelli}, S. and {Boquien}, M. and {Carraro}, R. and {Cimatti}, A. and {Dessauges-Zavadsky}, M. and {Ginolfi}, M. and {Fujimoto}, S. and {Hathi}, N.~P. and {Jones}, G.~C. and {Khusanova}, Y. and {Koekemoer}, A.~M. and {Lagache}, G. and {Lemaux}, B.~C. and {Oesch}, P.~A. and {Pozzi}, F. and {Riechers}, D.~A. and {Rodighiero}, G. and {Romano}, M. and {Talia}, M. and {Vallini}, L. and {Vergani}, D. and {Zamorani}, G. and {Zucca}, E.},
        title = "{The ALPINE-ALMA [CII] survey. The nature, luminosity function, and star formation history of dusty galaxies up to z ≃ 6}",
      journal = {\aap},
         year = 2020,
        month = nov,
       volume = {643},
          eid = {A8},
        pages = {A8},
          doi = {10.1051/0004-6361/202038487},
archivePrefix = {arXiv},
       eprint = {2006.04974},
 primaryClass = {astro-ph.GA},
       adsurl = {https://ui.adsabs.harvard.edu/abs/2020A&A...643A...8G}
}

@ARTICLE{2015MNRAS.452.2018O,
       author = {{Oteo}, I. and {Sobral}, D. and {Ivison}, R.~J. and {Smail}, I. and {Best}, P.~N. and {Cepa}, J. and {P{\'e}rez-Garc{\'\i}a}, A.~M.},
        title = "{On the nature of H{\ensuremath{\alpha}} emitters at z {\ensuremath{\sim}} 2 from the HiZELS survey: physical properties, Ly{\ensuremath{\alpha}} escape fraction and main sequence}",
      journal = {\mnras},
         year = 2015,
        month = sep,
       volume = {452},
       number = {2},
        pages = {2018-2033},
          doi = {10.1093/mnras/stv1284},
archivePrefix = {arXiv},
       eprint = {1506.02670},
 primaryClass = {astro-ph.GA},
       adsurl = {https://ui.adsabs.harvard.edu/abs/2015MNRAS.452.2018O}
}

@ARTICLE{2008MNRAS.388.1473G,
       author = {{Geach}, J.~E. and {Smail}, Ian and {Best}, P.~N. and {Kurk}, J. and {Casali}, M. and {Ivison}, R.~J. and {Coppin}, K.},
        title = "{HiZELS: a high-redshift survey of H{\ensuremath{\alpha}} emitters - I. The cosmic star formation rate and clustering at z = 2.23}",
      journal = {\mnras},
         year = 2008,
        month = aug,
       volume = {388},
       number = {4},
        pages = {1473-1486},
          doi = {10.1111/j.1365-2966.2008.13481.x},
archivePrefix = {arXiv},
       eprint = {0805.2861},
 primaryClass = {astro-ph},
       adsurl = {https://ui.adsabs.harvard.edu/abs/2008MNRAS.388.1473G}
}

@ARTICLE{2023ApJ...952..143R,
       author = {{Rinaldi}, P. and {Caputi}, K.~I. and {Costantin}, L. and {Gillman}, S. and {Iani}, E. and {P{\'e}rez-Gonz{\'a}lez}, P.~G. and {{\"O}stlin}, G. and {Colina}, L. and {Greve}, T.~R. and {Noorgard-Nielsen}, H.~U. and {Wright}, G.~S. and {Alonso-Herrero}, A. and {{\'A}lvarez-M{\'a}rquez}, J. and {Eckart}, A. and {Garc{\'\i}a-Mar{\'\i}n}, M. and {Hjorth}, J. and {Ilbert}, O. and {Kendrew}, S. and {Labiano}, A. and {Le F{\`e}vre}, O. and {Pye}, J. and {Tikkanen}, T. and {Walter}, F. and {van der Werf}, P. and {Ward}, M. and {Annunziatella}, M. and {Azzollini}, R. and {Bik}, A. and {Boogaard}, L. and {Bosman}, S.~E.~I. and {Crespo G{\'o}mez}, A. and {Jermann}, I. and {Langeroodi}, D. and {Melinder}, J. and {Meyer}, R.~A. and {Moutard}, T. and {Peissker}, F. and {Topinka}, M. and {van Dishoeck}, E. and {G{\"u}del}, M. and {Henning}, Th. and {Lagage}, P. -O. and {Ray}, T. and {Vandenbussche}, B. and {Waelkens}, C. and {Navarro-Carrera}, R. and {Kokorev}, V.},
        title = "{MIDIS: Strong (H{\ensuremath{\beta}}+[O III]) and H{\ensuremath{\alpha}} Emitters at Redshift z ≃ 7-8 Unveiled with JWST NIRCam and MIRI Imaging in the Hubble eXtreme Deep Field}",
      journal = {\apj},
         year = 2023,
        month = aug,
       volume = {952},
       number = {2},
          eid = {143},
        pages = {143},
          doi = {10.3847/1538-4357/acdc27},
archivePrefix = {arXiv},
       eprint = {2301.10717},
 primaryClass = {astro-ph.GA},
       adsurl = {https://ui.adsabs.harvard.edu/abs/2023ApJ...952..143R}
}

@ARTICLE{2023ApJ...950...67M,
       author = {{Matthee}, Jorryt and {Mackenzie}, Ruari and {Simcoe}, Robert A. and {Kashino}, Daichi and {Lilly}, Simon J. and {Bordoloi}, Rongmon and {Eilers}, Anna-Christina},
        title = "{EIGER. II. First Spectroscopic Characterization of the Young Stars and Ionized Gas Associated with Strong H{\ensuremath{\beta}} and [O III] Line Emission in Galaxies at z = 5-7 with JWST}",
      journal = {\apj},
         year = 2023,
        month = jun,
       volume = {950},
       number = {1},
          eid = {67},
        pages = {67},
          doi = {10.3847/1538-4357/acc846},
archivePrefix = {arXiv},
       eprint = {2211.08255},
 primaryClass = {astro-ph.GA},
       adsurl = {https://ui.adsabs.harvard.edu/abs/2023ApJ...950...67M}
}

@INPROCEEDINGS{2013ASSP...37..235B,
       author = {{Best}, Philip and {Smail}, Ian and {Sobral}, David and {Geach}, Jim and {Garn}, Tim and {Ivison}, Rob and {Kurk}, Jaron and {Dalton}, Gavin and {Cirasuolo}, Michele and {Casali}, Mark},
        title = "{HiZELS: The High Redshift Emission Line Survey with UKIRT}",
    booktitle = {Thirty Years of Astronomical Discovery with UKIRT},
         year = 2013,
       series = {Astrophysics and Space Science Proceedings},
       volume = {37},
        month = jan,
        pages = {235},
          doi = {10.1007/978-94-007-7432-2_22},
archivePrefix = {arXiv},
       eprint = {1003.5183},
 primaryClass = {astro-ph.CO},
       adsurl = {https://ui.adsabs.harvard.edu/abs/2013ASSP...37..235B}
}

@ARTICLE{2024MNRAS.533.3222D,
       author = {{Donnan}, C.~T. and {McLure}, R.~J. and {Dunlop}, J.~S. and {McLeod}, D.~J. and {Magee}, D. and {Arellano-C{\'o}rdova}, K.~Z. and {Barrufet}, L. and {Begley}, R. and {Bowler}, R.~A.~A. and {Carnall}, A.~C. and {Cullen}, F. and {Ellis}, R.~S. and {Fontana}, A. and {Illingworth}, G.~D. and {Grogin}, N.~A. and {Hamadouche}, M.~L. and {Koekemoer}, A.~M. and {Liu}, F. -Y. and {Mason}, C. and {Santini}, P. and {Stanton}, T.~M.},
        title = "{JWST PRIMER: a new multifield determination of the evolving galaxy UV luminosity function at redshifts z ≃ 9 - 15}",
      journal = {\mnras},
         year = 2024,
        month = sep,
       volume = {533},
       number = {3},
        pages = {3222-3237},
          doi = {10.1093/mnras/stae2037},
archivePrefix = {arXiv},
       eprint = {2403.03171},
 primaryClass = {astro-ph.GA},
       adsurl = {https://ui.adsabs.harvard.edu/abs/2024MNRAS.533.3222D}
}

@MISC{2021jwst.prop.1837D,
       author = {{Dunlop}, James S. and {Abraham}, Roberto G. and {Ashby}, Matthew L.~N. and {Bagley}, Micaela and {Best}, Philip N. and {Bongiorno}, Angela and {Bouwens}, Rychard and {Bowler}, Rebecca A.~A. and {Brammer}, Gabriel and {Bremer}, Malcolm and {Calabro'}, Antonello and {Carnall}, Adam and {Castellano}, Marco and {Cirasuolo}, Michele and {Conselice}, Christopher and {Cullen}, Fergus and {Dave}, Romeel and {Dayal}, Pratika and {Dekel}, Avishai and {Dickinson}, Mark and {Duncan}, Kenneth James and {Elbaz}, David and {Ellis}, Richard S. and {Ferguson}, Harry C. and {Ferrara}, Andrea and {Finkelstein}, Steven L. and {Fontana}, Adriano and {Furlanetto}, Steven and {Fynbo}, Johan P.~U. and {Gallerani}, Simona and {Gardner}, Jonathan P. and {Giavalisco}, Mauro and {Grazian}, Andrea and {Grogin}, Norman and {Harikane}, Yuichi and {Hopkins}, Philip F. and {Ilbert}, Olivier and {Illingworth}, Garth D. and {Juneau}, Stephanie and {Jung}, Intae and {Kartaltepe}, Jeyhan and {Kassin}, Susan and {Kauffmann}, Olivier Benjamin and {Khochfar}, Sadegh and {Kirkpatrick}, Allison and {Kocevski}, Dale D. and {Koekemoer}, Anton M. and {Labbe}, Ivo and {Laporte}, Nicolas and {Larson}, Rebecca L. and {Lucas}, Ray A. and {Magee}, Daniel K. and {Mason}, Charlotte and {McCracken}, Henry Joy and {McLeod}, Derek and {McLure}, Ross and {Merlin}, Emiliano and {Mesinger}, Andrei and {Milvang-Jensen}, Bo and {Newman}, Jeffrey Allen and {Oesch}, Pascal and {Ouchi}, Masami and {Pacifici}, Camilla and {Papovich}, Casey and {Peacock}, John and {Peeples}, Molly and {Pentericci}, Laura and {Perez-Gonzalez}, Pablo G. and {Pirzkal}, Norbert and {Pope}, Alexandra and {Pye}, John P. and {Reddy}, Naveen A. and {Robertson}, Brant and {Salvato}, Mara and {Santini}, Paola and {Schaerer}, Daniel and {Shapley}, Alice E. and {Simons}, Raymond and {Smit}, Renske and {Smith}, Britton D. and {Snyder}, Greg and {Somerville}, Rachel S. and {Stanway}, Elizabeth R. and {Stefanon}, Mauro and {Tasca}, Lidia and {Tikkanen}, Tuomo and {Tresse}, Laurence and {Trump}, Jonathan R. and {Whitaker}, Katherine E. and {Wilkins}, Stephen Matthew and {Wright}, Gillian and {Wyithe}, J. Stuart B. and {van Dokkum}, Pieter and {van der Werf}, Paul},
        title = "{PRIMER: Public Release IMaging for Extragalactic Research}",
 howpublished = {JWST Proposal. Cycle 1, ID. \#1837},
         year = 2021,
        month = mar,
        pages = {1837},
       adsurl = {https://ui.adsabs.harvard.edu/abs/2021jwst.prop.1837D}
}

@ARTICLE{2000ApJ...533..682C,
       author = {{Calzetti}, Daniela and {Armus}, Lee and {Bohlin}, Ralph C. and {Kinney}, Anne L. and {Koornneef}, Jan and {Storchi-Bergmann}, Thaisa},
        title = "{The Dust Content and Opacity of Actively Star-forming Galaxies}",
      journal = {\apj},
         year = 2000,
        month = apr,
       volume = {533},
       number = {2},
        pages = {682-695},
          doi = {10.1086/308692},
archivePrefix = {arXiv},
       eprint = {astro-ph/9911459},
 primaryClass = {astro-ph},
       adsurl = {https://ui.adsabs.harvard.edu/abs/2000ApJ...533..682C}
}

@ARTICLE{2018ApJ...859...11S,
       author = {{Salim}, Samir and {Boquien}, M{\'e}d{\'e}ric and {Lee}, Janice C.},
        title = "{Dust Attenuation Curves in the Local Universe: Demographics and New Laws for Star-forming Galaxies and High-redshift Analogs}",
      journal = {\apj},
         year = 2018,
        month = may,
       volume = {859},
       number = {1},
          eid = {11},
        pages = {11},
          doi = {10.3847/1538-4357/aabf3c},
archivePrefix = {arXiv},
       eprint = {1804.05850},
 primaryClass = {astro-ph.GA},
       adsurl = {https://ui.adsabs.harvard.edu/abs/2018ApJ...859...11S}
}

@ARTICLE{2023MNRAS.519.1526P,
       author = {{Popesso}, P. and {Concas}, A. and {Cresci}, G. and {Belli}, S. and {Rodighiero}, G. and {Inami}, H. and {Dickinson}, M. and {Ilbert}, O. and {Pannella}, M. and {Elbaz}, D.},
        title = "{The main sequence of star-forming galaxies across cosmic times}",
      journal = {\mnras},
         year = 2023,
        month = feb,
       volume = {519},
       number = {1},
        pages = {1526-1544},
          doi = {10.1093/mnras/stac3214},
archivePrefix = {arXiv},
       eprint = {2203.10487},
 primaryClass = {astro-ph.GA},
       adsurl = {https://ui.adsabs.harvard.edu/abs/2023MNRAS.519.1526P}
}

@ARTICLE{2011ApJS..197...35G,
       author = {{Grogin}, Norman A. and {Kocevski}, Dale D. and {Faber}, S.~M. and {Ferguson}, Henry C. and {Koekemoer}, Anton M. and {Riess}, Adam G. and {Acquaviva}, Viviana and {Alexander}, David M. and {Almaini}, Omar and {Ashby}, Matthew L.~N. and {Barden}, Marco and {Bell}, Eric F. and {Bournaud}, Fr{\'e}d{\'e}ric and {Brown}, Thomas M. and {Caputi}, Karina I. and {Casertano}, Stefano and {Cassata}, Paolo and {Castellano}, Marco and {Challis}, Peter and {Chary}, Ranga-Ram and {Cheung}, Edmond and {Cirasuolo}, Michele and {Conselice}, Christopher J. and {Roshan Cooray}, Asantha and {Croton}, Darren J. and {Daddi}, Emanuele and {Dahlen}, Tomas and {Dav{\'e}}, Romeel and {de Mello}, Du{\'\i}lia F. and {Dekel}, Avishai and {Dickinson}, Mark and {Dolch}, Timothy and {Donley}, Jennifer L. and {Dunlop}, James S. and {Dutton}, Aaron A. and {Elbaz}, David and {Fazio}, Giovanni G. and {Filippenko}, Alexei V. and {Finkelstein}, Steven L. and {Fontana}, Adriano and {Gardner}, Jonathan P. and {Garnavich}, Peter M. and {Gawiser}, Eric and {Giavalisco}, Mauro and {Grazian}, Andrea and {Guo}, Yicheng and {Hathi}, Nimish P. and {H{\"a}ussler}, Boris and {Hopkins}, Philip F. and {Huang}, Jia-Sheng and {Huang}, Kuang-Han and {Jha}, Saurabh W. and {Kartaltepe}, Jeyhan S. and {Kirshner}, Robert P. and {Koo}, David C. and {Lai}, Kamson and {Lee}, Kyoung-Soo and {Li}, Weidong and {Lotz}, Jennifer M. and {Lucas}, Ray A. and {Madau}, Piero and {McCarthy}, Patrick J. and {McGrath}, Elizabeth J. and {McIntosh}, Daniel H. and {McLure}, Ross J. and {Mobasher}, Bahram and {Moustakas}, Leonidas A. and {Mozena}, Mark and {Nandra}, Kirpal and {Newman}, Jeffrey A. and {Niemi}, Sami-Matias and {Noeske}, Kai G. and {Papovich}, Casey J. and {Pentericci}, Laura and {Pope}, Alexandra and {Primack}, Joel R. and {Rajan}, Abhijith and {Ravindranath}, Swara and {Reddy}, Naveen A. and {Renzini}, Alvio and {Rix}, Hans-Walter and {Robaina}, Aday R. and {Rodney}, Steven A. and {Rosario}, David J. and {Rosati}, Piero and {Salimbeni}, Sara and {Scarlata}, Claudia and {Siana}, Brian and {Simard}, Luc and {Smidt}, Joseph and {Somerville}, Rachel S. and {Spinrad}, Hyron and {Straughn}, Amber N. and {Strolger}, Louis-Gregory and {Telford}, Olivia and {Teplitz}, Harry I. and {Trump}, Jonathan R. and {van der Wel}, Arjen and {Villforth}, Carolin and {Wechsler}, Risa H. and {Weiner}, Benjamin J. and {Wiklind}, Tommy and {Wild}, Vivienne and {Wilson}, Grant and {Wuyts}, Stijn and {Yan}, Hao-Jing and {Yun}, Min S.},
        title = "{CANDELS: The Cosmic Assembly Near-infrared Deep Extragalactic Legacy Survey}",
      journal = {\apjs},
         year = 2011,
        month = dec,
       volume = {197},
       number = {2},
          eid = {35},
        pages = {35},
          doi = {10.1088/0067-0049/197/2/35},
archivePrefix = {arXiv},
       eprint = {1105.3753},
 primaryClass = {astro-ph.CO},
       adsurl = {https://ui.adsabs.harvard.edu/abs/2011ApJS..197...35G}
}

@ARTICLE{2011ApJS..197...36K,
       author = {{Koekemoer}, Anton M. and {Faber}, S.~M. and {Ferguson}, Henry C. and {Grogin}, Norman A. and {Kocevski}, Dale D. and {Koo}, David C. and {Lai}, Kamson and {Lotz}, Jennifer M. and {Lucas}, Ray A. and {McGrath}, Elizabeth J. and {Ogaz}, Sara and {Rajan}, Abhijith and {Riess}, Adam G. and {Rodney}, Steve A. and {Strolger}, Louis and {Casertano}, Stefano and {Castellano}, Marco and {Dahlen}, Tomas and {Dickinson}, Mark and {Dolch}, Timothy and {Fontana}, Adriano and {Giavalisco}, Mauro and {Grazian}, Andrea and {Guo}, Yicheng and {Hathi}, Nimish P. and {Huang}, Kuang-Han and {van der Wel}, Arjen and {Yan}, Hao-Jing and {Acquaviva}, Viviana and {Alexander}, David M. and {Almaini}, Omar and {Ashby}, Matthew L.~N. and {Barden}, Marco and {Bell}, Eric F. and {Bournaud}, Fr{\'e}d{\'e}ric and {Brown}, Thomas M. and {Caputi}, Karina I. and {Cassata}, Paolo and {Challis}, Peter J. and {Chary}, Ranga-Ram and {Cheung}, Edmond and {Cirasuolo}, Michele and {Conselice}, Christopher J. and {Roshan Cooray}, Asantha and {Croton}, Darren J. and {Daddi}, Emanuele and {Dav{\'e}}, Romeel and {de Mello}, Duilia F. and {de Ravel}, Loic and {Dekel}, Avishai and {Donley}, Jennifer L. and {Dunlop}, James S. and {Dutton}, Aaron A. and {Elbaz}, David and {Fazio}, Giovanni G. and {Filippenko}, Alexei V. and {Finkelstein}, Steven L. and {Frazer}, Chris and {Gardner}, Jonathan P. and {Garnavich}, Peter M. and {Gawiser}, Eric and {Gruetzbauch}, Ruth and {Hartley}, Will G. and {H{\"a}ussler}, Boris and {Herrington}, Jessica and {Hopkins}, Philip F. and {Huang}, Jia-Sheng and {Jha}, Saurabh W. and {Johnson}, Andrew and {Kartaltepe}, Jeyhan S. and {Khostovan}, Ali A. and {Kirshner}, Robert P. and {Lani}, Caterina and {Lee}, Kyoung-Soo and {Li}, Weidong and {Madau}, Piero and {McCarthy}, Patrick J. and {McIntosh}, Daniel H. and {McLure}, Ross J. and {McPartland}, Conor and {Mobasher}, Bahram and {Moreira}, Heidi and {Mortlock}, Alice and {Moustakas}, Leonidas A. and {Mozena}, Mark and {Nandra}, Kirpal and {Newman}, Jeffrey A. and {Nielsen}, Jennifer L. and {Niemi}, Sami and {Noeske}, Kai G. and {Papovich}, Casey J. and {Pentericci}, Laura and {Pope}, Alexandra and {Primack}, Joel R. and {Ravindranath}, Swara and {Reddy}, Naveen A. and {Renzini}, Alvio and {Rix}, Hans-Walter and {Robaina}, Aday R. and {Rosario}, David J. and {Rosati}, Piero and {Salimbeni}, Sara and {Scarlata}, Claudia and {Siana}, Brian and {Simard}, Luc and {Smidt}, Joseph and {Snyder}, Diana and {Somerville}, Rachel S. and {Spinrad}, Hyron and {Straughn}, Amber N. and {Telford}, Olivia and {Teplitz}, Harry I. and {Trump}, Jonathan R. and {Vargas}, Carlos and {Villforth}, Carolin and {Wagner}, Cory R. and {Wandro}, Pat and {Wechsler}, Risa H. and {Weiner}, Benjamin J. and {Wiklind}, Tommy and {Wild}, Vivienne and {Wilson}, Grant and {Wuyts}, Stijn and {Yun}, Min S.},
        title = "{CANDELS: The Cosmic Assembly Near-infrared Deep Extragalactic Legacy Survey{\textemdash}The Hubble Space Telescope Observations, Imaging Data Products, and Mosaics}",
      journal = {\apjs},
         year = 2011,
        month = dec,
       volume = {197},
       number = {2},
          eid = {36},
        pages = {36},
          doi = {10.1088/0067-0049/197/2/36},
archivePrefix = {arXiv},
       eprint = {1105.3754},
 primaryClass = {astro-ph.CO},
       adsurl = {https://ui.adsabs.harvard.edu/abs/2011ApJS..197...36K}
}

@ARTICLE{2023A&A...674A...1G,
       author = {{Gaia Collaboration} and {Vallenari}, A. and {Brown}, A.~G.~A. and {Prusti}, T. and {de Bruijne}, J.~H.~J. and {Arenou}, F. and {Babusiaux}, C. and {Biermann}, M. and {Creevey}, O.~L. and {Ducourant}, C. and {Evans}, D.~W. and {Eyer}, L. and {Guerra}, R. and {Hutton}, A. and {Jordi}, C. and {Klioner}, S.~A. and {Lammers}, U.~L. and {Lindegren}, L. and {Luri}, X. and {Mignard}, F. and {Panem}, C. and {Pourbaix}, D. and {Randich}, S. and {Sartoretti}, P. and {Soubiran}, C. and {Tanga}, P. and {Walton}, N.~A. and {Bailer-Jones}, C.~A.~L. and {Bastian}, U. and {Drimmel}, R. and {Jansen}, F. and {Katz}, D. and {Lattanzi}, M.~G. and {van Leeuwen}, F. and {Bakker}, J. and {Cacciari}, C. and {Casta{\~n}eda}, J. and {De Angeli}, F. and {Fabricius}, C. and {Fouesneau}, M. and {Fr{\'e}mat}, Y. and {Galluccio}, L. and {Guerrier}, A. and {Heiter}, U. and {Masana}, E. and {Messineo}, R. and {Mowlavi}, N. and {Nicolas}, C. and {Nienartowicz}, K. and {Pailler}, F. and {Panuzzo}, P. and {Riclet}, F. and {Roux}, W. and {Seabroke}, G.~M. and {Sordo}, R. and {Th{\'e}venin}, F. and {Gracia-Abril}, G. and {Portell}, J. and {Teyssier}, D. and {Altmann}, M. and {Andrae}, R. and {Audard}, M. and {Bellas-Velidis}, I. and {Benson}, K. and {Berthier}, J. and {Blomme}, R. and {Burgess}, P.~W. and {Busonero}, D. and {Busso}, G. and {C{\'a}novas}, H. and {Carry}, B. and {Cellino}, A. and {Cheek}, N. and {Clementini}, G. and {Damerdji}, Y. and {Davidson}, M. and {de Teodoro}, P. and {Nu{\~n}ez Campos}, M. and {Delchambre}, L. and {Dell'Oro}, A. and {Esquej}, P. and {Fern{\'a}ndez-Hern{\'a}ndez}, J. and {Fraile}, E. and {Garabato}, D. and {Garc{\'\i}a-Lario}, P. and {Gosset}, E. and {Haigron}, R. and {Halbwachs}, J. -L. and {Hambly}, N.~C. and {Harrison}, D.~L. and {Hern{\'a}ndez}, J. and {Hestroffer}, D. and {Hodgkin}, S.~T. and {Holl}, B. and {Jan{\ss}en}, K. and {Jevardat de Fombelle}, G. and {Jordan}, S. and {Krone-Martins}, A. and {Lanzafame}, A.~C. and {L{\"o}ffler}, W. and {Marchal}, O. and {Marrese}, P.~M. and {Moitinho}, A. and {Muinonen}, K. and {Osborne}, P. and {Pancino}, E. and {Pauwels}, T. and {Recio-Blanco}, A. and {Reyl{\'e}}, C. and {Riello}, M. and {Rimoldini}, L. and {Roegiers}, T. and {Rybizki}, J. and {Sarro}, L.~M. and {Siopis}, C. and {Smith}, M. and {Sozzetti}, A. and {Utrilla}, E. and {van Leeuwen}, M. and {Abbas}, U. and {{\'A}brah{\'a}m}, P. and {Abreu Aramburu}, A. and {Aerts}, C. and {Aguado}, J.~J. and {Ajaj}, M. and {Aldea-Montero}, F. and {Altavilla}, G. and {{\'A}lvarez}, M.~A. and {Alves}, J. and {Anders}, F. and {Anderson}, R.~I. and {Anglada Varela}, E. and {Antoja}, T. and {Baines}, D. and {Baker}, S.~G. and {Balaguer-N{\'u}{\~n}ez}, L. and {Balbinot}, E. and {Balog}, Z. and {Barache}, C. and {Barbato}, D. and {Barros}, M. and {Barstow}, M.~A. and {Bartolom{\'e}}, S. and {Bassilana}, J. -L. and {Bauchet}, N. and {Becciani}, U. and {Bellazzini}, M. and {Berihuete}, A. and {Bernet}, M. and {Bertone}, S. and {Bianchi}, L. and {Binnenfeld}, A. and {Blanco-Cuaresma}, S. and {Blazere}, A. and {Boch}, T. and {Bombrun}, A. and {Bossini}, D. and {Bouquillon}, S. and {Bragaglia}, A. and {Bramante}, L. and {Breedt}, E. and {Bressan}, A. and {Brouillet}, N. and {Brugaletta}, E. and {Bucciarelli}, B. and {Burlacu}, A. and {Butkevich}, A.~G. and {Buzzi}, R. and {Caffau}, E. and {Cancelliere}, R. and {Cantat-Gaudin}, T. and {Carballo}, R. and {Carlucci}, T. and {Carnerero}, M.~I. and {Carrasco}, J.~M. and {Casamiquela}, L. and {Castellani}, M. and {Castro-Ginard}, A. and {Chaoul}, L. and {Charlot}, P. and {Chemin}, L. and {Chiaramida}, V. and {Chiavassa}, A. and {Chornay}, N. and {Comoretto}, G. and {Contursi}, G. and {Cooper}, W.~J. and {Cornez}, T. and {Cowell}, S. and {Crifo}, F. and {Cropper}, M. and {Crosta}, M. and {Crowley}, C. and {Dafonte}, C. and {Dapergolas}, A. and {David}, M. and {David}, P. and {de Laverny}, P. and {De Luise}, F. and {De March}, R. and {De Ridder}, J. and {de Souza}, R. and {de Torres}, A. and {del Peloso}, E.~F. and {del Pozo}, E. and {Delbo}, M. and {Delgado}, A. and {Delisle}, J. -B. and {Demouchy}, C. and {Dharmawardena}, T.~E. and {Di Matteo}, P. and {Diakite}, S. and {Diener}, C. and {Distefano}, E. and {Dolding}, C. and {Edvardsson}, B. and {Enke}, H. and {Fabre}, C. and {Fabrizio}, M. and {Faigler}, S. and {Fedorets}, G. and {Fernique}, P. and {Fienga}, A. and {Figueras}, F. and {Fournier}, Y. and {Fouron}, C. and {Fragkoudi}, F. and {Gai}, M. and {Garcia-Gutierrez}, A. and {Garcia-Reinaldos}, M. and {Garc{\'\i}a-Torres}, M. and {Garofalo}, A. and {Gavel}, A. and {Gavras}, P. and {Gerlach}, E. and {Geyer}, R. and {Giacobbe}, P. and {Gilmore}, G. and {Girona}, S. and {Giuffrida}, G. and {Gomel}, R. and {Gomez}, A. and {Gonz{\'a}lez-N{\'u}{\~n}ez}, J. and {Gonz{\'a}lez-Santamar{\'\i}a}, I. and {Gonz{\'a}lez-Vidal}, J.~J. and {Granvik}, M. and {Guillout}, P. and {Guiraud}, J. and {Guti{\'e}rrez-S{\'a}nchez}, R. and {Guy}, L.~P. and {Hatzidimitriou}, D. and {Hauser}, M. and {Haywood}, M. and {Helmer}, A. and {Helmi}, A. and {Sarmiento}, M.~H. and {Hidalgo}, S.~L. and {Hilger}, T. and {H{\l}adczuk}, N. and {Hobbs}, D. and {Holland}, G. and {Huckle}, H.~E. and {Jardine}, K. and {Jasniewicz}, G. and {Jean-Antoine Piccolo}, A. and {Jim{\'e}nez-Arranz}, {\'O}. and {Jorissen}, A. and {Juaristi Campillo}, J. and {Julbe}, F. and {Karbevska}, L. and {Kervella}, P. and {Khanna}, S. and {Kontizas}, M. and {Kordopatis}, G. and {Korn}, A.~J. and {K{\'o}sp{\'a}l}, {\'A}. and {Kostrzewa-Rutkowska}, Z. and {Kruszy{\'n}ska}, K. and {Kun}, M. and {Laizeau}, P. and {Lambert}, S. and {Lanza}, A.~F. and {Lasne}, Y. and {Le Campion}, J. -F. and {Lebreton}, Y. and {Lebzelter}, T. and {Leccia}, S. and {Leclerc}, N. and {Lecoeur-Taibi}, I. and {Liao}, S. and {Licata}, E.~L. and {Lindstr{\o}m}, H.~E.~P. and {Lister}, T.~A. and {Livanou}, E. and {Lobel}, A. and {Lorca}, A. and {Loup}, C. and {Madrero Pardo}, P. and {Magdaleno Romeo}, A. and {Managau}, S. and {Mann}, R.~G. and {Manteiga}, M. and {Marchant}, J.~M. and {Marconi}, M. and {Marcos}, J. and {Marcos Santos}, M.~M.~S. and {Mar{\'\i}n Pina}, D. and {Marinoni}, S. and {Marocco}, F. and {Marshall}, D.~J. and {Martin Polo}, L. and {Mart{\'\i}n-Fleitas}, J.~M. and {Marton}, G. and {Mary}, N. and {Masip}, A. and {Massari}, D. and {Mastrobuono-Battisti}, A. and {Mazeh}, T. and {McMillan}, P.~J. and {Messina}, S. and {Michalik}, D. and {Millar}, N.~R. and {Mints}, A. and {Molina}, D. and {Molinaro}, R. and {Moln{\'a}r}, L. and {Monari}, G. and {Mongui{\'o}}, M. and {Montegriffo}, P. and {Montero}, A. and {Mor}, R. and {Mora}, A. and {Morbidelli}, R. and {Morel}, T. and {Morris}, D. and {Muraveva}, T. and {Murphy}, C.~P. and {Musella}, I. and {Nagy}, Z. and {Noval}, L. and {Oca{\~n}a}, F. and {Ogden}, A. and {Ordenovic}, C. and {Osinde}, J.~O. and {Pagani}, C. and {Pagano}, I. and {Palaversa}, L. and {Palicio}, P.~A. and {Pallas-Quintela}, L. and {Panahi}, A. and {Payne-Wardenaar}, S. and {Pe{\~n}alosa Esteller}, X. and {Penttil{\"a}}, A. and {Pichon}, B. and {Piersimoni}, A.~M. and {Pineau}, F. -X. and {Plachy}, E. and {Plum}, G. and {Poggio}, E. and {Pr{\v{s}}a}, A. and {Pulone}, L. and {Racero}, E. and {Ragaini}, S. and {Rainer}, M. and {Raiteri}, C.~M. and {Rambaux}, N. and {Ramos}, P. and {Ramos-Lerate}, M. and {Re Fiorentin}, P. and {Regibo}, S. and {Richards}, P.~J. and {Rios Diaz}, C. and {Ripepi}, V. and {Riva}, A. and {Rix}, H. -W. and {Rixon}, G. and {Robichon}, N. and {Robin}, A.~C. and {Robin}, C. and {Roelens}, M. and {Rogues}, H.~R.~O. and {Rohrbasser}, L. and {Romero-G{\'o}mez}, M. and {Rowell}, N. and {Royer}, F. and {Ruz Mieres}, D. and {Rybicki}, K.~A. and {Sadowski}, G. and {S{\'a}ez N{\'u}{\~n}ez}, A. and {Sagrist{\`a} Sell{\'e}s}, A. and {Sahlmann}, J. and {Salguero}, E. and {Samaras}, N. and {Sanchez Gimenez}, V. and {Sanna}, N. and {Santove{\~n}a}, R. and {Sarasso}, M. and {Schultheis}, M. and {Sciacca}, E. and {Segol}, M. and {Segovia}, J.~C. and {S{\'e}gransan}, D. and {Semeux}, D. and {Shahaf}, S. and {Siddiqui}, H.~I. and {Siebert}, A. and {Siltala}, L. and {Silvelo}, A. and {Slezak}, E. and {Slezak}, I. and {Smart}, R.~L. and {Snaith}, O.~N. and {Solano}, E. and {Solitro}, F. and {Souami}, D. and {Souchay}, J. and {Spagna}, A. and {Spina}, L. and {Spoto}, F. and {Steele}, I.~A. and {Steidelm{\"u}ller}, H. and {Stephenson}, C.~A. and {S{\"u}veges}, M. and {Surdej}, J. and {Szabados}, L. and {Szegedi-Elek}, E. and {Taris}, F. and {Taylor}, M.~B. and {Teixeira}, R. and {Tolomei}, L. and {Tonello}, N. and {Torra}, F. and {Torra}, J. and {Torralba Elipe}, G. and {Trabucchi}, M. and {Tsounis}, A.~T. and {Turon}, C. and {Ulla}, A. and {Unger}, N. and {Vaillant}, M.~V. and {van Dillen}, E. and {van Reeven}, W. and {Vanel}, O. and {Vecchiato}, A. and {Viala}, Y. and {Vicente}, D. and {Voutsinas}, S. and {Weiler}, M. and {Wevers}, T. and {Wyrzykowski}, {\L}. and {Yoldas}, A. and {Yvard}, P. and {Zhao}, H. and {Zorec}, J. and {Zucker}, S. and {Zwitter}, T.},
        title = "{Gaia Data Release 3. Summary of the content and survey properties}",
      journal = {\aap},
         year = 2023,
        month = jun,
       volume = {674},
          eid = {A1},
        pages = {A1},
          doi = {10.1051/0004-6361/202243940},
archivePrefix = {arXiv},
       eprint = {2208.00211},
 primaryClass = {astro-ph.GA},
       adsurl = {https://ui.adsabs.harvard.edu/abs/2023A&A...674A...1G}
}

@ARTICLE{2017MNRAS.469.2913C,
       author = {{Cochrane}, R.~K. and {Best}, P.~N. and {Sobral}, D. and {Smail}, I. and {Wake}, D.~A. and {Stott}, J.~P. and {Geach}, J.~E.},
        title = "{The H {\ensuremath{\alpha}} luminosity-dependent clustering of star-forming galaxies from z {\ensuremath{\sim}} 0.8 to {\ensuremath{\sim}}2.2 with HiZELS}",
      journal = {\mnras},
         year = 2017,
        month = aug,
       volume = {469},
       number = {3},
        pages = {2913-2932},
          doi = {10.1093/mnras/stx957},
archivePrefix = {arXiv},
       eprint = {1704.05472},
 primaryClass = {astro-ph.GA},
       adsurl = {https://ui.adsabs.harvard.edu/abs/2017MNRAS.469.2913C}
}

@ARTICLE{2018MNRAS.475.3730C,
       author = {{Cochrane}, R.~K. and {Best}, P.~N. and {Sobral}, D. and {Smail}, I. and {Geach}, J.~E. and {Stott}, J.~P. and {Wake}, D.~A.},
        title = "{The dependence of galaxy clustering on stellar mass, star-formation rate and redshift at z = 0.8-2.2, with HiZELS}",
      journal = {\mnras},
         year = 2018,
        month = apr,
       volume = {475},
       number = {3},
        pages = {3730-3745},
          doi = {10.1093/mnras/stx3345},
archivePrefix = {arXiv},
       eprint = {1801.04933},
 primaryClass = {astro-ph.GA},
       adsurl = {https://ui.adsabs.harvard.edu/abs/2018MNRAS.475.3730C}
}

@ARTICLE{2021MNRAS.503.2622C,
       author = {{Cochrane}, R.~K. and {Best}, P.~N. and {Smail}, I. and {Ibar}, E. and {Cheng}, C. and {Swinbank}, A.~M. and {Molina}, J. and {Sobral}, D. and {Dudzevi{\v{c}}i{\={u}}t{\.{e}}}, U.},
        title = "{Resolving a dusty, star-forming SHiZELS galaxy at z = 2.2 with HST, ALMA, and SINFONI on kiloparsec scales}",
      journal = {\mnras},
         year = 2021,
        month = may,
       volume = {503},
       number = {2},
        pages = {2622-2638},
          doi = {10.1093/mnras/stab467},
archivePrefix = {arXiv},
       eprint = {2102.07791},
 primaryClass = {astro-ph.GA},
       adsurl = {https://ui.adsabs.harvard.edu/abs/2021MNRAS.503.2622C}
}

@ARTICLE{2023ApJS..268...64W,
       author = {{Williams}, Christina C. and {Tacchella}, Sandro and {Maseda}, Michael V. and {Robertson}, Brant E. and {Johnson}, Benjamin D. and {Willott}, Chris J. and {Eisenstein}, Daniel J. and {Willmer}, Christopher N.~A. and {Ji}, Zhiyuan and {Hainline}, Kevin N. and {Helton}, Jakob M. and {Alberts}, Stacey and {Baum}, Stefi and {Bhatawdekar}, Rachana and {Boyett}, Kristan and {Bunker}, Andrew J. and {Carniani}, Stefano and {Charlot}, Stephane and {Chevallard}, Jacopo and {Curtis-Lake}, Emma and {de Graaff}, Anna and {Egami}, Eiichi and {Franx}, Marijn and {Kumari}, Nimisha and {Maiolino}, Roberto and {Nelson}, Erica J. and {Rieke}, Marcia J. and {Sandles}, Lester and {Shivaei}, Irene and {Simmonds}, Charlotte and {Smit}, Renske and {Suess}, Katherine A. and {Sun}, Fengwu and {{\"U}bler}, Hannah and {Witstok}, Joris},
        title = "{JEMS: A Deep Medium-band Imaging Survey in the Hubble Ultra Deep Field with JWST NIRCam and NIRISS}",
      journal = {\apjs},
         year = 2023,
        month = oct,
       volume = {268},
       number = {2},
          eid = {64},
        pages = {64},
          doi = {10.3847/1538-4365/acf130},
archivePrefix = {arXiv},
       eprint = {2301.09780},
 primaryClass = {astro-ph.GA},
       adsurl = {https://ui.adsabs.harvard.edu/abs/2023ApJS..268...64W}
}

@ARTICLE{1983ApJ...266..713O,
       author = {{Oke}, J.~B. and {Gunn}, J.~E.},
        title = "{Secondary standard stars for absolute spectrophotometry.}",
      journal = {\apj},
         year = 1983,
        month = mar,
       volume = {266},
        pages = {713-717},
          doi = {10.1086/160817},
       adsurl = {https://ui.adsabs.harvard.edu/abs/1983ApJ...266..713O}
}

@ARTICLE{2008ApJ...686.1503B,
       author = {{Brammer}, Gabriel B. and {van Dokkum}, Pieter G. and {Coppi}, Paolo},
        title = "{EAZY: A Fast, Public Photometric Redshift Code}",
      journal = {\apj},
         year = 2008,
        month = oct,
       volume = {686},
       number = {2},
        pages = {1503-1513},
          doi = {10.1086/591786},
archivePrefix = {arXiv},
       eprint = {0807.1533},
 primaryClass = {astro-ph},
       adsurl = {https://ui.adsabs.harvard.edu/abs/2008ApJ...686.1503B}
}

@ARTICLE{2024A&A...681A.118T,
       author = {{Traina}, A. and {Gruppioni}, C. and {Delvecchio}, I. and {Calura}, F. and {Bisigello}, L. and {Feltre}, A. and {Magnelli}, B. and {Schinnerer}, E. and {Liu}, D. and {Adscheid}, S. and {Behiri}, M. and {Gentile}, F. and {Pozzi}, F. and {Talia}, M. and {Zamorani}, G. and {Algera}, H. and {Gillman}, S. and {Lambrides}, E. and {Symeonidis}, M.},
        title = "{A$^{3}$COSMOS: The infrared luminosity function and dust-obscured star formation rate density at 0.5 < z < 6}",
      journal = {\aap},
         year = 2024,
        month = jan,
       volume = {681},
          eid = {A118},
        pages = {A118},
          doi = {10.1051/0004-6361/202347048},
archivePrefix = {arXiv},
       eprint = {2309.15150},
 primaryClass = {astro-ph.GA},
       adsurl = {https://ui.adsabs.harvard.edu/abs/2024A&A...681A.118T}
}

@ARTICLE{2013ApJ...763L...7E,
       author = {{Ellis}, Richard S. and {McLure}, Ross J. and {Dunlop}, James S. and {Robertson}, Brant E. and {Ono}, Yoshiaki and {Schenker}, Matthew A. and {Koekemoer}, Anton and {Bowler}, Rebecca A.~A. and {Ouchi}, Masami and {Rogers}, Alexander B. and {Curtis-Lake}, Emma and {Schneider}, Evan and {Charlot}, Stephane and {Stark}, Daniel P. and {Furlanetto}, Steven R. and {Cirasuolo}, Michele},
        title = "{The Abundance of Star-forming Galaxies in the Redshift Range 8.5-12: New Results from the 2012 Hubble Ultra Deep Field Campaign}",
      journal = {\apjl},
         year = 2013,
        month = jan,
       volume = {763},
       number = {1},
          eid = {L7},
        pages = {L7},
          doi = {10.1088/2041-8205/763/1/L7},
archivePrefix = {arXiv},
       eprint = {1211.6804},
 primaryClass = {astro-ph.CO},
       adsurl = {https://ui.adsabs.harvard.edu/abs/2013ApJ...763L...7E}
}

@ARTICLE{2013MNRAS.432.2696M,
       author = {{McLure}, R.~J. and {Dunlop}, J.~S. and {Bowler}, R.~A.~A. and {Curtis-Lake}, E. and {Schenker}, M. and {Ellis}, R.~S. and {Robertson}, B.~E. and {Koekemoer}, A.~M. and {Rogers}, A.~B. and {Ono}, Y. and {Ouchi}, M. and {Charlot}, S. and {Wild}, V. and {Stark}, D.~P. and {Furlanetto}, S.~R. and {Cirasuolo}, M. and {Targett}, T.~A.},
        title = "{A new multifield determination of the galaxy luminosity function at z = 7-9 incorporating the 2012 Hubble Ultra-Deep Field imaging}",
      journal = {\mnras},
         year = 2013,
        month = jul,
       volume = {432},
       number = {4},
        pages = {2696-2716},
          doi = {10.1093/mnras/stt627},
archivePrefix = {arXiv},
       eprint = {1212.5222},
 primaryClass = {astro-ph.CO},
       adsurl = {https://ui.adsabs.harvard.edu/abs/2013MNRAS.432.2696M}
}

@ARTICLE{2015ApJ...810...71F,
       author = {{Finkelstein}, Steven L. and {Ryan}, Russell E., Jr. and {Papovich}, Casey and {Dickinson}, Mark and {Song}, Mimi and {Somerville}, Rachel S. and {Ferguson}, Henry C. and {Salmon}, Brett and {Giavalisco}, Mauro and {Koekemoer}, Anton M. and {Ashby}, Matthew L.~N. and {Behroozi}, Peter and {Castellano}, Marco and {Dunlop}, James S. and {Faber}, Sandy M. and {Fazio}, Giovanni G. and {Fontana}, Adriano and {Grogin}, Norman A. and {Hathi}, Nimish and {Jaacks}, Jason and {Kocevski}, Dale D. and {Livermore}, Rachael and {McLure}, Ross J. and {Merlin}, Emiliano and {Mobasher}, Bahram and {Newman}, Jeffrey A. and {Rafelski}, Marc and {Tilvi}, Vithal and {Willner}, S.~P.},
        title = "{The Evolution of the Galaxy Rest-frame Ultraviolet Luminosity Function over the First Two Billion Years}",
      journal = {\apj},
         year = 2015,
        month = sep,
       volume = {810},
       number = {1},
          eid = {71},
        pages = {71},
          doi = {10.1088/0004-637X/810/1/71},
archivePrefix = {arXiv},
       eprint = {1410.5439},
 primaryClass = {astro-ph.GA},
       adsurl = {https://ui.adsabs.harvard.edu/abs/2015ApJ...810...71F}
}

@ARTICLE{2015MNRAS.450.3032M,
       author = {{McLeod}, D.~J. and {McLure}, R.~J. and {Dunlop}, J.~S. and {Robertson}, B.~E. and {Ellis}, R.~S. and {Targett}, T.~A.},
        title = "{New redshift z ≃ 9 galaxies in the Hubble Frontier Fields: implications for early evolution of the UV luminosity density}",
      journal = {\mnras},
         year = 2015,
        month = jul,
       volume = {450},
       number = {3},
        pages = {3032-3044},
          doi = {10.1093/mnras/stv780},
archivePrefix = {arXiv},
       eprint = {1412.1472},
 primaryClass = {astro-ph.GA},
       adsurl = {https://ui.adsabs.harvard.edu/abs/2015MNRAS.450.3032M}
}

@ARTICLE{2016MNRAS.459.3812M,
       author = {{McLeod}, D.~J. and {McLure}, R.~J. and {Dunlop}, J.~S.},
        title = "{The z = 9-10 galaxy population in the Hubble Frontier Fields and CLASH surveys: the z = 9 luminosity function and further evidence for a smooth decline in ultraviolet luminosity density at z{\ensuremath{\geq}} 8}",
      journal = {\mnras},
         year = 2016,
        month = jul,
       volume = {459},
       number = {4},
        pages = {3812-3824},
          doi = {10.1093/mnras/stw904},
archivePrefix = {arXiv},
       eprint = {1602.05199},
 primaryClass = {astro-ph.GA},
       adsurl = {https://ui.adsabs.harvard.edu/abs/2016MNRAS.459.3812M}
}

@ARTICLE{2018ApJ...855..105O,
       author = {{Oesch}, P.~A. and {Bouwens}, R.~J. and {Illingworth}, G.~D. and {Labb{\'e}}, I. and {Stefanon}, M.},
        title = "{The Dearth of z {\ensuremath{\sim}} 10 Galaxies in All HST Legacy Fields{\textemdash}The Rapid Evolution of the Galaxy Population in the First 500 Myr}",
      journal = {\apj},
         year = 2018,
        month = mar,
       volume = {855},
       number = {2},
          eid = {105},
        pages = {105},
          doi = {10.3847/1538-4357/aab03f},
archivePrefix = {arXiv},
       eprint = {1710.11131},
 primaryClass = {astro-ph.GA},
       adsurl = {https://ui.adsabs.harvard.edu/abs/2018ApJ...855..105O}
}

@ARTICLE{2021AJ....162...47B,
       author = {{Bouwens}, R.~J. and {Oesch}, P.~A. and {Stefanon}, M. and {Illingworth}, G. and {Labb{\'e}}, I. and {Reddy}, N. and {Atek}, H. and {Montes}, M. and {Naidu}, R. and {Nanayakkara}, T. and {Nelson}, E. and {Wilkins}, S.},
        title = "{New Determinations of the UV Luminosity Functions from z   9 to 2 Show a Remarkable Consistency with Halo Growth and a Constant Star Formation Efficiency}",
      journal = {\aj},
         year = 2021,
        month = aug,
       volume = {162},
       number = {2},
          eid = {47},
        pages = {47},
          doi = {10.3847/1538-3881/abf83e},
archivePrefix = {arXiv},
       eprint = {2102.07775},
 primaryClass = {astro-ph.GA},
       adsurl = {https://ui.adsabs.harvard.edu/abs/2021AJ....162...47B}
}

@ARTICLE{2022ApJ...940...55B,
       author = {{Bouwens}, R.~J. and {Illingworth}, G. and {Ellis}, R.~S. and {Oesch}, P. and {Stefanon}, M.},
        title = "{z   2-9 Galaxies Magnified by the Hubble Frontier Field Clusters. II. Luminosity Functions and Constraints on a Faint-end Turnover}",
      journal = {\apj},
         year = 2022,
        month = nov,
       volume = {940},
       number = {1},
          eid = {55},
        pages = {55},
          doi = {10.3847/1538-4357/ac86d1},
archivePrefix = {arXiv},
       eprint = {2205.11526},
 primaryClass = {astro-ph.GA},
       adsurl = {https://ui.adsabs.harvard.edu/abs/2022ApJ...940...55B}
}

@INPROCEEDINGS{2023AAS...24110001R,
       author = {{Rigby}, Jane},
        title = "{The science performance of JWST}",
    booktitle = {American Astronomical Society Meeting Abstracts},
         year = 2023,
       series = {American Astronomical Society Meeting Abstracts},
       volume = {241},
        month = jan,
          eid = {100.01},
        pages = {100.01},
       adsurl = {https://ui.adsabs.harvard.edu/abs/2023AAS...24110001R}
}

@ARTICLE{2023MNRAS.520.4554D,
       author = {{Donnan}, C.~T. and {McLeod}, D.~J. and {McLure}, R.~J. and {Dunlop}, J.~S. and {Carnall}, A.~C. and {Cullen}, F. and {Magee}, D.},
        title = "{The abundance of z {\ensuremath{\gtrsim}} 10 galaxy candidates in the HUDF using deep JWST NIRCam medium-band imaging}",
      journal = {\mnras},
         year = 2023,
        month = apr,
       volume = {520},
       number = {3},
        pages = {4554-4561},
          doi = {10.1093/mnras/stad471},
archivePrefix = {arXiv},
       eprint = {2212.10126},
 primaryClass = {astro-ph.GA},
       adsurl = {https://ui.adsabs.harvard.edu/abs/2023MNRAS.520.4554D}
}

@ARTICLE{2023MNRAS.518.6011D,
       author = {{Donnan}, C.~T. and {McLeod}, D.~J. and {Dunlop}, J.~S. and {McLure}, R.~J. and {Carnall}, A.~C. and {Begley}, R. and {Cullen}, F. and {Hamadouche}, M.~L. and {Bowler}, R.~A.~A. and {Magee}, D. and {McCracken}, H.~J. and {Milvang-Jensen}, B. and {Moneti}, A. and {Targett}, T.},
        title = "{The evolution of the galaxy UV luminosity function at redshifts z ≃ 8 - 15 from deep JWST and ground-based near-infrared imaging}",
      journal = {\mnras},
         year = 2023,
        month = feb,
       volume = {518},
       number = {4},
        pages = {6011-6040},
          doi = {10.1093/mnras/stac3472},
archivePrefix = {arXiv},
       eprint = {2207.12356},
 primaryClass = {astro-ph.GA},
       adsurl = {https://ui.adsabs.harvard.edu/abs/2023MNRAS.518.6011D}
}

@ARTICLE{2024MNRAS.527.5004M,
       author = {{McLeod}, D.~J. and {Donnan}, C.~T. and {McLure}, R.~J. and {Dunlop}, J.~S. and {Magee}, D. and {Begley}, R. and {Carnall}, A.~C. and {Cullen}, F. and {Ellis}, R.~S. and {Hamadouche}, M.~L. and {Stanton}, T.~M.},
        title = "{The galaxy UV luminosity function at z ≃ 11 from a suite of public JWST ERS, ERO, and Cycle-1 programs}",
      journal = {\mnras},
         year = 2024,
        month = jan,
       volume = {527},
       number = {3},
        pages = {5004-5022},
          doi = {10.1093/mnras/stad3471},
archivePrefix = {arXiv},
       eprint = {2304.14469},
 primaryClass = {astro-ph.GA},
       adsurl = {https://ui.adsabs.harvard.edu/abs/2024MNRAS.527.5004M}
}

@ARTICLE{2024ApJ...965..169A,
       author = {{Adams}, Nathan J. and {Conselice}, Christopher J. and {Austin}, Duncan and {Harvey}, Thomas and {Ferreira}, Leonardo and {Trussler}, James and {Juod{\v{z}}balis}, Ignas and {Li}, Qiong and {Windhorst}, Rogier and {Cohen}, Seth H. and {Jansen}, Rolf A. and {Summers}, Jake and {Tompkins}, Scott and {Driver}, Simon P. and {Robotham}, Aaron and {D'Silva}, Jordan C.~J. and {Yan}, Haojing and {Coe}, Dan and {Frye}, Brenda and {Grogin}, Norman A. and {Koekemoer}, Anton M. and {Marshall}, Madeline A. and {Pirzkal}, Nor and {Ryan}, Russell E. and {Maksym}, W. Peter and {Rutkowski}, Michael J. and {Willmer}, Christopher N.~A. and {Hammel}, Heidi B. and {Nonino}, Mario and {Bhatawdekar}, Rachana and {Wilkins}, Stephen M. and {Bradley}, Larry D. and {Broadhurst}, Tom and {Cheng}, Cheng and {Dole}, Herv{\'e} and {Hathi}, Nimish P. and {Zitrin}, Adi},
        title = "{EPOCHS. II. The Ultraviolet Luminosity Function from 7.5 < z < 13.5 Using 180 arcmin$^{2}$ of Deep, Blank Fields from the PEARLS Survey and Public JWST Data}",
      journal = {\apj},
         year = 2024,
        month = apr,
       volume = {965},
       number = {2},
          eid = {169},
        pages = {169},
          doi = {10.3847/1538-4357/ad2a7b},
archivePrefix = {arXiv},
       eprint = {2304.13721},
 primaryClass = {astro-ph.GA},
       adsurl = {https://ui.adsabs.harvard.edu/abs/2024ApJ...965..169A}
}

@ARTICLE{2022ApJ...940L..14N,
       author = {{Naidu}, Rohan P. and {Oesch}, Pascal A. and {van Dokkum}, Pieter and {Nelson}, Erica J. and {Suess}, Katherine A. and {Brammer}, Gabriel and {Whitaker}, Katherine E. and {Illingworth}, Garth and {Bouwens}, Rychard and {Tacchella}, Sandro and {Matthee}, Jorryt and {Allen}, Natalie and {Bezanson}, Rachel and {Conroy}, Charlie and {Labbe}, Ivo and {Leja}, Joel and {Leonova}, Ecaterina and {Magee}, Dan and {Price}, Sedona H. and {Setton}, David J. and {Strait}, Victoria and {Stefanon}, Mauro and {Toft}, Sune and {Weaver}, John R. and {Weibel}, Andrea},
        title = "{Two Remarkably Luminous Galaxy Candidates at z {\ensuremath{\approx}} 10-12 Revealed by JWST}",
      journal = {\apjl},
         year = 2022,
        month = nov,
       volume = {940},
       number = {1},
          eid = {L14},
        pages = {L14},
          doi = {10.3847/2041-8213/ac9b22},
archivePrefix = {arXiv},
       eprint = {2207.09434},
 primaryClass = {astro-ph.GA},
       adsurl = {https://ui.adsabs.harvard.edu/abs/2022ApJ...940L..14N}
}

@ARTICLE{2023ApJS..265....5H,
       author = {{Harikane}, Yuichi and {Ouchi}, Masami and {Oguri}, Masamune and {Ono}, Yoshiaki and {Nakajima}, Kimihiko and {Isobe}, Yuki and {Umeda}, Hiroya and {Mawatari}, Ken and {Zhang}, Yechi},
        title = "{A Comprehensive Study of Galaxies at z   9-16 Found in the Early JWST Data: Ultraviolet Luminosity Functions and Cosmic Star Formation History at the Pre-reionization Epoch}",
      journal = {\apjs},
         year = 2023,
        month = mar,
       volume = {265},
       number = {1},
          eid = {5},
        pages = {5},
          doi = {10.3847/1538-4365/acaaa9},
archivePrefix = {arXiv},
       eprint = {2208.01612},
 primaryClass = {astro-ph.GA},
       adsurl = {https://ui.adsabs.harvard.edu/abs/2023ApJS..265....5H}
}

@ARTICLE{2025ApJ...980..242S,
       author = {{Shapley}, Alice E. and {Sanders}, Ryan L. and {Topping}, Michael W. and {Reddy}, Naveen A. and {Berg}, Danielle A. and {Bouwens}, Rychard J. and {Brammer}, Gabriel and {Carnall}, Adam C. and {Cullen}, Fergus and {Dav{\'e}}, Romeel and {Dunlop}, James S. and {Ellis}, Richard S. and {F{\"o}rster Schreiber}, N.~M. and {Furlanetto}, Steven R. and {Glazebrook}, Karl and {Illingworth}, Garth D. and {Jones}, Tucker and {Kriek}, Mariska and {McLeod}, Derek J. and {McLure}, Ross J. and {Narayanan}, Desika and {Oesch}, Pascal and {Pahl}, Anthony J. and {Pettini}, Max and {Schaerer}, Daniel and {Stark}, Daniel P. and {Steidel}, Charles C. and {Tang}, Mengtao and {Clarke}, Leonardo and {Donnan}, Callum T. and {Kehoe}, Emily},
        title = "{The AURORA Survey: A New Era of Emission-line Diagrams with JWST/NIRSpec}",
      journal = {\apj},
         year = 2025,
        month = feb,
       volume = {980},
       number = {2},
          eid = {242},
        pages = {242},
          doi = {10.3847/1538-4357/adad68},
archivePrefix = {arXiv},
       eprint = {2407.00157},
 primaryClass = {astro-ph.GA},
       adsurl = {https://ui.adsabs.harvard.edu/abs/2025ApJ...980..242S}
}

@ARTICLE{2023ApJ...950L...1S,
       author = {{Shapley}, Alice E. and {Reddy}, Naveen A. and {Sanders}, Ryan L. and {Topping}, Michael W. and {Brammer}, Gabriel B.},
        title = "{JWST/NIRSpec Measurements of the Relationships between Nebular Emission-line Ratios and Stellar Mass at z   3-6}",
      journal = {\apjl},
         year = 2023,
        month = jun,
       volume = {950},
       number = {1},
          eid = {L1},
        pages = {L1},
          doi = {10.3847/2041-8213/acd939},
archivePrefix = {arXiv},
       eprint = {2303.00410},
 primaryClass = {astro-ph.GA},
       adsurl = {https://ui.adsabs.harvard.edu/abs/2023ApJ...950L...1S}
}

@ARTICLE{2023ApJ...955...54S,
       author = {{Sanders}, Ryan L. and {Shapley}, Alice E. and {Topping}, Michael W. and {Reddy}, Naveen A. and {Brammer}, Gabriel B.},
        title = "{Excitation and Ionization Properties of Star-forming Galaxies at z = 2.0-9.3 with JWST/NIRSpec}",
      journal = {\apj},
         year = 2023,
        month = sep,
       volume = {955},
       number = {1},
          eid = {54},
        pages = {54},
          doi = {10.3847/1538-4357/acedad},
archivePrefix = {arXiv},
       eprint = {2301.06696},
 primaryClass = {astro-ph.GA},
       adsurl = {https://ui.adsabs.harvard.edu/abs/2023ApJ...955...54S}
}

@ARTICLE{2024ApJ...962...24S,
       author = {{Sanders}, Ryan L. and {Shapley}, Alice E. and {Topping}, Michael W. and {Reddy}, Naveen A. and {Brammer}, Gabriel B.},
        title = "{Direct T $_{e}$-based Metallicities of z = 2{\textendash}9 Galaxies with JWST/NIRSpec: Empirical Metallicity Calibrations Applicable from Reionization to Cosmic Noon}",
      journal = {\apj},
         year = 2024,
        month = feb,
       volume = {962},
       number = {1},
          eid = {24},
        pages = {24},
          doi = {10.3847/1538-4357/ad15fc},
archivePrefix = {arXiv},
       eprint = {2303.08149},
 primaryClass = {astro-ph.GA},
       adsurl = {https://ui.adsabs.harvard.edu/abs/2024ApJ...962...24S}
}

@ARTICLE{2023A&A...677A.115C,
       author = {{Cameron}, Alex J. and {Saxena}, Aayush and {Bunker}, Andrew J. and {D'Eugenio}, Francesco and {Carniani}, Stefano and {Maiolino}, Roberto and {Curtis-Lake}, Emma and {Ferruit}, Pierre and {Jakobsen}, Peter and {Arribas}, Santiago and {Bonaventura}, Nina and {Charlot}, Stephane and {Chevallard}, Jacopo and {Curti}, Mirko and {Looser}, Tobias J. and {Maseda}, Michael V. and {Rawle}, Tim and {Rodr{\'\i}guez Del Pino}, Bruno and {Smit}, Renske and {{\"U}bler}, Hannah and {Willott}, Chris and {Witstok}, Joris and {Egami}, Eiichi and {Eisenstein}, Daniel J. and {Johnson}, Benjamin D. and {Hainline}, Kevin and {Rieke}, Marcia and {Robertson}, Brant E. and {Stark}, Daniel P. and {Tacchella}, Sandro and {Williams}, Christina C. and {Willmer}, Christopher N.~A. and {Bhatawdekar}, Rachana and {Bowler}, Rebecca and {Boyett}, Kristan and {Circosta}, Chiara and {Helton}, Jakob M. and {Jones}, Gareth C. and {Kumari}, Nimisha and {Ji}, Zhiyuan and {Nelson}, Erica and {Parlanti}, Eleonora and {Sandles}, Lester and {Scholtz}, Jan and {Sun}, Fengwu},
        title = "{JADES: Probing interstellar medium conditions at z {\ensuremath{\sim}} 5.5-9.5 with ultra-deep JWST/NIRSpec spectroscopy}",
      journal = {\aap},
         year = 2023,
        month = sep,
       volume = {677},
          eid = {A115},
        pages = {A115},
          doi = {10.1051/0004-6361/202346107},
archivePrefix = {arXiv},
       eprint = {2302.04298},
 primaryClass = {astro-ph.GA},
       adsurl = {https://ui.adsabs.harvard.edu/abs/2023A&A...677A.115C}
}

@ARTICLE{2023ApJS..269...33N,
       author = {{Nakajima}, Kimihiko and {Ouchi}, Masami and {Isobe}, Yuki and {Harikane}, Yuichi and {Zhang}, Yechi and {Ono}, Yoshiaki and {Umeda}, Hiroya and {Oguri}, Masamune},
        title = "{JWST Census for the Mass-Metallicity Star Formation Relations at z = 4-10 with Self-consistent Flux Calibration and Proper Metallicity Calibrators}",
      journal = {\apjs},
         year = 2023,
        month = dec,
       volume = {269},
       number = {2},
          eid = {33},
        pages = {33},
          doi = {10.3847/1538-4365/acd556},
archivePrefix = {arXiv},
       eprint = {2301.12825},
 primaryClass = {astro-ph.GA},
       adsurl = {https://ui.adsabs.harvard.edu/abs/2023ApJS..269...33N}
}

@ARTICLE{2024ApJ...976..193R,
       author = {{Roberts-Borsani}, Guido and {Treu}, Tommaso and {Shapley}, Alice and {Fontana}, Adriano and {Pentericci}, Laura and {Castellano}, Marco and {Morishita}, Takahiro and {Bergamini}, Pietro and {Rosati}, Piero},
        title = "{Between the Extremes: A JWST Spectroscopic Benchmark for High-redshift Galaxies Using {\ensuremath{\sim}}500 Confirmed Sources at z {\ensuremath{\geq}} 5}",
      journal = {\apj},
         year = 2024,
        month = dec,
       volume = {976},
       number = {2},
          eid = {193},
        pages = {193},
          doi = {10.3847/1538-4357/ad85d3},
archivePrefix = {arXiv},
       eprint = {2403.07103},
 primaryClass = {astro-ph.GA},
       adsurl = {https://ui.adsabs.harvard.edu/abs/2024ApJ...976..193R}
}

@ARTICLE{2023MNRAS.525.2864O,
       author = {{Oesch}, P.~A. and {Brammer}, G. and {Naidu}, R.~P. and {Bouwens}, R.~J. and {Chisholm}, J. and {Illingworth}, G.~D. and {Matthee}, J. and {Nelson}, E. and {Qin}, Y. and {Reddy}, N. and {Shapley}, A. and {Shivaei}, I. and {van Dokkum}, P. and {Weibel}, A. and {Whitaker}, K. and {Wuyts}, S. and {Covelo-Paz}, A. and {Endsley}, R. and {Fudamoto}, Y. and {Giovinazzo}, E. and {Herard-Demanche}, T. and {Kerutt}, J. and {Kramarenko}, I. and {Labbe}, I. and {Leonova}, E. and {Lin}, J. and {Magee}, D. and {Marchesini}, D. and {Maseda}, M. and {Mason}, C. and {Matharu}, J. and {Meyer}, R.~A. and {Neufeld}, C. and {Prieto Lyon}, G. and {Schaerer}, D. and {Sharma}, R. and {Shuntov}, M. and {Smit}, R. and {Stefanon}, M. and {Wyithe}, J.~S.~B. and {Xiao}, M.},
        title = "{The JWST FRESCO survey: legacy NIRCam/grism spectroscopy and imaging in the two GOODS fields}",
      journal = {\mnras},
         year = 2023,
        month = oct,
       volume = {525},
       number = {2},
        pages = {2864-2874},
          doi = {10.1093/mnras/stad2411},
archivePrefix = {arXiv},
       eprint = {2304.02026},
 primaryClass = {astro-ph.GA},
       adsurl = {https://ui.adsabs.harvard.edu/abs/2023MNRAS.525.2864O}
}

@ARTICLE{2019ApJ...871..128T,
       author = {{Theios}, Rachel L. and {Steidel}, Charles C. and {Strom}, Allison L. and {Rudie}, Gwen C. and {Trainor}, Ryan F. and {Reddy}, Naveen A.},
        title = "{Dust Attenuation, Star Formation, and Metallicity in z {\ensuremath{\sim}} 2-3 Galaxies from KBSS-MOSFIRE}",
      journal = {\apj},
         year = 2019,
        month = jan,
       volume = {871},
       number = {1},
          eid = {128},
        pages = {128},
          doi = {10.3847/1538-4357/aaf386},
archivePrefix = {arXiv},
       eprint = {1805.00016},
 primaryClass = {astro-ph.GA},
       adsurl = {https://ui.adsabs.harvard.edu/abs/2019ApJ...871..128T}
}

@ARTICLE{1989ApJ...345..245C,
       author = {{Cardelli}, Jason A. and {Clayton}, Geoffrey C. and {Mathis}, John S.},
        title = "{The Relationship between Infrared, Optical, and Ultraviolet Extinction}",
      journal = {\apj},
         year = 1989,
        month = oct,
       volume = {345},
        pages = {245},
          doi = {10.1086/167900},
       adsurl = {https://ui.adsabs.harvard.edu/abs/1989ApJ...345..245C}
}

@ARTICLE{2023Natur.622..707A,
       author = {{Arrabal Haro}, Pablo and {Dickinson}, Mark and {Finkelstein}, Steven L. and {Kartaltepe}, Jeyhan S. and {Donnan}, Callum T. and {Burgarella}, Denis and {Carnall}, Adam C. and {Cullen}, Fergus and {Dunlop}, James S. and {Fern{\'a}ndez}, Vital and {Fujimoto}, Seiji and {Jung}, Intae and {Krips}, Melanie and {Larson}, Rebecca L. and {Papovich}, Casey and {P{\'e}rez-Gonz{\'a}lez}, Pablo G. and {Amor{\'\i}n}, Ricardo O. and {Bagley}, Micaela B. and {Buat}, V{\'e}ronique and {Casey}, Caitlin M. and {Chworowsky}, Katherine and {Cohen}, Seth H. and {Ferguson}, Henry C. and {Giavalisco}, Mauro and {Huertas-Company}, Marc and {Hutchison}, Taylor A. and {Kocevski}, Dale D. and {Koekemoer}, Anton M. and {Lucas}, Ray A. and {McLeod}, Derek J. and {McLure}, Ross J. and {Pirzkal}, Norbert and {Seill{\'e}}, Lise-Marie and {Trump}, Jonathan R. and {Weiner}, Benjamin J. and {Wilkins}, Stephen M. and {Zavala}, Jorge A.},
        title = "{Confirmation and refutation of very luminous galaxies in the early Universe}",
      journal = {\nat},
         year = 2023,
        month = oct,
       volume = {622},
       number = {7984},
        pages = {707-711},
          doi = {10.1038/s41586-023-06521-7},
archivePrefix = {arXiv},
       eprint = {2303.15431},
 primaryClass = {astro-ph.GA},
       adsurl = {https://ui.adsabs.harvard.edu/abs/2023Natur.622..707A}
}

@ARTICLE{2023ApJ...958..141L,
       author = {{Larson}, Rebecca L. and {Hutchison}, Taylor A. and {Bagley}, Micaela and {Finkelstein}, Steven L. and {Yung}, L.~Y. Aaron and {Somerville}, Rachel S. and {Hirschmann}, Michaela and {Brammer}, Gabriel and {Holwerda}, Benne W. and {Papovich}, Casey and {Morales}, Alexa M. and {Wilkins}, Stephen M.},
        title = "{Spectral Templates Optimal for Selecting Galaxies at z > 8 with the JWST}",
      journal = {\apj},
         year = 2023,
        month = dec,
       volume = {958},
       number = {2},
          eid = {141},
        pages = {141},
          doi = {10.3847/1538-4357/acfed4},
archivePrefix = {arXiv},
       eprint = {2211.10035},
 primaryClass = {astro-ph.GA},
       adsurl = {https://ui.adsabs.harvard.edu/abs/2023ApJ...958..141L}
}

@INPROCEEDINGS{2005SPIE.5904....1R,
       author = {{Rieke}, Marcia J. and {Kelly}, Douglas and {Horner}, Scott},
        title = "{Overview of James Webb Space Telescope and NIRCam's Role}",
    booktitle = {Cryogenic Optical Systems and Instruments XI},
         year = 2005,
       editor = {{Heaney}, James B. and {Burriesci}, Lawrence G.},
       series = {Society of Photo-Optical Instrumentation Engineers (SPIE) Conference Series},
       volume = {5904},
        month = aug,
        pages = {1-8},
          doi = {10.1117/12.615554},
       adsurl = {https://ui.adsabs.harvard.edu/abs/2005SPIE.5904....1R}
}

@ARTICLE{2023PASP..135b8001R,
       author = {{Rieke}, Marcia J. and {Kelly}, Douglas M. and {Misselt}, Karl and {Stansberry}, John and {Boyer}, Martha and {Beatty}, Thomas and {Egami}, Eiichi and {Florian}, Michael and {Greene}, Thomas P. and {Hainline}, Kevin and {Leisenring}, Jarron and {Roellig}, Thomas and {Schlawin}, Everett and {Sun}, Fengwu and {Tinnin}, Lee and {Williams}, Christina C. and {Willmer}, Christopher N.~A. and {Wilson}, Debra and {Clark}, Charles R. and {Rohrbach}, Scott and {Brooks}, Brian and {Canipe}, Alicia and {Correnti}, Matteo and {DiFelice}, Audrey and {Gennaro}, Mario and {Girard}, Julien H. and {Hartig}, George and {Hilbert}, Bryan and {Koekemoer}, Anton M. and {Nikolov}, Nikolay K. and {Pirzkal}, Norbert and {Rest}, Armin and {Robberto}, Massimo and {Sunnquist}, Ben and {Telfer}, Randal and {Wu}, Chi Rai and {Ferry}, Malcolm and {Lewis}, Dan and {Baum}, Stefi and {Beichman}, Charles and {Doyon}, Ren{\'e} and {Dressler}, Alan and {Eisenstein}, Daniel J. and {Ferrarese}, Laura and {Hodapp}, Klaus and {Horner}, Scott and {Jaffe}, Daniel T. and {Johnstone}, Doug and {Krist}, John and {Martin}, Peter and {McCarthy}, Donald W. and {Meyer}, Michael and {Rieke}, George H. and {Trauger}, John and {Young}, Erick T.},
        title = "{Performance of NIRCam on JWST in Flight}",
      journal = {\pasp},
         year = 2023,
        month = feb,
       volume = {135},
       number = {1044},
          eid = {028001},
        pages = {028001},
          doi = {10.1088/1538-3873/acac53},
archivePrefix = {arXiv},
       eprint = {2212.12069},
 primaryClass = {astro-ph.IM},
       adsurl = {https://ui.adsabs.harvard.edu/abs/2023PASP..135b8001R}
}

@ARTICLE{2022A&A...661A..80J,
       author = {{Jakobsen}, P. and {Ferruit}, P. and {Alves de Oliveira}, C. and {Arribas}, S. and {Bagnasco}, G. and {Barho}, R. and {Beck}, T.~L. and {Birkmann}, S. and {B{\"o}ker}, T. and {Bunker}, A.~J. and {Charlot}, S. and {de Jong}, P. and {de Marchi}, G. and {Ehrenwinkler}, R. and {Falcolini}, M. and {Fels}, R. and {Franx}, M. and {Franz}, D. and {Funke}, M. and {Giardino}, G. and {Gnata}, X. and {Holota}, W. and {Honnen}, K. and {Jensen}, P.~L. and {Jentsch}, M. and {Johnson}, T. and {Jollet}, D. and {Karl}, H. and {Kling}, G. and {K{\"o}hler}, J. and {Kolm}, M. -G. and {Kumari}, N. and {Lander}, M.~E. and {Lemke}, R. and {L{\'o}pez-Caniego}, M. and {L{\"u}tzgendorf}, N. and {Maiolino}, R. and {Manjavacas}, E. and {Marston}, A. and {Maschmann}, M. and {Maurer}, R. and {Messerschmidt}, B. and {Moseley}, S.~H. and {Mosner}, P. and {Mott}, D.~B. and {Muzerolle}, J. and {Pirzkal}, N. and {Pittet}, J. -F. and {Plitzke}, A. and {Posselt}, W. and {Rapp}, B. and {Rauscher}, B.~J. and {Rawle}, T. and {Rix}, H. -W. and {R{\"o}del}, A. and {Rumler}, P. and {Sabbi}, E. and {Salvignol}, J. -C. and {Schmid}, T. and {Sirianni}, M. and {Smith}, C. and {Strada}, P. and {te Plate}, M. and {Valenti}, J. and {Wettemann}, T. and {Wiehe}, T. and {Wiesmayer}, M. and {Willott}, C.~J. and {Wright}, R. and {Zeidler}, P. and {Zincke}, C.},
        title = "{The Near-Infrared Spectrograph (NIRSpec) on the James Webb Space Telescope. I. Overview of the instrument and its capabilities}",
      journal = {\aap},
         year = 2022,
        month = may,
       volume = {661},
          eid = {A80},
        pages = {A80},
          doi = {10.1051/0004-6361/202142663},
archivePrefix = {arXiv},
       eprint = {2202.03305},
 primaryClass = {astro-ph.IM},
       adsurl = {https://ui.adsabs.harvard.edu/abs/2022A&A...661A..80J}
}

@ARTICLE{2020MNRAS.491..944C,
       author = {{Curti}, Mirko and {Mannucci}, Filippo and {Cresci}, Giovanni and {Maiolino}, Roberto},
        title = "{The mass-metallicity and the fundamental metallicity relation revisited on a fully T$_{e}$-based abundance scale for galaxies}",
      journal = {\mnras},
         year = 2020,
        month = jan,
       volume = {491},
       number = {1},
        pages = {944-964},
          doi = {10.1093/mnras/stz2910},
archivePrefix = {arXiv},
       eprint = {1910.00597},
 primaryClass = {astro-ph.GA},
       adsurl = {https://ui.adsabs.harvard.edu/abs/2020MNRAS.491..944C}
}

@ARTICLE{2017PASA...34...58E,
       author = {{Eldridge}, J.~J. and {Stanway}, E.~R. and {Xiao}, L. and {McClelland}, L.~A.~S. and {Taylor}, G. and {Ng}, M. and {Greis}, S.~M.~L. and {Bray}, J.~C.},
        title = "{Binary Population and Spectral Synthesis Version 2.1: Construction, Observational Verification, and New Results}",
      journal = {\pasa},
         year = 2017,
        month = nov,
       volume = {34},
          eid = {e058},
        pages = {e058},
          doi = {10.1017/pasa.2017.51},
archivePrefix = {arXiv},
       eprint = {1710.02154},
 primaryClass = {astro-ph.SR},
       adsurl = {https://ui.adsabs.harvard.edu/abs/2017PASA...34...58E}
}

@ARTICLE{2018MNRAS.479...75S,
       author = {{Stanway}, E.~R. and {Eldridge}, J.~J.},
        title = "{Re-evaluating old stellar populations}",
      journal = {\mnras},
         year = 2018,
        month = sep,
       volume = {479},
       number = {1},
        pages = {75-93},
          doi = {10.1093/mnras/sty1353},
archivePrefix = {arXiv},
       eprint = {1805.08784},
 primaryClass = {astro-ph.GA},
       adsurl = {https://ui.adsabs.harvard.edu/abs/2018MNRAS.479...75S}
}

@ARTICLE{2013ApJ...777L...8K,
       author = {{Kashino}, D. and {Silverman}, J.~D. and {Rodighiero}, G. and {Renzini}, A. and {Arimoto}, N. and {Daddi}, E. and {Lilly}, S.~J. and {Sanders}, D.~B. and {Kartaltepe}, J. and {Zahid}, H.~J. and {Nagao}, T. and {Sugiyama}, N. and {Capak}, P. and {Carollo}, C.~M. and {Chu}, J. and {Hasinger}, G. and {Ilbert}, O. and {Kajisawa}, M. and {Kewley}, L.~J. and {Koekemoer}, A.~M. and {Kova{\v{c}}}, K. and {Le F{\`e}vre}, O. and {Masters}, D. and {McCracken}, H.~J. and {Onodera}, M. and {Scoville}, N. and {Strazzullo}, V. and {Symeonidis}, M. and {Taniguchi}, Y.},
        title = "{The FMOS-COSMOS Survey of Star-forming Galaxies at z \raisebox{-0.5ex}\textasciitilde 1.6. I. H{\ensuremath{\alpha}}-based Star Formation Rates and Dust Extinction}",
      journal = {\apjl},
         year = 2013,
        month = nov,
       volume = {777},
       number = {1},
          eid = {L8},
        pages = {L8},
          doi = {10.1088/2041-8205/777/1/L8},
archivePrefix = {arXiv},
       eprint = {1309.4774},
 primaryClass = {astro-ph.CO},
       adsurl = {https://ui.adsabs.harvard.edu/abs/2013ApJ...777L...8K}
}

@ARTICLE{2015ApJ...806..259R,
       author = {{Reddy}, Naveen A. and {Kriek}, Mariska and {Shapley}, Alice E. and {Freeman}, William R. and {Siana}, Brian and {Coil}, Alison L. and {Mobasher}, Bahram and {Price}, Sedona H. and {Sanders}, Ryan L. and {Shivaei}, Irene},
        title = "{The MOSDEF Survey: Measurements of Balmer Decrements and the Dust Attenuation Curve at Redshifts z \raisebox{-0.5ex}\textasciitilde 1.4-2.6}",
      journal = {\apj},
         year = 2015,
        month = jun,
       volume = {806},
       number = {2},
          eid = {259},
        pages = {259},
          doi = {10.1088/0004-637X/806/2/259},
archivePrefix = {arXiv},
       eprint = {1504.02782},
 primaryClass = {astro-ph.GA},
       adsurl = {https://ui.adsabs.harvard.edu/abs/2015ApJ...806..259R}
}

@ARTICLE{2020ApJ...902..123R,
       author = {{Reddy}, Naveen A. and {Shapley}, Alice E. and {Kriek}, Mariska and {Steidel}, Charles C. and {Shivaei}, Irene and {Sanders}, Ryan L. and {Mobasher}, Bahram and {Coil}, Alison L. and {Siana}, Brian and {Freeman}, William R. and {Azadi}, Mojegan and {Fetherolf}, Tara and {Leung}, Gene and {Price}, Sedona H. and {Zick}, Tom},
        title = "{The MOSDEF Survey: The First Direct Measurements of the Nebular Dust Attenuation Curve at High Redshift}",
      journal = {\apj},
         year = 2020,
        month = oct,
       volume = {902},
       number = {2},
          eid = {123},
        pages = {123},
          doi = {10.3847/1538-4357/abb674},
archivePrefix = {arXiv},
       eprint = {2009.10085},
 primaryClass = {astro-ph.GA},
       adsurl = {https://ui.adsabs.harvard.edu/abs/2020ApJ...902..123R}
}

@ARTICLE{2021ApJ...914...19S,
       author = {{Sanders}, Ryan L. and {Shapley}, Alice E. and {Jones}, Tucker and {Reddy}, Naveen A. and {Kriek}, Mariska and {Siana}, Brian and {Coil}, Alison L. and {Mobasher}, Bahram and {Shivaei}, Irene and {Dav{\'e}}, Romeel and {Azadi}, Mojegan and {Price}, Sedona H. and {Leung}, Gene and {Freeman}, William R. and {Fetherolf}, Tara and {de Groot}, Laura and {Zick}, Tom and {Barro}, Guillermo},
        title = "{The MOSDEF Survey: The Evolution of the Mass-Metallicity Relation from z = 0 to z 3.3}",
      journal = {\apj},
         year = 2021,
        month = jun,
       volume = {914},
       number = {1},
          eid = {19},
        pages = {19},
          doi = {10.3847/1538-4357/abf4c1},
archivePrefix = {arXiv},
       eprint = {2009.07292},
 primaryClass = {astro-ph.GA},
       adsurl = {https://ui.adsabs.harvard.edu/abs/2021ApJ...914...19S}
}

@ARTICLE{2022ApJ...926...31R,
       author = {{Reddy}, Naveen A. and {Topping}, Michael W. and {Shapley}, Alice E. and {Steidel}, Charles C. and {Sanders}, Ryan L. and {Du}, Xinnan and {Coil}, Alison L. and {Mobasher}, Bahram and {Price}, Sedona H. and {Shivaei}, Irene},
        title = "{The Effects of Stellar Population and Gas Covering Fraction on the Emergent Ly{\ensuremath{\alpha}} Emission of High-redshift Galaxies}",
      journal = {\apj},
         year = 2022,
        month = feb,
       volume = {926},
       number = {1},
          eid = {31},
        pages = {31},
          doi = {10.3847/1538-4357/ac3b4c},
archivePrefix = {arXiv},
       eprint = {2108.05363},
 primaryClass = {astro-ph.GA},
       adsurl = {https://ui.adsabs.harvard.edu/abs/2022ApJ...926...31R}
}

@ARTICLE{2018MNRAS.476.3218C,
       author = {{Cullen}, F. and {McLure}, R.~J. and {Khochfar}, S. and {Dunlop}, J.~S. and {Dalla Vecchia}, C. and {Carnall}, A.~C. and {Bourne}, N. and {Castellano}, M. and {Cimatti}, A. and {Cirasuolo}, M. and {Elbaz}, D. and {Fynbo}, J.~P.~U. and {Garilli}, B. and {Koekemoer}, A. and {Marchi}, F. and {Pentericci}, L. and {Talia}, M. and {Zamorani}, G.},
        title = "{The VANDELS survey: dust attenuation in star-forming galaxies at z = 3-4}",
      journal = {\mnras},
         year = 2018,
        month = may,
       volume = {476},
       number = {3},
        pages = {3218-3232},
          doi = {10.1093/mnras/sty469},
archivePrefix = {arXiv},
       eprint = {1712.01292},
 primaryClass = {astro-ph.GA},
       adsurl = {https://ui.adsabs.harvard.edu/abs/2018MNRAS.476.3218C}
}

@ARTICLE{2023OJAp....6E..44K,
       author = {{Katz}, Harley and {Rosdahl}, Joki and {Kimm}, Taysun and {Blaizot}, Jeremy and {Choustikov}, Nicholas and {Farcy}, Marion and {Garel}, Thibault and {Haehnelt}, Martin G. and {Michel-Dansac}, Leo and {Ocvirk}, Pierre},
        title = "{The SPHINX Public Data Release: Forward Modelling High-Redshift JWST Observations with Cosmological Radiation Hydrodynamics Simulations}",
      journal = {The Open Journal of Astrophysics},
         year = 2023,
        month = dec,
       volume = {6},
          eid = {44},
        pages = {44},
          doi = {10.21105/astro.2309.03269},
archivePrefix = {arXiv},
       eprint = {2309.03269},
 primaryClass = {astro-ph.GA},
       adsurl = {https://ui.adsabs.harvard.edu/abs/2023OJAp....6E..44K}
}

@INCOLLECTION{2013seg..book..419C,
       author = {{Calzetti}, Daniela},
        title = "{Star Formation Rate Indicators}",
    booktitle = {Secular Evolution of Galaxies},
         year = 2013,
       editor = {{Falc{\'o}n-Barroso}, Jes{\'u}s and {Knapen}, Johan H.},
        pages = {419},
          doi = {10.48550/arXiv.1208.2997},
       adsurl = {https://ui.adsabs.harvard.edu/abs/2013seg..book..419C}
}

@ARTICLE{2012ApJS..200...13B,
       author = {{Brammer}, Gabriel B. and {van Dokkum}, Pieter G. and {Franx}, Marijn and {Fumagalli}, Mattia and {Patel}, Shannon and {Rix}, Hans-Walter and {Skelton}, Rosalind E. and {Kriek}, Mariska and {Nelson}, Erica and {Schmidt}, Kasper B. and {Bezanson}, Rachel and {da Cunha}, Elisabete and {Erb}, Dawn K. and {Fan}, Xiaohui and {F{\"o}rster Schreiber}, Natascha and {Illingworth}, Garth D. and {Labb{\'e}}, Ivo and {Leja}, Joel and {Lundgren}, Britt and {Magee}, Dan and {Marchesini}, Danilo and {McCarthy}, Patrick and {Momcheva}, Ivelina and {Muzzin}, Adam and {Quadri}, Ryan and {Steidel}, Charles C. and {Tal}, Tomer and {Wake}, David and {Whitaker}, Katherine E. and {Williams}, Anna},
        title = "{3D-HST: A Wide-field Grism Spectroscopic Survey with the Hubble Space Telescope}",
      journal = {\apjs},
         year = 2012,
        month = jun,
       volume = {200},
       number = {2},
          eid = {13},
        pages = {13},
          doi = {10.1088/0067-0049/200/2/13},
archivePrefix = {arXiv},
       eprint = {1204.2829},
 primaryClass = {astro-ph.CO},
       adsurl = {https://ui.adsabs.harvard.edu/abs/2012ApJS..200...13B}
}

@MISC{2018hst..prop15647T,
       author = {{Teplitz}, Harry},
        title = "{Ultraviolet Imaging of the Cosmic Assembly Near-infrared Deep Extragalactic Legacy Survey Fields (UVCANDELS)}",
 howpublished = {HST Proposal. Cycle 26, ID. \#15647},
         year = 2018,
        month = nov,
        pages = {15647},
       adsurl = {https://ui.adsabs.harvard.edu/abs/2018hst..prop15647T}
}

@ARTICLE{2019A&A...622A...3D,
       author = {{Duncan}, K.~J. and {Sabater}, J. and {R{\"o}ttgering}, H.~J.~A. and {Jarvis}, M.~J. and {Smith}, D.~J.~B. and {Best}, P.~N. and {Callingham}, J.~R. and {Cochrane}, R. and {Croston}, J.~H. and {Hardcastle}, M.~J. and {Mingo}, B. and {Morabito}, L. and {Nisbet}, D. and {Prandoni}, I. and {Shimwell}, T.~W. and {Tasse}, C. and {White}, G.~J. and {Williams}, W.~L. and {Alegre}, L. and {Chy{\.z}y}, K.~T. and {G{\"u}rkan}, G. and {Hoeft}, M. and {Kondapally}, R. and {Mechev}, A.~P. and {Miley}, G.~K. and {Schwarz}, D.~J. and {van Weeren}, R.~J.},
        title = "{The LOFAR Two-metre Sky Survey. IV. First Data Release: Photometric redshifts and rest-frame magnitudes}",
      journal = {\aap},
         year = 2019,
        month = feb,
       volume = {622},
          eid = {A3},
        pages = {A3},
          doi = {10.1051/0004-6361/201833562},
archivePrefix = {arXiv},
       eprint = {1811.07928},
 primaryClass = {astro-ph.GA},
       adsurl = {https://ui.adsabs.harvard.edu/abs/2019A&A...622A...3D}
}

@ARTICLE{2021ApJ...909..165Z,
       author = {{Zavala}, J.~A. and {Casey}, C.~M. and {Manning}, S.~M. and {Aravena}, M. and {Bethermin}, M. and {Caputi}, K.~I. and {Clements}, D.~L. and {Cunha}, E. da and {Drew}, P. and {Finkelstein}, S.~L. and {Fujimoto}, S. and {Hayward}, C. and {Hodge}, J. and {Kartaltepe}, J.~S. and {Knudsen}, K. and {Koekemoer}, A.~M. and {Long}, A.~S. and {Magdis}, G.~E. and {Man}, A.~W.~S. and {Popping}, G. and {Sanders}, D. and {Scoville}, N. and {Sheth}, K. and {Staguhn}, J. and {Toft}, S. and {Treister}, E. and {Vieira}, J.~D. and {Yun}, M.~S.},
        title = "{The Evolution of the IR Luminosity Function and Dust-obscured Star Formation over the Past 13 Billion Years}",
      journal = {\apj},
         year = 2021,
        month = mar,
       volume = {909},
       number = {2},
          eid = {165},
        pages = {165},
          doi = {10.3847/1538-4357/abdb27},
archivePrefix = {arXiv},
       eprint = {2101.04734},
 primaryClass = {astro-ph.GA},
       adsurl = {https://ui.adsabs.harvard.edu/abs/2021ApJ...909..165Z}
}

@ARTICLE{2013ApJ...775...93D,
       author = {{Dahlen}, Tomas and {Mobasher}, Bahram and {Faber}, Sandra M. and {Ferguson}, Henry C. and {Barro}, Guillermo and {Finkelstein}, Steven L. and {Finlator}, Kristian and {Fontana}, Adriano and {Gruetzbauch}, Ruth and {Johnson}, Seth and {Pforr}, Janine and {Salvato}, Mara and {Wiklind}, Tommy and {Wuyts}, Stijn and {Acquaviva}, Viviana and {Dickinson}, Mark E. and {Guo}, Yicheng and {Huang}, Jiasheng and {Huang}, Kuang-Han and {Newman}, Jeffrey A. and {Bell}, Eric F. and {Conselice}, Christopher J. and {Galametz}, Audrey and {Gawiser}, Eric and {Giavalisco}, Mauro and {Grogin}, Norman A. and {Hathi}, Nimish and {Kocevski}, Dale and {Koekemoer}, Anton M. and {Koo}, David C. and {Lee}, Kyoung-Soo and {McGrath}, Elizabeth J. and {Papovich}, Casey and {Peth}, Michael and {Ryan}, Russell and {Somerville}, Rachel and {Weiner}, Benjamin and {Wilson}, Grant},
        title = "{A Critical Assessment of Photometric Redshift Methods: A CANDELS Investigation}",
      journal = {\apj},
         year = 2013,
        month = oct,
       volume = {775},
       number = {2},
          eid = {93},
        pages = {93},
          doi = {10.1088/0004-637X/775/2/93},
archivePrefix = {arXiv},
       eprint = {1308.5353},
 primaryClass = {astro-ph.CO},
       adsurl = {https://ui.adsabs.harvard.edu/abs/2013ApJ...775...93D}
}

@ARTICLE{1974ApJS...27...21O,
       author = {{Oke}, J.~B.},
        title = "{Absolute Spectral Energy Distributions for White Dwarfs}",
      journal = {\apjs},
         year = 1974,
        month = feb,
       volume = {27},
        pages = {21},
          doi = {10.1086/190287},
       adsurl = {https://ui.adsabs.harvard.edu/abs/1974ApJS...27...21O}
}

@ARTICLE{2025A&A...694A.178C,
       author = {{Covelo-Paz}, Alba and {Giovinazzo}, Emma and {Oesch}, Pascal A. and {Meyer}, Romain A. and {Weibel}, Andrea and {Brammer}, Gabriel and {Fudamoto}, Yoshinobu and {Kerutt}, Josephine and {Lin}, Jamie and {Matharu}, Jasleen and {Naidu}, Rohan P. and {Velichko}, Anna and {Bollo}, Victoria and {Bouwens}, Rychard and {Chisholm}, John and {Illingworth}, Garth D. and {Kramarenko}, Ivan and {Magee}, Daniel and {Maseda}, Michael and {Matthee}, Jorryt and {Nelson}, Erica and {Reddy}, Naveen and {Schaerer}, Daniel and {Stefanon}, Mauro and {Xiao}, Mengyuan},
        title = "{An H{\ensuremath{\alpha}} view of galaxy buildup in the first 2 Gyr: Luminosity functions at z {\ensuremath{\sim}} 4‑6.5 from NIRCam/grism spectroscopy}",
      journal = {\aap},
         year = 2025,
        month = feb,
       volume = {694},
          eid = {A178},
        pages = {A178},
          doi = {10.1051/0004-6361/202452363},
archivePrefix = {arXiv},
       eprint = {2409.17241},
 primaryClass = {astro-ph.GA},
       adsurl = {https://ui.adsabs.harvard.edu/abs/2025A&A...694A.178C}
}

@ARTICLE{2015MNRAS.451.2303S,
       author = {{Sobral}, D. and {Matthee}, J. and {Best}, P.~N. and {Smail}, I. and {Khostovan}, A.~A. and {Milvang-Jensen}, B. and {Kim}, J. -W. and {Stott}, J. and {Calhau}, J. and {Nayyeri}, H. and {Mobasher}, B.},
        title = "{CF-HiZELS, an {\ensuremath{\sim}}10 deg$^{2}$ emission-line survey with spectroscopic follow-up: H{\ensuremath{\alpha}}, [O III] + H{\ensuremath{\beta}} and [O II] luminosity functions at z = 0.8, 1.4 and 2.2}",
      journal = {\mnras},
         year = 2015,
        month = aug,
       volume = {451},
       number = {3},
        pages = {2303-2323},
          doi = {10.1093/mnras/stv1076},
archivePrefix = {arXiv},
       eprint = {1502.06602},
 primaryClass = {astro-ph.GA},
       adsurl = {https://ui.adsabs.harvard.edu/abs/2015MNRAS.451.2303S}
}

@ARTICLE{2024MNRAS.533.1111E,
       author = {{Endsley}, Ryan and {Stark}, Daniel P. and {Whitler}, Lily and {Topping}, Michael W. and {Johnson}, Benjamin D. and {Robertson}, Brant and {Tacchella}, Sandro and {Alberts}, Stacey and {Baker}, William M. and {Bhatawdekar}, Rachana and {Boyett}, Kristan and {Bunker}, Andrew J. and {Cameron}, Alex J. and {Carniani}, Stefano and {Charlot}, Stephane and {Chen}, Zuyi and {Chevallard}, Jacopo and {Curtis-Lake}, Emma and {Danhaive}, A. Lola and {Egami}, Eiichi and {Eisenstein}, Daniel J. and {Hainline}, Kevin and {Helton}, Jakob M. and {Ji}, Zhiyuan and {Looser}, Tobias J. and {Maiolino}, Roberto and {Nelson}, Erica and {Pusk{\'a}s}, D{\'a}vid and {Rieke}, George and {Rieke}, Marcia and {Rix}, Hans-Walter and {Sandles}, Lester and {Saxena}, Aayush and {Simmonds}, Charlotte and {Smit}, Renske and {Sun}, Fengwu and {Williams}, Christina C. and {Willmer}, Christopher N.~A. and {Willott}, Chris and {Witstok}, Joris},
        title = "{The star-forming and ionizing properties of dwarf z 6-9 galaxies in JADES: insights on bursty star formation and ionized bubble growth}",
      journal = {\mnras},
         year = 2024,
        month = sep,
       volume = {533},
       number = {1},
        pages = {1111-1142},
          doi = {10.1093/mnras/stae1857},
archivePrefix = {arXiv},
       eprint = {2306.05295},
 primaryClass = {astro-ph.GA},
       adsurl = {https://ui.adsabs.harvard.edu/abs/2024MNRAS.533.1111E}
}

@ARTICLE{2023MNRAS.518.6142A,
       author = {{Algera}, Hiddo S.~B. and {Inami}, Hanae and {Oesch}, Pascal A. and {Sommovigo}, Laura and {Bouwens}, Rychard J. and {Topping}, Michael W. and {Schouws}, Sander and {Stefanon}, Mauro and {Stark}, Daniel P. and {Aravena}, Manuel and {Barrufet}, Laia and {da Cunha}, Elisabete and {Dayal}, Pratika and {Endsley}, Ryan and {Ferrara}, Andrea and {Fudamoto}, Yoshinobu and {Gonzalez}, Valentino and {Graziani}, Luca and {Hodge}, Jacqueline A. and {Hygate}, Alexander P.~S. and {de Looze}, Ilse and {Nanayakkara}, Themiya and {Schneider}, Raffaella and {van der Werf}, Paul P.},
        title = "{The ALMA REBELS survey: the dust-obscured cosmic star formation rate density at redshift 7}",
      journal = {\mnras},
         year = 2023,
        month = feb,
       volume = {518},
       number = {4},
        pages = {6142-6157},
          doi = {10.1093/mnras/stac3195},
archivePrefix = {arXiv},
       eprint = {2208.08243},
 primaryClass = {astro-ph.GA},
       adsurl = {https://ui.adsabs.harvard.edu/abs/2023MNRAS.518.6142A}
}

@ARTICLE{1983ApJ...272...54K,
       author = {{Kennicutt}, Jr., R.~C.},
        title = "{The rate of star formation in normal disk galaxies.}",
      journal = {\apj},
         year = 1983,
        month = sep,
       volume = {272},
        pages = {54-67},
          doi = {10.1086/161261},
       adsurl = {https://ui.adsabs.harvard.edu/abs/1983ApJ...272...54K}
}

@ARTICLE{2023ApJ...953...53S,
       author = {{Sun}, Fengwu and {Egami}, Eiichi and {Pirzkal}, Nor and {Rieke}, Marcia and {Baum}, Stefi and {Boyer}, Martha and {Boyett}, Kristan and {Bunker}, Andrew J. and {Cameron}, Alex J. and {Curti}, Mirko and {Eisenstein}, Daniel J. and {Gennaro}, Mario and {Greene}, Thomas P. and {Jaffe}, Daniel and {Kelly}, Doug and {Koekemoer}, Anton M. and {Kumari}, Nimisha and {Maiolino}, Roberto and {Maseda}, Michael and {Perna}, Michele and {Rest}, Armin and {Robertson}, Brant E. and {Schlawin}, Everett and {Smit}, Renske and {Stansberry}, John and {Sunnquist}, Ben and {Tacchella}, Sandro and {Williams}, Christina C. and {Willmer}, Christopher N.~A.},
        title = "{First Sample of H{\ensuremath{\alpha}}+[O III]{\ensuremath{\lambda}}5007 Line Emitters at z > 6 Through JWST/NIRCam Slitless Spectroscopy: Physical Properties and Line-luminosity Functions}",
      journal = {\apj},
         year = 2023,
        month = aug,
       volume = {953},
       number = {1},
          eid = {53},
        pages = {53},
          doi = {10.3847/1538-4357/acd53c},
archivePrefix = {arXiv},
       eprint = {2209.03374},
 primaryClass = {astro-ph.GA},
       adsurl = {https://ui.adsabs.harvard.edu/abs/2023ApJ...953...53S}
}

@ARTICLE{2013PASP..125..306F,
       author = {{Foreman-Mackey}, Daniel and {Hogg}, David W. and {Lang}, Dustin and {Goodman}, Jonathan},
        title = "{emcee: The MCMC Hammer}",
      journal = {\pasp},
         year = 2013,
        month = mar,
       volume = {125},
       number = {925},
        pages = {306},
          doi = {10.1086/670067},
archivePrefix = {arXiv},
       eprint = {1202.3665},
 primaryClass = {astro-ph.IM},
       adsurl = {https://ui.adsabs.harvard.edu/abs/2013PASP..125..306F}
}

@ARTICLE{1986ApJ...303..336G,
       author = {{Gehrels}, N.},
        title = "{Confidence Limits for Small Numbers of Events in Astrophysical Data}",
      journal = {\apj},
         year = 1986,
        month = apr,
       volume = {303},
        pages = {336},
          doi = {10.1086/164079},
       adsurl = {https://ui.adsabs.harvard.edu/abs/1986ApJ...303..336G}
}

@ARTICLE{2023ApJ...946..117B,
       author = {{Bollo}, Victoria and {Gonz{\'a}lez}, Valentino and {Stefanon}, Mauro and {Oesch}, Pascal A. and {Bouwens}, Rychard J. and {Smit}, Renske and {Illingworth}, Garth D. and {Labb{\'e}}, Ivo},
        title = "{The H{\ensuremath{\alpha}} Luminosity Function of Galaxies at z 4.5}",
      journal = {\apj},
         year = 2023,
        month = apr,
       volume = {946},
       number = {2},
          eid = {117},
        pages = {117},
          doi = {10.3847/1538-4357/acbc79},
archivePrefix = {arXiv},
       eprint = {2304.05034},
 primaryClass = {astro-ph.GA},
       adsurl = {https://ui.adsabs.harvard.edu/abs/2023ApJ...946..117B}
}

@ARTICLE{2015ApJ...803...34B,
       author = {{Bouwens}, R.~J. and {Illingworth}, G.~D. and {Oesch}, P.~A. and {Trenti}, M. and {Labb{\'e}}, I. and {Bradley}, L. and {Carollo}, M. and {van Dokkum}, P.~G. and {Gonzalez}, V. and {Holwerda}, B. and {Franx}, M. and {Spitler}, L. and {Smit}, R. and {Magee}, D.},
        title = "{UV Luminosity Functions at Redshifts z {\ensuremath{\sim}} 4 to z {\ensuremath{\sim}} 10: 10,000 Galaxies from HST Legacy Fields}",
      journal = {\apj},
         year = 2015,
        month = apr,
       volume = {803},
       number = {1},
          eid = {34},
        pages = {34},
          doi = {10.1088/0004-637X/803/1/34},
archivePrefix = {arXiv},
       eprint = {1403.4295},
 primaryClass = {astro-ph.CO},
       adsurl = {https://ui.adsabs.harvard.edu/abs/2015ApJ...803...34B}
}

@ARTICLE{2016MNRAS.456.3194P,
       author = {{Parsa}, Shaghayegh and {Dunlop}, James S. and {McLure}, Ross J. and {Mortlock}, Alice},
        title = "{The galaxy UV luminosity function at z ≃ 2-4; new results on faint-end slope and the evolution of luminosity density}",
      journal = {\mnras},
         year = 2016,
        month = mar,
       volume = {456},
       number = {3},
        pages = {3194-3211},
          doi = {10.1093/mnras/stv2857},
archivePrefix = {arXiv},
       eprint = {1507.05629},
 primaryClass = {astro-ph.GA},
       adsurl = {https://ui.adsabs.harvard.edu/abs/2016MNRAS.456.3194P}
}

@ARTICLE{2016ApJ...833..254S,
       author = {{Smit}, Renske and {Bouwens}, Rychard J. and {Labb{\'e}}, Ivo and {Franx}, Marijn and {Wilkins}, Stephen M. and {Oesch}, Pascal A.},
        title = "{Inferred H⍺ Flux as a Star Formation Rate Indicator at z \raisebox{-0.5ex}\textasciitilde 4-5: Implications for Dust Properties, Burstiness, and the z = 4-8 Star Formation Rate Functions}",
      journal = {\apj},
         year = 2016,
        month = dec,
       volume = {833},
       number = {2},
          eid = {254},
        pages = {254},
          doi = {10.3847/1538-4357/833/2/254},
archivePrefix = {arXiv},
       eprint = {1511.08808},
 primaryClass = {astro-ph.GA},
       adsurl = {https://ui.adsabs.harvard.edu/abs/2016ApJ...833..254S}
}

@ARTICLE{2024MNRAS.527.7337V,
       author = {{Vijayan}, Aswin P. and {Thomas}, Peter A. and {Lovell}, Christopher C. and {Wilkins}, Stephen M. and {Greve}, Thomas R. and {Irodotou}, Dimitrios and {Roper}, William J. and {Seeyave}, Louise T.~C.},
        title = "{First Light And Reionisation Epoch Simulations (FLARES) - XII: The consequences of star-dust geometry on galaxies in the EoR}",
      journal = {\mnras},
         year = 2024,
        month = jan,
       volume = {527},
       number = {3},
        pages = {7337-7354},
          doi = {10.1093/mnras/stad3594},
archivePrefix = {arXiv},
       eprint = {2303.04177},
 primaryClass = {astro-ph.GA},
       adsurl = {https://ui.adsabs.harvard.edu/abs/2024MNRAS.527.7337V}
}

@ARTICLE{2021MNRAS.500.2127L,
       author = {{Lovell}, Christopher C. and {Vijayan}, Aswin P. and {Thomas}, Peter A. and {Wilkins}, Stephen M. and {Barnes}, David J. and {Irodotou}, Dimitrios and {Roper}, Will},
        title = "{First Light And Reionization Epoch Simulations (FLARES) - I. Environmental dependence of high-redshift galaxy evolution}",
      journal = {\mnras},
         year = 2021,
        month = jan,
       volume = {500},
       number = {2},
        pages = {2127-2145},
          doi = {10.1093/mnras/staa3360},
archivePrefix = {arXiv},
       eprint = {2004.07283},
 primaryClass = {astro-ph.GA},
       adsurl = {https://ui.adsabs.harvard.edu/abs/2021MNRAS.500.2127L}
}

@ARTICLE{2022MNRAS.514.3857K,
       author = {{Kannan}, Rahul and {Smith}, Aaron and {Garaldi}, Enrico and {Shen}, Xuejian and {Vogelsberger}, Mark and {Pakmor}, R{\"u}diger and {Springel}, Volker and {Hernquist}, Lars},
        title = "{The THESAN project: predictions for multitracer line intensity mapping in the epoch of reionization}",
      journal = {\mnras},
         year = 2022,
        month = aug,
       volume = {514},
       number = {3},
        pages = {3857-3878},
          doi = {10.1093/mnras/stac1557},
archivePrefix = {arXiv},
       eprint = {2111.02411},
 primaryClass = {astro-ph.CO},
       adsurl = {https://ui.adsabs.harvard.edu/abs/2022MNRAS.514.3857K}
}

@MISC{2023jwst.prop.3577E,
       author = {{Egami}, Eiichi and {Sun}, Fengwu and {Alberts}, Stacey and {Baum}, Stefi A. and {Boyett}, Kristan and {Bunker}, Andrew and {Cameron}, Alex James and {Carniani}, Stefano and {Charlot}, Stephane and {Chen}, Zuyi and {Chevallard}, Jacopo and {Curti}, Mirko and {D'Eugenio}, Francesco and {Danhaive}, Lola and {DeCoursey}, Christa Noel and {Dudzeviciute}, Ugne and {Eisenstein}, Daniel J. and {Hainline}, Kevin and {Helton}, Jakob and {Ji}, Zhiyuan and {Johnson}, Benjamin D. and {Kumari}, Nimisha and {Looser}, Tobias Jakob and {Lyu}, Jianwei and {Ma}, Zheng and {Maiolino}, Roberto and {Maseda}, Michael and {Nelson}, Erica and {Rawle}, Tim and {Rieke}, Marcia J. and {Robertson}, Brant and {Sandles}, Lester and {Shivaei}, Irene and {Smit}, Renske and {Suess}, Katherine and {Tacchella}, Sandro and {Uebler}, Hannah and {Whitler}, Lily and {Williams}, Christina C. and {Willmer}, Christopher Nicholas Andrew and {Willott}, Chris J. and {Witstok}, Joris and {de Graaff}, Anna G.},
        title = "{Complete NIRCam Grism Redshift Survey (CONGRESS)}",
 howpublished = {JWST Proposal. Cycle 2, ID. \#3577},
         year = 2023,
        month = aug,
        pages = {3577},
       adsurl = {https://ui.adsabs.harvard.edu/abs/2023jwst.prop.3577E}
}

@ARTICLE{2003PASP..115..763C,
       author = {{Chabrier}, Gilles},
        title = "{Galactic Stellar and Substellar Initial Mass Function}",
      journal = {\pasp},
         year = 2003,
        month = jul,
       volume = {115},
       number = {809},
        pages = {763-795},
          doi = {10.1086/376392},
archivePrefix = {arXiv},
       eprint = {astro-ph/0304382},
 primaryClass = {astro-ph},
       adsurl = {https://ui.adsabs.harvard.edu/abs/2003PASP..115..763C}
}

@ARTICLE{2024A&A...684A..75C,
       author = {{Curti}, Mirko and {Maiolino}, Roberto and {Curtis-Lake}, Emma and {Chevallard}, Jacopo and {Carniani}, Stefano and {D'Eugenio}, Francesco and {Looser}, Tobias J. and {Scholtz}, Jan and {Charlot}, Stephane and {Cameron}, Alex and {{\"U}bler}, Hannah and {Witstok}, Joris and {Boyett}, Kristian and {Laseter}, Isaac and {Sandles}, Lester and {Arribas}, Santiago and {Bunker}, Andrew and {Giardino}, Giovanna and {Maseda}, Michael V. and {Rawle}, Tim and {Rodr{\'\i}guez Del Pino}, Bruno and {Smit}, Renske and {Willott}, Chris J. and {Eisenstein}, Daniel J. and {Hausen}, Ryan and {Johnson}, Benjamin and {Rieke}, Marcia and {Robertson}, Brant and {Tacchella}, Sandro and {Williams}, Christina C. and {Willmer}, Christopher and {Baker}, William M. and {Bhatawdekar}, Rachana and {Egami}, Eiichi and {Helton}, Jakob M. and {Ji}, Zhiyuan and {Kumari}, Nimisha and {Perna}, Michele and {Shivaei}, Irene and {Sun}, Fengwu},
        title = "{JADES: Insights into the low-mass end of the mass-metallicity-SFR relation at 3 < z < 10 from deep JWST/NIRSpec spectroscopy}",
      journal = {\aap},
         year = 2024,
        month = apr,
       volume = {684},
          eid = {A75},
        pages = {A75},
          doi = {10.1051/0004-6361/202346698},
archivePrefix = {arXiv},
       eprint = {2304.08516},
 primaryClass = {astro-ph.GA},
       adsurl = {https://ui.adsabs.harvard.edu/abs/2024A&A...684A..75C}
}

@ARTICLE{2016A&A...591A.151G,
       author = {{G{\'o}mez-Guijarro}, Carlos and {Gallego}, Jes{\'u}s and {Villar}, V{\'\i}ctor and {Rodr{\'\i}guez-Mu{\~n}oz}, Luc{\'\i}a and {Cl{\'e}ment}, Benjamin and {Cuby}, Jean-Gabriel},
        title = "{Properties of galaxies at the faint end of the H{\ensuremath{\alpha}} luminosity function at z \raisebox{-0.5ex}\textasciitilde 0.62}",
      journal = {\aap},
         year = 2016,
        month = jul,
       volume = {591},
          eid = {A151},
        pages = {A151},
          doi = {10.1051/0004-6361/201526746},
archivePrefix = {arXiv},
       eprint = {1604.04632},
 primaryClass = {astro-ph.GA},
       adsurl = {https://ui.adsabs.harvard.edu/abs/2016A&A...591A.151G}
}

@ARTICLE{2019A&A...631A..10R,
       author = {{Ram{\'o}n-P{\'e}rez}, Marina and {Bongiovanni}, {\'A}ngel and {P{\'e}rez Garc{\'\i}a}, Ana Mar{\'\i}a and {Cepa}, Jordi and {Lara-L{\'o}pez}, Maritza A. and {de Diego}, Jos{\'e} A. and {Alfaro}, Emilio and {Casta{\~n}eda}, H{\'e}ctor O. and {Cervi{\~n}o}, Miguel and {Fern{\'a}ndez-Lorenzo}, Mirian and {Gallego}, Jes{\'u}s and {Gonz{\'a}lez}, J. Jes{\'u}s and {Gonz{\'a}lez-Serrano}, J. Ignacio and {Nadolny}, Jakub and {Oteo G{\'o}mez}, Iv{\'a}n and {P{\'e}rez Mart{\'\i}nez}, Ricardo and {Pintos-Castro}, Irene and {Povi{\'c}}, Mirjana and {S{\'a}nchez-Portal}, Miguel},
        title = "{The OTELO survey. II. The faint-end of the H{\ensuremath{\alpha}} luminosity function at z {\ensuremath{\sim}} 0.40}",
      journal = {\aap},
         year = 2019,
        month = nov,
       volume = {631},
          eid = {A10},
        pages = {A10},
          doi = {10.1051/0004-6361/201833295},
archivePrefix = {arXiv},
       eprint = {2002.02449},
 primaryClass = {astro-ph.GA},
       adsurl = {https://ui.adsabs.harvard.edu/abs/2019A&A...631A..10R}
}

@ARTICLE{2018PASA...35...39H,
       author = {{Hopkins}, A.~M.},
        title = "{The Dawes Review 8: Measuring the Stellar Initial Mass Function}",
      journal = {\pasa},
         year = 2018,
        month = nov,
       volume = {35},
          eid = {e039},
        pages = {e039},
          doi = {10.1017/pasa.2018.29},
archivePrefix = {arXiv},
       eprint = {1807.09949},
 primaryClass = {astro-ph.GA},
       adsurl = {https://ui.adsabs.harvard.edu/abs/2018PASA...35...39H}
}

@ARTICLE{2017ApJ...835..113L,
       author = {{Livermore}, R.~C. and {Finkelstein}, S.~L. and {Lotz}, J.~M.},
        title = "{Directly Observing the Galaxies Likely Responsible for Reionization}",
      journal = {\apj},
         year = 2017,
        month = feb,
       volume = {835},
       number = {2},
          eid = {113},
        pages = {113},
          doi = {10.3847/1538-4357/835/2/113},
archivePrefix = {arXiv},
       eprint = {1604.06799},
 primaryClass = {astro-ph.GA},
       adsurl = {https://ui.adsabs.harvard.edu/abs/2017ApJ...835..113L}
}

@ARTICLE{2024ApJ...977..133C,
       author = {{Clarke}, Leonardo and {Shapley}, Alice E. and {Sanders}, Ryan L. and {Topping}, Michael W. and {Brammer}, Gabriel B. and {Bento}, Trinity and {Reddy}, Naveen A. and {Kehoe}, Emily},
        title = "{The Star-forming Main Sequence in JADES and CEERS at z > 1.4: Investigating the Burstiness of Star Formation}",
      journal = {\apj},
         year = 2024,
        month = dec,
       volume = {977},
       number = {1},
          eid = {133},
        pages = {133},
          doi = {10.3847/1538-4357/ad8ba4},
archivePrefix = {arXiv},
       eprint = {2406.05178},
 primaryClass = {astro-ph.GA},
       adsurl = {https://ui.adsabs.harvard.edu/abs/2024ApJ...977..133C}
}

@ARTICLE{2025arXiv250307774L,
       author = {{Liu}, Feng-Yuan and {Dunlop}, James S. and {McLure}, Ross J. and {McLeod}, Derek J. and {Barrufet}, Laia and {Carnall}, Adam C. and {Begley}, Ryan and {P{\'e}rez-Gonz{\'a}lez}, Pablo G. and {Donnan}, Callum T. and {Ellis}, Richard S. and {Grogin}, Norman A. and {Magee}, Dan and {Illingworth}, Garth D. and {Cullen}, Fergus and {Stevenson}, Struan D. and {Koekemoer}, Anton M. and {Fontana}, Adriano and {Bowler}, Rebecca A.~A.},
        title = "{JWST PRIMER: A deep JWST study of all ALMA-detected galaxies in PRIMER COSMOS -- dust-obscured star-formation history back to z $\simeq$ 7}",
      journal = {arXiv e-prints},
         year = 2025,
        month = mar,
          eid = {arXiv:2503.07774},
        pages = {arXiv:2503.07774},
          doi = {10.48550/arXiv.2503.07774},
archivePrefix = {arXiv},
       eprint = {2503.07774},
 primaryClass = {astro-ph.GA},
       adsurl = {https://ui.adsabs.harvard.edu/abs/2025arXiv250307774L}
}

@ARTICLE{2011ApJ...730...61K,
       author = {{Karim}, A. and {Schinnerer}, E. and {Mart{\'\i}nez-Sansigre}, A. and {Sargent}, M.~T. and {van der Wel}, A. and {Rix}, H. -W. and {Ilbert}, O. and {Smol{\v{c}}i{\'c}}, V. and {Carilli}, C. and {Pannella}, M. and {Koekemoer}, A.~M. and {Bell}, E.~F. and {Salvato}, M.},
        title = "{The Star Formation History of Mass-selected Galaxies in the COSMOS Field}",
      journal = {\apj},
         year = 2011,
        month = apr,
       volume = {730},
       number = {2},
          eid = {61},
        pages = {61},
          doi = {10.1088/0004-637X/730/2/61},
archivePrefix = {arXiv},
       eprint = {1011.6370},
 primaryClass = {astro-ph.CO},
       adsurl = {https://ui.adsabs.harvard.edu/abs/2011ApJ...730...61K}
}

@ARTICLE{2017A&A...602A...5N,
       author = {{Novak}, M. and {Smol{\v{c}}i{\'c}}, V. and {Delhaize}, J. and {Delvecchio}, I. and {Zamorani}, G. and {Baran}, N. and {Bondi}, M. and {Capak}, P. and {Carilli}, C.~L. and {Ciliegi}, P. and {Civano}, F. and {Ilbert}, O. and {Karim}, A. and {Laigle}, C. and {Le F{\`e}vre}, O. and {Marchesi}, S. and {McCracken}, H. and {Miettinen}, O. and {Salvato}, M. and {Sargent}, M. and {Schinnerer}, E. and {Tasca}, L.},
        title = "{The VLA-COSMOS 3 GHz Large Project: Cosmic star formation history since z   5}",
      journal = {\aap},
         year = 2017,
        month = jun,
       volume = {602},
          eid = {A5},
        pages = {A5},
          doi = {10.1051/0004-6361/201629436},
archivePrefix = {arXiv},
       eprint = {1703.09724},
 primaryClass = {astro-ph.GA},
       adsurl = {https://ui.adsabs.harvard.edu/abs/2017A&A...602A...5N}
}

@ARTICLE{2022ApJ...927..204E,
       author = {{Enia}, Andrea and {Talia}, Margherita and {Pozzi}, Francesca and {Cimatti}, Andrea and {Delvecchio}, Ivan and {Zamorani}, Gianni and {D'Amato}, Quirino and {Bisigello}, Laura and {Gruppioni}, Carlotta and {Rodighiero}, Giulia and {Calura}, Francesco and {Dallacasa}, Daniele and {Giulietti}, Marika and {Barchiesi}, Luigi and {Behiri}, Meriem and {Romano}, Michael},
        title = "{A New Estimate of the Cosmic Star Formation Density from a Radio-selected Sample, and the Contribution of H-dark Galaxies at z {\ensuremath{\geq}} 3}",
      journal = {\apj},
         year = 2022,
        month = mar,
       volume = {927},
       number = {2},
          eid = {204},
        pages = {204},
          doi = {10.3847/1538-4357/ac51ca},
archivePrefix = {arXiv},
       eprint = {2202.00019},
 primaryClass = {astro-ph.CO},
       adsurl = {https://ui.adsabs.harvard.edu/abs/2022ApJ...927..204E}
}

@ARTICLE{2023MNRAS.523.6082C,
       author = {{Cochrane}, R.~K. and {Kondapally}, R. and {Best}, P.~N. and {Sabater}, J. and {Duncan}, K.~J. and {Smith}, D.~J.~B. and {Hardcastle}, M.~J. and {R{\"o}ttgering}, H.~J.~A. and {Prandoni}, I. and {Haskell}, P. and {G{\"u}rkan}, G. and {Miley}, G.~K.},
        title = "{The LOFAR Two-metre Sky Survey: the radio view of the cosmic star formation history}",
      journal = {\mnras},
         year = 2023,
        month = aug,
       volume = {523},
       number = {4},
        pages = {6082-6102},
          doi = {10.1093/mnras/stad1602},
archivePrefix = {arXiv},
       eprint = {2305.15510},
 primaryClass = {astro-ph.GA},
       adsurl = {https://ui.adsabs.harvard.edu/abs/2023MNRAS.523.6082C}
}

@ARTICLE{2021MNRAS.501.3289V,
       author = {{Vijayan}, Aswin P. and {Lovell}, Christopher C. and {Wilkins}, Stephen M. and {Thomas}, Peter A. and {Barnes}, David J. and {Irodotou}, Dimitrios and {Kuusisto}, Jussi and {Roper}, William J.},
        title = "{First Light And Reionization Epoch Simulations (FLARES) - II: The photometric properties of high-redshift galaxies}",
      journal = {\mnras},
         year = 2021,
        month = mar,
       volume = {501},
       number = {3},
        pages = {3289-3308},
          doi = {10.1093/mnras/staa3715},
archivePrefix = {arXiv},
       eprint = {2008.06057},
 primaryClass = {astro-ph.GA},
       adsurl = {https://ui.adsabs.harvard.edu/abs/2021MNRAS.501.3289V}
}

@ARTICLE{2022MNRAS.511.4005K,
       author = {{Kannan}, R. and {Garaldi}, E. and {Smith}, A. and {Pakmor}, R. and {Springel}, V. and {Vogelsberger}, M. and {Hernquist}, L.},
        title = "{Introducing the THESAN project: radiation-magnetohydrodynamic simulations of the epoch of reionization}",
      journal = {\mnras},
         year = 2022,
        month = apr,
       volume = {511},
       number = {3},
        pages = {4005-4030},
          doi = {10.1093/mnras/stab3710},
archivePrefix = {arXiv},
       eprint = {2110.00584},
 primaryClass = {astro-ph.GA},
       adsurl = {https://ui.adsabs.harvard.edu/abs/2022MNRAS.511.4005K}
}

@ARTICLE{2024MNRAS.534..361S,
       author = {{Scharr{\'e}}, Lucie and {Sorini}, Daniele and {Dav{\'e}}, Romeel},
        title = "{The effects of stellar and AGN feedback on the cosmic star formation history in the SIMBA simulations}",
      journal = {\mnras},
         year = 2024,
        month = oct,
       volume = {534},
       number = {1},
        pages = {361-383},
          doi = {10.1093/mnras/stae2098},
archivePrefix = {arXiv},
       eprint = {2404.07252},
 primaryClass = {astro-ph.GA},
       adsurl = {https://ui.adsabs.harvard.edu/abs/2024MNRAS.534..361S}
}

@ARTICLE{2018MNRAS.473.4077P,
       author = {{Pillepich}, Annalisa and {Springel}, Volker and {Nelson}, Dylan and {Genel}, Shy and {Naiman}, Jill and {Pakmor}, R{\"u}diger and {Hernquist}, Lars and {Torrey}, Paul and {Vogelsberger}, Mark and {Weinberger}, Rainer and {Marinacci}, Federico},
        title = "{Simulating galaxy formation with the IllustrisTNG model}",
      journal = {\mnras},
         year = 2018,
        month = jan,
       volume = {473},
       number = {3},
        pages = {4077-4106},
          doi = {10.1093/mnras/stx2656},
archivePrefix = {arXiv},
       eprint = {1703.02970},
 primaryClass = {astro-ph.GA},
       adsurl = {https://ui.adsabs.harvard.edu/abs/2018MNRAS.473.4077P}
}

@ARTICLE{2017MNRAS.472..919K,
       author = {{Katsianis}, A. and {Blanc}, G. and {Lagos}, C.~P. and {Tejos}, N. and {Bower}, R.~G. and {Alavi}, A. and {Gonzalez}, V. and {Theuns}, T. and {Schaller}, M. and {Lopez}, S.},
        title = "{The evolution of the star formation rate function in the EAGLE simulations: a comparison with UV, IR and H{\ensuremath{\alpha}} observations from z {\ensuremath{\sim}} 8 to z {\ensuremath{\sim}} 0}",
      journal = {\mnras},
         year = 2017,
        month = nov,
       volume = {472},
       number = {1},
        pages = {919-939},
          doi = {10.1093/mnras/stx2020},
archivePrefix = {arXiv},
       eprint = {1708.01913},
 primaryClass = {astro-ph.GA},
       adsurl = {https://ui.adsabs.harvard.edu/abs/2017MNRAS.472..919K}
}

@ARTICLE{2025MNRAS.541.1348P,
       author = {{Pirie}, C.~A. and {Best}, P.~N. and {Duncan}, K.~J. and {McLeod}, D.~J. and {Cochrane}, R.~K. and {Clausen}, M. and {Dunlop}, J.~S. and {Flury}, S.~R. and {Geach}, J.~E. and {Hale}, C.~L. and {Ibar}, E. and {Kondapally}, R. and {Li}, Zefeng and {Matthee}, J. and {McLure}, R.~J. and {Ossa-Fuentes}, L. and {Patrick}, A.~L. and {Smail}, Ian and {Sobral}, D. and {Stephenson}, H.~M.~O. and {Stott}, J.~P. and {Swinbank}, A.~M.},
        title = "{The JWST Emission Line Survey (JELS): an untargeted search for H {\ensuremath{\alpha}} emission line galaxies at z > 6 and their physical properties}",
      journal = {\mnras},
         year = 2025,
        month = aug,
       volume = {541},
       number = {2},
        pages = {1348-1376},
          doi = {10.1093/mnras/staf1006},
archivePrefix = {arXiv},
       eprint = {2410.11808},
 primaryClass = {astro-ph.GA},
       adsurl = {https://ui.adsabs.harvard.edu/abs/2025MNRAS.541.1348P},
      alias  = {P25}
}

@ARTICLE{2025MNRAS.541.1329D,
       author = {{Duncan}, K.~J. and {McLeod}, D.~J. and {Best}, P.~N. and {Pirie}, C.~A. and {Clausen}, M. and {Cochrane}, R.~K. and {Dunlop}, J.~S. and {Flury}, S.~R. and {Geach}, J.~E. and {Grogin}, N.~A. and {Hale}, C.~L. and {Ibar}, E. and {Kondapally}, R. and {Li}, Zefeng and {Matthee}, J. and {McLure}, R.~J. and {Ossa-Fuentes}, Luis and {Patrick}, A.~L. and {Smail}, Ian and {Sobral}, D. and {Stephenson}, H.~M.~O. and {Stott}, J.~P. and {Swinbank}, A.~M.},
        title = "{The JWST Emission-Line Survey: extending rest-optical narrow-band emission-line selection into the Epoch of Reionization}",
      journal = {\mnras},
         year = 2025,
        month = aug,
       volume = {541},
       number = {2},
        pages = {1329-1347},
          doi = {10.1093/mnras/staf1061},
archivePrefix = {arXiv},
       eprint = {2410.09000},
 primaryClass = {astro-ph.GA},
       adsurl = {https://ui.adsabs.harvard.edu/abs/2025MNRAS.541.1329D}
}

@ARTICLE{2023MNRAS.523.1036B,
       author = {{Bouwens}, Rychard J. and {Stefanon}, Mauro and {Brammer}, Gabriel and {Oesch}, Pascal A. and {Herard-Demanche}, Thomas and {Illingworth}, Garth D. and {Matthee}, Jorryt and {Naidu}, Rohan P. and {van Dokkum}, Pieter G. and {van Leeuwen}, Ivana F.},
        title = "{Evolution of the UV LF from z   15 to z   8 using new JWST NIRCam medium-band observations over the HUDF/XDF}",
      journal = {\mnras},
         year = 2023,
        month = jul,
       volume = {523},
       number = {1},
        pages = {1036-1055},
          doi = {10.1093/mnras/stad1145},
archivePrefix = {arXiv},
       eprint = {2211.02607},
 primaryClass = {astro-ph.GA},
       adsurl = {https://ui.adsabs.harvard.edu/abs/2023MNRAS.523.1036B}
}

@ARTICLE{2021Natur.597..489F,
       author = {{Fudamoto}, Y. and {Oesch}, P.~A. and {Schouws}, S. and {Stefanon}, M. and {Smit}, R. and {Bouwens}, R.~J. and {Bowler}, R.~A.~A. and {Endsley}, R. and {Gonzalez}, V. and {Inami}, H. and {Labbe}, I. and {Stark}, D. and {Aravena}, M. and {Barrufet}, L. and {da Cunha}, E. and {Dayal}, P. and {Ferrara}, A. and {Graziani}, L. and {Hodge}, J. and {Hutter}, A. and {Li}, Y. and {De Looze}, I. and {Nanayakkara}, T. and {Pallottini}, A. and {Riechers}, D. and {Schneider}, R. and {Ucci}, G. and {van der Werf}, P. and {White}, C.},
        title = "{Normal, dust-obscured galaxies in the epoch of reionization}",
      journal = {\nat},
         year = 2021,
        month = sep,
       volume = {597},
       number = {7877},
        pages = {489-492},
          doi = {10.1038/s41586-021-03846-z},
archivePrefix = {arXiv},
       eprint = {2109.10378},
 primaryClass = {astro-ph.GA},
       adsurl = {https://ui.adsabs.harvard.edu/abs/2021Natur.597..489F}
}

@ARTICLE{2023A&A...672A.108A,
       author = {{{\'A}lvarez-M{\'a}rquez}, J. and {Labiano}, A. and {Guillard}, P. and {Dicken}, D. and {Argyriou}, I. and {Patapis}, P. and {Law}, D.~R. and {Kavanagh}, P.~J. and {Larson}, K.~L. and {Gasman}, D. and {Mueller}, M. and {Alberts}, S. and {Brandl}, B.~R. and {Colina}, L. and {Garc{\'\i}a-Mar{\'\i}n}, M. and {Jones}, O.~C. and {Noriega-Crespo}, A. and {Shivaei}, I. and {Temim}, T. and {Wright}, G.~S.},
        title = "{Nuclear high-ionisation outflow in the Compton-thick AGN NGC 6552 as seen by the JWST mid-infrared instrument}",
      journal = {\aap},
         year = 2023,
        month = apr,
       volume = {672},
          eid = {A108},
        pages = {A108},
          doi = {10.1051/0004-6361/202244880},
archivePrefix = {arXiv},
       eprint = {2209.01695},
 primaryClass = {astro-ph.GA},
       adsurl = {https://ui.adsabs.harvard.edu/abs/2023A&A...672A.108A}
}

@ARTICLE{2024MNRAS.527.5808B,
       author = {{Bowler}, R.~A.~A. and {Inami}, H. and {Sommovigo}, L. and {Smit}, R. and {Algera}, H.~S.~B. and {Aravena}, M. and {Barrufet}, L. and {Bouwens}, R. and {da Cunha}, E. and {Cullen}, F. and {Dayal}, P. and {De Looze}, I. and {Dunlop}, J.~S. and {Fudamoto}, Y. and {Mauerhofer}, V. and {McLure}, R.~J. and {Stefanon}, M. and {Schneider}, R. and {Ferrara}, A. and {Graziani}, L. and {Hodge}, J.~A. and {Nanayakkara}, T. and {Palla}, M. and {Schouws}, S. and {Stark}, D.~P. and {van der Werf}, P.~P.},
        title = "{The ALMA REBELS survey: obscured star formation in massive Lyman-break galaxies at z= 4-8 revealed by the IRX-{\ensuremath{\beta}} and M$_{{\ensuremath{\star}}}$ relations}",
      journal = {\mnras},
         year = 2024,
        month = jan,
       volume = {527},
       number = {3},
        pages = {5808-5828},
          doi = {10.1093/mnras/stad3578},
archivePrefix = {arXiv},
       eprint = {2309.17386},
 primaryClass = {astro-ph.GA},
       adsurl = {https://ui.adsabs.harvard.edu/abs/2024MNRAS.527.5808B}
}

@ARTICLE{2021A&A...649A.152K,
       author = {{Khusanova}, Y. and {Bethermin}, M. and {Le F{\`e}vre}, O. and {Capak}, P. and {Faisst}, A.~L. and {Schaerer}, D. and {Silverman}, J.~D. and {Cassata}, P. and {Yan}, L. and {Ginolfi}, M. and {Fudamoto}, Y. and {Loiacono}, F. and {Amorin}, R. and {Bardelli}, S. and {Boquien}, M. and {Cimatti}, A. and {Dessauges-Zavadsky}, M. and {Gruppioni}, C. and {Hathi}, N.~P. and {Jones}, G.~C. and {Koekemoer}, A.~M. and {Lagache}, G. and {Maiolino}, R. and {Lemaux}, B.~C. and {Oesch}, P. and {Pozzi}, F. and {Riechers}, D.~A. and {Romano}, M. and {Talia}, M. and {Toft}, S. and {Vergani}, D. and {Zamorani}, G. and {Zucca}, E.},
        title = "{The ALPINE-ALMA [CII] survey. Obscured star formation rate density and main sequence of star-forming galaxies at z > 4}",
      journal = {\aap},
         year = 2021,
        month = may,
       volume = {649},
          eid = {A152},
        pages = {A152},
          doi = {10.1051/0004-6361/202038944},
archivePrefix = {arXiv},
       eprint = {2007.08384},
 primaryClass = {astro-ph.GA},
       adsurl = {https://ui.adsabs.harvard.edu/abs/2021A&A...649A.152K}
}

@ARTICLE{2025MNRAS.537..788H,
       author = {{Herard-Demanche}, Thomas and {Bouwens}, Rychard J. and {Oesch}, Pascal A. and {Naidu}, Rohan P. and {Decarli}, Roberto and {Nelson}, Erica J. and {Brammer}, Gabriel and {Weibel}, Andrea and {Xiao}, Mengyuan and {Stefanon}, Mauro and {Walter}, Fabian and {Matthee}, Jorryt and {Meyer}, Romain A. and {Wuyts}, Stijn and {Reddy}, Naveen and {Rowland}, Lucie and {van Leeuwen}, Ivana and {Haro}, Pablo Arrabal and {Dannerbauer}, Helmut and {Shapley}, Alice E. and {Chisholm}, John and {van Dokkum}, Pieter and {Labbe}, Ivo and {Illingworth}, Garth and {Schaerer}, Daniel and {Shivaei}, Irene},
        title = "{Mapping dusty galaxy growth at z > 5 with FRESCO: detection of H{\ensuremath{\alpha}} in submm galaxy HDF850.1 and the surrounding overdense structures}",
      journal = {\mnras},
         year = 2025,
        month = feb,
       volume = {537},
       number = {2},
        pages = {788-808},
          doi = {10.1093/mnras/staf030},
archivePrefix = {arXiv},
       eprint = {2309.04525},
 primaryClass = {astro-ph.GA},
       adsurl = {https://ui.adsabs.harvard.edu/abs/2025MNRAS.537..788H}
}

@ARTICLE{2014ApJ...795..165S,
       author = {{Steidel}, Charles C. and {Rudie}, Gwen C. and {Strom}, Allison L. and {Pettini}, Max and {Reddy}, Naveen A. and {Shapley}, Alice E. and {Trainor}, Ryan F. and {Erb}, Dawn K. and {Turner}, Monica L. and {Konidaris}, Nicholas P. and {Kulas}, Kristin R. and {Mace}, Gregory and {Matthews}, Keith and {McLean}, Ian S.},
        title = "{Strong Nebular Line Ratios in the Spectra of z \raisebox{-0.5ex}\textasciitilde 2-3 Star Forming Galaxies: First Results from KBSS-MOSFIRE}",
      journal = {\apj},
         year = 2014,
        month = nov,
       volume = {795},
       number = {2},
          eid = {165},
        pages = {165},
          doi = {10.1088/0004-637X/795/2/165},
archivePrefix = {arXiv},
       eprint = {1405.5473},
 primaryClass = {astro-ph.GA},
       adsurl = {https://ui.adsabs.harvard.edu/abs/2014ApJ...795..165S}
}

@ARTICLE{2023ApJ...950...66K,
       author = {{Kashino}, Daichi and {Lilly}, Simon J. and {Matthee}, Jorryt and {Eilers}, Anna-Christina and {Mackenzie}, Ruari and {Bordoloi}, Rongmon and {Simcoe}, Robert A.},
        title = "{EIGER. I. A Large Sample of [O III]-emitting Galaxies at 5.3 < z < 6.9 and Direct Evidence for Local Reionization by Galaxies}",
      journal = {\apj},
         year = 2023,
        month = jun,
       volume = {950},
       number = {1},
          eid = {66},
        pages = {66},
          doi = {10.3847/1538-4357/acc588},
archivePrefix = {arXiv},
       eprint = {2211.08254},
 primaryClass = {astro-ph.GA},
       adsurl = {https://ui.adsabs.harvard.edu/abs/2023ApJ...950...66K}
}

@MISC{2023jwst.prop.2883S,
       author = {{Sun}, Fengwu and {Bauer}, Franz and {Bian}, Fuyan and {Cai}, Zheng and {Caputi}, Karina and {Chen (Tc)}, Chian-Chou and {Chen}, Wenlei and {Cheng}, Cheng and {Coe}, Dan and {Danhaive}, Lola and {Dessauges-Zavadsky}, Miroslava and {Egami}, Eiichi and {Eilers}, Anna-Christina and {Espada}, Daniel and {Fan}, Xiaohui and {Farina}, Emanuele Paolo and {Frye}, Brenda Louise and {Fujimoto}, Seiji and {Furtak}, Lukas Jonathan and {Huang}, Shuo and {Hainline}, Kevin and {Helton}, Jakob and {Huang}, Jiasheng and {Hughes}, David and {Jauzac}, Mathilde and {Ji}, Zhiyuan and {Jin}, Xiangyu and {Jolly}, Jean-Baptiste and {Kneib}, Jean-Paul Richard and {Knudsen}, Kirsten Kraiberg and {Koekemoer}, Anton M. and {Kohno}, Kotaro and {Kokorev}, Vasily and {Lee}, Minju and {Li}, Mingyu and {Li}, Zihao and {Lin}, Xiaojing and {Liu}, Daizhong and {Liu}, Weizhe and {Magdis}, Georgios and {Maseda}, Michael and {Oguri}, Masamune and {Pascale}, Massimo and {Perez-Gonzalez}, Pablo G. and {Richard}, Johan Pierre and {Schaerer}, Daniel and {Steinhardt}, Charles Louis and {Sun}, Bangzheng and {Sun}, Zechang and {Tacchella}, Sandro and {Tee}, Wei Leong and {Trebitsch}, Maxime and {Tsujita}, Akiyoshi and {Ueda}, Yoshihiro and {Uematsu}, Ryosuke and {Umehata}, Hideki and {Walth}, Gregory and {Wang}, Feige and {Wang}, Wei-Hao and {Weaver}, John Raymond and {Williams}, Christina C. and {Willmer}, Christopher Nicholas Andrew and {Windhorst}, Rogier A. and {Wu}, Yunjing and {Yan}, Haojing and {Yang}, Jinyi and {Zhang}, Haowen and {Zitrin}, Adi and {Zou}, Siwei},
        title = "{MAGNIF: Medium-band Astrophysics with the Grism of NIRCam in Frontier Fields}",
 howpublished = {JWST Proposal. Cycle 2, ID. \#2883},
         year = 2023,
        month = may,
        pages = {2883},
       adsurl = {https://ui.adsabs.harvard.edu/abs/2023jwst.prop.2883S}
}

@ARTICLE{2024ApJ...976..101S,
       author = {{Suess}, Katherine A. and {Weaver}, John R. and {Price}, Sedona H. and {Pan}, Richard and {Wang}, Bingjie and {Bezanson}, Rachel and {Brammer}, Gabriel and {Cutler}, Sam E. and {Labb{\'e}}, Ivo and {Leja}, Joel and {Williams}, Christina C. and {Whitaker}, Katherine E. and {Atek}, Hakim and {Dayal}, Pratika and {de Graaff}, Anna and {Feldmann}, Robert and {Franx}, Marijn and {Fudamoto}, Yoshinobu and {Fujimoto}, Seiji and {Furtak}, Lukas J. and {Goulding}, Andy D. and {Greene}, Jenny E. and {Khullar}, Gourav and {Kokorev}, Vasily and {Kriek}, Mariska and {Lorenz}, Brian and {Marchesini}, Danilo and {Maseda}, Michael V. and {Matthee}, Jorryt and {Miller}, Tim B. and {Mitsuhashi}, Ikki and {Mowla}, Lamiya A. and {Muzzin}, Adam and {Naidu}, Rohan P. and {Nanayakkara}, Themiya and {Nelson}, Erica J. and {Oesch}, Pascal A. and {Setton}, David J. and {Shipley}, Heath and {Smit}, Renske and {Spilker}, Justin S. and {van Dokkum}, Pieter and {Zitrin}, Adi},
        title = "{Medium Bands, Mega Science: A JWST/NIRCam Medium-band Imaging Survey of A2744}",
      journal = {\apj},
         year = 2024,
        month = nov,
       volume = {976},
       number = {1},
          eid = {101},
        pages = {101},
          doi = {10.3847/1538-4357/ad75fe},
archivePrefix = {arXiv},
       eprint = {2404.13132},
 primaryClass = {astro-ph.GA},
       adsurl = {https://ui.adsabs.harvard.edu/abs/2024ApJ...976..101S}
}

@ARTICLE{2023arXiv231012340E,
       author = {{Eisenstein}, Daniel J. and {Johnson}, Benjamin D. and {Robertson}, Brant and {Tacchella}, Sandro and {Hainline}, Kevin and {Jakobsen}, Peter and {Maiolino}, Roberto and {Bonaventura}, Nina and {Bunker}, Andrew J. and {Cameron}, Alex J. and {Cargile}, Phillip A. and {Curtis-Lake}, Emma and {Hausen}, Ryan and {Pusk{\'a}s}, D{\'a}vid and {Rieke}, Marcia and {Sun}, Fengwu and {Willmer}, Christopher N.~A. and {Willott}, Chris and {Alberts}, Stacey and {Arribas}, Santiago and {Baker}, William M. and {Baum}, Stefi and {Bhatawdekar}, Rachana and {Carniani}, Stefano and {Charlot}, Stephane and {Chen}, Zuyi and {Chevallard}, Jacopo and {Curti}, Mirko and {DeCoursey}, Christa and {D'Eugenio}, Francesco and {de Graaff}, Anna and {Egami}, Eiichi and {Helton}, Jakob M. and {Ji}, Zhiyuan and {Jones}, Gareth C. and {Kumari}, Nimisha and {L{\"u}tzgendorf}, Nora and {Laseter}, Isaac and {Looser}, Tobias J. and {Lyu}, Jianwei and {Maseda}, Michael V. and {Nelson}, Erica and {Parlanti}, Eleonora and {Rauscher}, Bernard J. and {Rawle}, Tim and {Rieke}, George and {Rix}, Hans-Walter and {Rujopakarn}, Wiphu and {Sandles}, Lester and {Saxena}, Aayush and {Scholtz}, Jan and {Sharpe}, Katherine and {Shivaei}, Irene and {Simmonds}, Charlotte and {Smit}, Renske and {Topping}, Michael W. and {{\"U}bler}, Hannah and {Venturi}, Giacomo and {Williams}, Christina C. and {Witstok}, Joris and {Woodrum}, Charity},
        title = "{The JADES Origins Field: A New JWST Deep Field in the JADES Second NIRCam Data Release}",
      journal = {arXiv e-prints},
         year = 2023,
        month = oct,
          eid = {arXiv:2310.12340},
        pages = {arXiv:2310.12340},
          doi = {10.48550/arXiv.2310.12340},
archivePrefix = {arXiv},
       eprint = {2310.12340},
 primaryClass = {astro-ph.GA},
       adsurl = {https://ui.adsabs.harvard.edu/abs/2023arXiv231012340E}
}

@ARTICLE{2025arXiv250719706M,
       author = {{Muzzin}, Adam and {Suess}, Katherine A. and {Marchesini}, Danilo and {Robbins}, Luke and {Willott}, Chris J. and {Alberts}, Stacey and {Antwi-Danso}, Jacqueline and {Asada}, Yoshihisa and {Brammer}, Gabriel and {Cutler}, Sam E. and {Iyer}, Kartheik G. and {Labbe}, Ivo and {Martis}, Nicholas S. and {Miller}, Tim B. and {Mitsuhashi}, Ikki and {Pope}, Alexandra and {Sajina}, Anna and {Sarrouh}, Ghassan T.~E. and {Sharma}, Monu and {Stefanon}, Mauro and {Whitaker}, Katherine E. and {Abraham}, Roberto and {Atek}, Hakim and {Bradac}, Marusa and {Berek}, Samantha and {Bezanson}, Rachel and {Brown}, Westley and {Burgasser}, Adam J. and {Chicoine}, Nathalie and {Cloonan}, Aidan P. and {Cooper}, Olivia R. and {Dayal}, Pratika and {de Graaff}, Anna and {Desprez}, Guillaume and {Feldmann}, Robert and {Forrest}, Ben and {Franx}, Marijn and {Fudamoto}, Yoshinobu and {Fujimoto}, Seiji and {Furtak}, Lukas J. and {Glazebrook}, Karl and {Goovaerts}, Ilias and {Greene}, Jenny E. and {Jagga}, Naadiyah and {Jarvis}, William W.~H. and {Kriek}, Mariska and {Khullar}, Gourav and {La Torre}, Valentina and {Leja}, Joel and {Lin}, Jamie and {Lorenz}, Brian and {Lyon}, Daniel and {Markov}, Vladan and {Maseda}, Michael V. and {McConachie}, Ian and {Merchant}, Maya and {Merida}, Rosa M. and {Mowla}, Lamiya and {Myers}, Katherine and {Naidu}, Rohan P. and {Nanayakkara}, Themiya and {Nelson}, Erica J. and {Noirot}, Gael and {Oesch}, Pascal A. and {Omori}, Kiyoaki C. and {Pan}, Richard and {Porraz Barrera}, Natalia and {Price}, Sedona H. and {Ravindranath}, Swara and {Sawicki}, Marcin and {Setton}, David J. and {Smit}, Renske and {Sok}, Visal and {Speagle}, Joshua S. and {Taylor}, Edward N. and {Tan}, Vivian Yun Yan and {Tripodi}, Roberta and {van der Wel}, Arjen and {Perez Vidal}, Edgar and {Wang}, Bingjie and {Weaver}, John R. and {Williams}, Christina C. and {Withers}, Sunna and {Zaidi}, Kumail},
        title = "{MINERVA: A NIRCam Medium Band and MIRI Imaging Survey to Unlock the Hidden Gems of the Distant Universe}",
      journal = {arXiv e-prints},
         year = 2025,
        month = jul,
          eid = {arXiv:2507.19706},
        pages = {arXiv:2507.19706},
          doi = {10.48550/arXiv.2507.19706},
archivePrefix = {arXiv},
       eprint = {2507.19706},
 primaryClass = {astro-ph.GA},
       adsurl = {https://ui.adsabs.harvard.edu/abs/2025arXiv250719706M}
}

@ARTICLE{2007ApJ...657..738L,
       author = {{Ly}, Chun and {Malkan}, Matt A. and {Kashikawa}, Nobunari and {Shimasaku}, Kazuhiro and {Doi}, Mamoru and {Nagao}, Tohru and {Iye}, Masanori and {Kodama}, Tadayuki and {Morokuma}, Tomoki and {Motohara}, Kentaro},
        title = "{The Luminosity Function and Star Formation Rate between Redshifts of 0.07 and 1.47 for Narrowband Emitters in the Subaru Deep Field}",
      journal = {\apj},
         year = 2007,
        month = mar,
       volume = {657},
       number = {2},
        pages = {738-759},
          doi = {10.1086/510828},
archivePrefix = {arXiv},
       eprint = {astro-ph/0610846},
 primaryClass = {astro-ph},
       adsurl = {https://ui.adsabs.harvard.edu/abs/2007ApJ...657..738L}
}

@MISC{2024jwst.prop.5893K,
       author = {{Kakiichi}, Koki and {Egami}, Eiichi and {Fan}, Xiaohui and {Lyu}, Jianwei and {Wang}, Feige and {Yang}, Jinyi and {Bechtel}, Shane and {Behroozi}, Peter and {Bosman}, Sarah E.~I. and {Cai}, Zheng and {Champagne}, Jaclyn and {Davies}, Frederick and {De Rosa}, Gisella and {Decarli}, Roberto and {Eilers}, Anna-Christina and {Ellis}, Richard S. and {Endsley}, Ryan and {Farina}, Emanuele Paolo and {Finkelstein}, Steven L. and {Fujimoto}, Seiji and {Hennawi}, Joseph and {Inoue}, Akio and {Jiang}, Linhua and {Jin}, Xiangyu and {Khusanova}, Yana and {Kirkpatrick}, Allison and {Kocevski}, Dale D. and {Kulkarni}, Girish and {Lee}, Khee-Gan and {Liu}, Weizhe and {Meyer}, Romain Alexis and {Ono}, Yoshiaki and {Onoue}, Masafusa and {Ouchi}, Masami and {Papovich}, Casey and {Satyavolu}, Sindhu and {Schindler}, Jan-Torge and {Sun}, Fengwu and {Tee}, Wei Leong and {Vestergaard}, Marianne and {Zhang}, Haowen and {Zou}, Siwei},
        title = "{COSMOS-3D: A Legacy Spectroscopic/Imaging Survey of the Early Universe}",
 howpublished = {JWST Proposal. Cycle 3, ID. \#5893},
         year = 2024,
        month = feb,
        pages = {5893},
       adsurl = {https://ui.adsabs.harvard.edu/abs/2024jwst.prop.5893K}
}

@ARTICLE{2017ApJ...843..129B,
       author = {{Bouwens}, R.~J. and {Oesch}, P.~A. and {Illingworth}, G.~D. and {Ellis}, R.~S. and {Stefanon}, M.},
        title = "{The z {\ensuremath{\sim}} 6 Luminosity Function Fainter than -15 mag from the Hubble Frontier Fields: The Impact of Magnification Uncertainties}",
      journal = {\apj},
         year = 2017,
        month = jul,
       volume = {843},
       number = {2},
          eid = {129},
        pages = {129},
          doi = {10.3847/1538-4357/aa70a4},
archivePrefix = {arXiv},
       eprint = {1610.00283},
 primaryClass = {astro-ph.GA},
       adsurl = {https://ui.adsabs.harvard.edu/abs/2017ApJ...843..129B}
}

@ARTICLE{2025MNRAS.544.1412S,
       author = {{Stephenson}, H.~M.~O. and {Stott}, J.~P. and {Pirie}, C.~A. and {Duncan}, K.~J. and {McLeod}, D.~J. and {Best}, P.~N. and {Brinch}, M. and {Clausen}, M. and {Cochrane}, R.~K. and {Dunlop}, J.~S. and {Flury}, S.~R. and {Geach}, J.~E. and {Hale}, C.~L. and {Ibar}, E. and {Li}, Zefeng and {Matthee}, J. and {McLure}, R.~J. and {Ossa-Fuentes}, L. and {Patrick}, A.~L. and {Sobral}, D. and {Swinbank}, A.~M.},
        title = "{The JWST Emission Line Survey (JELS): the sizes and merger fraction of star-forming galaxies during the Epoch of Reionization}",
      journal = {\mnras},
         year = 2025,
        month = dec,
       volume = {544},
       number = {2},
        pages = {1412-1431},
          doi = {10.1093/mnras/staf1725},
archivePrefix = {arXiv},
       eprint = {2509.08045},
 primaryClass = {astro-ph.GA},
       adsurl = {https://ui.adsabs.harvard.edu/abs/2025MNRAS.544.1412S}
}

@ARTICLE{2025ApJ...987..186F,
       author = {{Fu}, Shuqi and {Sun}, Fengwu and {Jiang}, Linhua and {Lin}, Xiaojing and {Diego}, Jose M. and {Furtak}, Lukas J. and {Jauzac}, Mathilde and {Koekemoer}, Anton M. and {Li}, Mingyu and {Oguri}, Masamune and {Patel}, Nency R. and {Willmer}, Christopher N.~A. and {Windhorst}, Rogier A. and {Zitrin}, Adi and {Bauer}, Franz E. and {Chen}, Chian-Chou and {Chen}, Wenlei and {Cheng}, Cheng and {Conselice}, Christopher J. and {Eisenstein}, Daniel J. and {Egami}, Eiichi and {Espada}, Daniel and {Fan}, Xiaohui and {Fujimoto}, Seiji and {Hsiao}, Tiger Yu-Yang and {Jin}, Xiangyu and {Kohno}, Kotaro and {Lagattuta}, David J. and {Li}, Zihao and {Liu}, Weizhe and {Miralda-Escud{\'e}}, Jordi and {Ning}, Yuanhang and {Tacchella}, Sandro and {Tee}, Wei Leong and {Umehata}, Hideki and {Wang}, Feige and {Yan}, Haojing and {Zhu}, Yongda},
        title = "{Medium-band Astrophysics with the Grism of NIRCam In Frontier Fields (MAGNIF): Spectroscopic Census of H{\ensuremath{\alpha}} Luminosity Functions and Cosmic Star Formation at z {\ensuremath{\sim}} 4.5 and 6.3}",
      journal = {\apj},
         year = 2025,
        month = jul,
       volume = {987},
       number = {2},
          eid = {186},
        pages = {186},
          doi = {10.3847/1538-4357/adddb1},
archivePrefix = {arXiv},
       eprint = {2503.03829},
 primaryClass = {astro-ph.GA},
       adsurl = {https://ui.adsabs.harvard.edu/abs/2025ApJ...987..186F}
}

@ARTICLE{1980ApJS...43..305K,
       author = {{Kron}, R.~G.},
        title = "{Photometry of a complete sample of faint galaxies.}",
      journal = {\apjs},
         year = 1980,
        month = jun,
       volume = {43},
        pages = {305-325},
          doi = {10.1086/190669},
       adsurl = {https://ui.adsabs.harvard.edu/abs/1980ApJS...43..305K}
}

@ARTICLE{2024ARA&A..62...63H,
       author = {{Hennebelle}, P. and {Grudi{\'c}}, M.~Y.},
        title = "{The Physical Origin of the Stellar Initial Mass Function}",
      journal = {\araa},
         year = 2024,
        month = sep,
       volume = {62},
       number = {1},
        pages = {63-111},
          doi = {10.1146/annurev-astro-052622-031748},
archivePrefix = {arXiv},
       eprint = {2404.07301},
 primaryClass = {astro-ph.GA},
       adsurl = {https://ui.adsabs.harvard.edu/abs/2024ARA&A..62...63H}
}

@ARTICLE{2025ApJ...984..188K,
       author = {{Korhonen Cuestas}, Nathalie A. and {Strom}, Allison L. and {Miller}, Tim B. and {Steidel}, Charles C. and {Trainor}, Ryan F. and {Rudie}, Gwen C. and {Nu{\~n}ez}, Evan Haze},
        title = "{Exploring the Relationship between Stellar Mass, Metallicity, and Star Formation Rate at z {\ensuremath{\sim}} 2.3 in KBSS-MOSFIRE}",
      journal = {\apj},
         year = 2025,
        month = may,
       volume = {984},
       number = {2},
          eid = {188},
        pages = {188},
          doi = {10.3847/1538-4357/adc5f7},
archivePrefix = {arXiv},
       eprint = {2503.10800},
 primaryClass = {astro-ph.GA},
       adsurl = {https://ui.adsabs.harvard.edu/abs/2025ApJ...984..188K}
}

@ARTICLE{2025arXiv250810099S,
       author = {{Sanders}, Ryan L. and {Shapley}, Alice E. and {Topping}, Michael W. and {Reddy}, Naveen A. and {Berg}, Danielle A. and {Khostovan}, Ali Ahmad and {Bouwens}, Rychard J. and {Brammer}, Gabriel and {Carnall}, Adam C. and {Cullen}, Fergus and {Dav{\'e}}, Romeel and {Dunlop}, James S. and {Ellis}, Richard S. and {F{\"o}rster Schreiber}, N.~M. and {Furlanetto}, Steven R. and {Glazebrook}, Karl and {Illingworth}, Garth D. and {Jones}, Tucker and {Kriek}, Mariska and {McLeod}, Derek J. and {McLure}, Ross J. and {Narayanan}, Desika and {Oesch}, Pascal A. and {Pahl}, Anthony J. and {Pettini}, Max and {Schaerer}, Daniel and {Stark}, Daniel P. and {Steidel}, Charles C. and {Tang}, Mengtao and {Clarke}, Leonardo and {Donnan}, Callum T. and {Kehoe}, Emily},
        title = "{The AURORA Survey: High-Redshift Empirical Metallicity Calibrations from Electron Temperature Measurements at z=2-10}",
      journal = {arXiv e-prints},
         year = 2025,
        month = aug,
          eid = {arXiv:2508.10099},
        pages = {arXiv:2508.10099},
          doi = {10.48550/arXiv.2508.10099},
archivePrefix = {arXiv},
       eprint = {2508.10099},
 primaryClass = {astro-ph.GA},
       adsurl = {https://ui.adsabs.harvard.edu/abs/2025arXiv250810099S}
}

@ARTICLE{2024MNRAS.532.3102S,
       author = {{Stanton}, T.~M. and {Cullen}, F. and {McLure}, R.~J. and {Shapley}, A.~E. and {Arellano-C{\'o}rdova}, K.~Z. and {Begley}, R. and {Amor{\'\i}n}, R. and {Barrufet}, L. and {Calabr{\`o}}, A. and {Carnall}, A.~C. and {Cirasuolo}, M. and {Dunlop}, J.~S. and {Donnan}, C.~T. and {Hamadouche}, M.~L. and {Liu}, F.~Y. and {McLeod}, D.~J. and {Pentericci}, L. and {Pozzetti}, L. and {Sanders}, R.~L. and {Scholte}, D. and {Topping}, M.~W.},
        title = "{The NIRVANDELS survey: the stellar and gas-phase mass-metallicity relations of star-forming galaxies at z = 3.5}",
      journal = {\mnras},
         year = 2024,
        month = aug,
       volume = {532},
       number = {3},
        pages = {3102-3119},
          doi = {10.1093/mnras/stae1705},
archivePrefix = {arXiv},
       eprint = {2405.00774},
 primaryClass = {astro-ph.GA},
       adsurl = {https://ui.adsabs.harvard.edu/abs/2024MNRAS.532.3102S}
}

@ARTICLE{2019MNRAS.487.2038C,
       author = {{Cullen}, F. and {McLure}, R.~J. and {Dunlop}, J.~S. and {Khochfar}, S. and {Dav{\'e}}, R. and {Amor{\'\i}n}, R. and {Bolzonella}, M. and {Carnall}, A.~C. and {Castellano}, M. and {Cimatti}, A. and {Cirasuolo}, M. and {Cresci}, G. and {Fynbo}, J.~P.~U. and {Fontanot}, F. and {Gargiulo}, A. and {Garilli}, B. and {Guaita}, L. and {Hathi}, N. and {Hibon}, P. and {Mannucci}, F. and {Marchi}, F. and {McLeod}, D.~J. and {Pentericci}, L. and {Pozzetti}, L. and {Shapley}, A.~E. and {Talia}, M. and {Zamorani}, G.},
        title = "{The VANDELS survey: the stellar metallicities of star-forming galaxies at 2.5 < z < 5.0}",
      journal = {\mnras},
         year = 2019,
        month = aug,
       volume = {487},
       number = {2},
        pages = {2038-2060},
          doi = {10.1093/mnras/stz1402},
archivePrefix = {arXiv},
       eprint = {1903.11081},
 primaryClass = {astro-ph.GA},
       adsurl = {https://ui.adsabs.harvard.edu/abs/2019MNRAS.487.2038C}
}

@ARTICLE{2011ApJ...731..113M,
       author = {{Moster}, Benjamin P. and {Somerville}, Rachel S. and {Newman}, Jeffrey A. and {Rix}, Hans-Walter},
        title = "{A Cosmic Variance Cookbook}",
      journal = {\apj},
         year = 2011,
        month = apr,
       volume = {731},
       number = {2},
          eid = {113},
        pages = {113},
          doi = {10.1088/0004-637X/731/2/113},
archivePrefix = {arXiv},
       eprint = {1001.1737},
 primaryClass = {astro-ph.CO},
       adsurl = {https://ui.adsabs.harvard.edu/abs/2011ApJ...731..113M}
}

@ARTICLE{2024MNRAS.527.6591B,
       author = {{Brinch}, Malte and {Greve}, Thomas R. and {Sanders}, David B. and {McPartland}, Conor J.~R. and {Chartab}, Nima and {Gillman}, Steven and {Vijayan}, Aswin P. and {Lee}, Minju M. and {Brammer}, Gabriel and {Casey}, Caitlin M. and {Ilbert}, Olivier and {Jin}, Shuowen and {Magdis}, Georgios and {McCracken}, H.~J. and {Sillassen}, Nikolaj B. and {Toft}, Sune and {Zavala}, Jorge A.},
        title = "{DEIMOS spectroscopy of z = 6 protocluster candidate in COSMOS - a massive protocluster embedded in a large-scale structure?}",
      journal = {\mnras},
         year = 2024,
        month = jan,
       volume = {527},
       number = {3},
        pages = {6591-6615},
          doi = {10.1093/mnras/stad3409},
archivePrefix = {arXiv},
       eprint = {2311.00511},
 primaryClass = {astro-ph.GA},
       adsurl = {https://ui.adsabs.harvard.edu/abs/2024MNRAS.527.6591B}
}

@ARTICLE{2021MNRAS.505..903C,
       author = {{Cullen}, F. and {Shapley}, A.~E. and {McLure}, R.~J. and {Dunlop}, J.~S. and {Sanders}, R.~L. and {Topping}, M.~W. and {Reddy}, N.~A. and {Amor{\'\i}n}, R. and {Begley}, R. and {Bolzonella}, M. and {Calabr{\`o}}, A. and {Carnall}, A.~C. and {Castellano}, M. and {Cimatti}, A. and {Cirasuolo}, M. and {Cresci}, G. and {Fontana}, A. and {Fontanot}, F. and {Garilli}, B. and {Guaita}, L. and {Hamadouche}, M. and {Hathi}, N.~P. and {Mannucci}, F. and {McLeod}, D.~J. and {Pentericci}, L. and {Saxena}, A. and {Talia}, M. and {Zamorani}, G.},
        title = "{The NIRVANDELS Survey: a robust detection of {\ensuremath{\alpha}}-enhancement in star-forming galaxies at z ≃ 3.4}",
      journal = {\mnras},
         year = 2021,
        month = jul,
       volume = {505},
       number = {1},
        pages = {903-920},
          doi = {10.1093/mnras/stab1340},
archivePrefix = {arXiv},
       eprint = {2103.06300},
 primaryClass = {astro-ph.GA},
       adsurl = {https://ui.adsabs.harvard.edu/abs/2021MNRAS.505..903C}
}

@ARTICLE{2022ApJ...925...82K,
       author = {{Kashino}, Daichi and {Lilly}, Simon J. and {Renzini}, Alvio and {Daddi}, Emanuele and {Zamorani}, Giovanni and {Silverman}, John D. and {Ilbert}, Olivier and {Peng}, Ying-jie and {Mainieri}, Vincenzo and {Bardelli}, Sandro and {Zucca}, Elena and {Kartaltepe}, Jeyhan S. and {Sanders}, David B.},
        title = "{The Stellar Mass versus Stellar Metallicity Relation of Star-forming Galaxies at 1.6 {\ensuremath{\leq}} z {\ensuremath{\leq}} 3.0 and Implications for the Evolution of the {\ensuremath{\alpha}}-enhancement}",
      journal = {\apj},
         year = 2022,
        month = jan,
       volume = {925},
       number = {1},
          eid = {82},
        pages = {82},
          doi = {10.3847/1538-4357/ac399e},
archivePrefix = {arXiv},
       eprint = {2109.06044},
 primaryClass = {astro-ph.GA},
       adsurl = {https://ui.adsabs.harvard.edu/abs/2022ApJ...925...82K}
}

@ARTICLE{2001ApJ...548..681B,
       author = {{Bell}, Eric F. and {Kennicutt}, Jr., Robert C.},
        title = "{A Comparison of Ultraviolet Imaging Telescope Far-Ultraviolet and H{\ensuremath{\alpha}} Star Formation Rates}",
      journal = {\apj},
         year = 2001,
        month = feb,
       volume = {548},
       number = {2},
        pages = {681-693},
          doi = {10.1086/319025},
archivePrefix = {arXiv},
       eprint = {astro-ph/0010340},
 primaryClass = {astro-ph},
       adsurl = {https://ui.adsabs.harvard.edu/abs/2001ApJ...548..681B}
}

@ARTICLE{1976ApJ...203..587C,
       author = {{Cohen}, J.~G.},
        title = "{Halpha emission from the disk of spiral galaxies.}",
      journal = {\apj},
         year = 1976,
        month = feb,
       volume = {203},
        pages = {587-592},
          doi = {10.1086/154115},
       adsurl = {https://ui.adsabs.harvard.edu/abs/1976ApJ...203..587C}
}

@ARTICLE{2025ApJ...989..209S,
       author = {{Sanders}, Ryan L. and {Shapley}, Alice E. and {Topping}, Michael W. and {Reddy}, Naveen A. and {Berg}, Danielle A. and {Bouwens}, Rychard J. and {Brammer}, Gabriel and {Carnall}, Adam C. and {Cullen}, Fergus and {Dav{\'e}}, Romeel and {Dunlop}, James S. and {Ellis}, Richard S. and {F{\"o}rster Schreiber}, N.~M. and {Furlanetto}, Steven R. and {Glazebrook}, Karl and {Illingworth}, Garth D. and {Jones}, Tucker and {Kriek}, Mariska and {McLeod}, Derek J. and {McLure}, Ross J. and {Narayanan}, Desika and {Oesch}, Pascal A. and {Pahl}, Anthony J. and {Pettini}, Max and {Schaerer}, Daniel and {Stark}, Daniel P. and {Steidel}, Charles C. and {Tang}, Mengtao and {Clarke}, Leonardo and {Donnan}, Callum T. and {Kehoe}, Emily},
        title = "{The AURORA Survey: The Nebular Attenuation Curve of a Galaxy at z = 4.41 from Ultraviolet to Near-infrared Wavelengths}",
      journal = {\apj},
         year = 2025,
        month = aug,
       volume = {989},
       number = {2},
          eid = {209},
        pages = {209},
          doi = {10.3847/1538-4357/adf066},
archivePrefix = {arXiv},
       eprint = {2408.05273},
 primaryClass = {astro-ph.GA},
       adsurl = {https://ui.adsabs.harvard.edu/abs/2025ApJ...989..209S}
}

@ARTICLE{2020ARA&A..58..529S,
       author = {{Salim}, Samir and {Narayanan}, Desika},
        title = "{The Dust Attenuation Law in Galaxies}",
      journal = {\araa},
         year = 2020,
        month = aug,
       volume = {58},
        pages = {529-575},
          doi = {10.1146/annurev-astro-032620-021933},
archivePrefix = {arXiv},
       eprint = {2001.03181},
 primaryClass = {astro-ph.GA},
       adsurl = {https://ui.adsabs.harvard.edu/abs/2020ARA&A..58..529S}
}

@ARTICLE{2025arXiv251006681C,
       author = {{Clarke}, Leonardo and {Shapley}, Alice E. and {Lam}, Natalie and {Topping}, Michael W. and {Brammer}, Gabriel B. and {Sanders}, Ryan L. and {Reddy}, Naveen A. and {Karthikeyan}, Shreya},
        title = "{The Star-forming Main Sequence and Bursty Star-formation Histories at $z>1.4$ in JADES and AURORA}",
      journal = {arXiv e-prints},
         year = 2025,
        month = oct,
          eid = {arXiv:2510.06681},
        pages = {arXiv:2510.06681},
          doi = {10.48550/arXiv.2510.06681},
archivePrefix = {arXiv},
       eprint = {2510.06681},
 primaryClass = {astro-ph.GA},
       adsurl = {https://ui.adsabs.harvard.edu/abs/2025arXiv251006681C}
}

@ARTICLE{2025arXiv251100705S,
       author = {{Stanton}, T.~M. and {Cullen}, F. and {Carnall}, A.~C. and {Scholte}, D. and {Arellano-C{\'o}rdova}, K.~Z. and {Shapley}, A.~E. and {McLeod}, D.~J. and {Donnan}, C.~T. and {Begley}, R. and {Dav{\'e}}, R. and {Dunlop}, J.~S. and {McLure}, R.~J. and {Rowlands}, K. and {Bondestam}, C. and {Hamadouche}, M.~L. and {Leung}, H.-H. and {Stevenson}, S.~D. and {Taylor}, E.},
        title = "{The JWST EXCELS Survey: gas-phase metallicity evolution at 2 < z < 8}",
      journal = {arXiv e-prints},
         year = 2025,
        month = nov,
          eid = {arXiv:2511.00705},
        pages = {arXiv:2511.00705},
          doi = {10.48550/arXiv.2511.00705},
archivePrefix = {arXiv},
       eprint = {2511.00705},
 primaryClass = {astro-ph.GA},
       adsurl = {https://ui.adsabs.harvard.edu/abs/2025arXiv251100705S}
}

@ARTICLE{2025arXiv250818369J,
       author = {{Jain}, Shweta and {Sanders}, Ryan L. and {Khostovan}, Ali Ahmad and {Jones}, Tucker and {Shapley}, Alice E. and {Reddy}, Naveen A. and {Garcia}, Alex M. and {Torrey}, Paul and {Coil}, Alison},
        title = "{A Uniform Analysis of Gas-phase Metallicity Evolution with 1-3 Gyr Time Sampling over the Past 12 Billion Years}",
      journal = {arXiv e-prints},
         year = 2025,
        month = aug,
          eid = {arXiv:2508.18369},
        pages = {arXiv:2508.18369},
          doi = {10.48550/arXiv.2508.18369},
archivePrefix = {arXiv},
       eprint = {2508.18369},
 primaryClass = {astro-ph.GA},
       adsurl = {https://ui.adsabs.harvard.edu/abs/2025arXiv250818369J}
}

@ARTICLE{2025ApJ...980L..27I,
       author = {{Inayoshi}, Kohei and {Maiolino}, Roberto},
        title = "{Extremely Dense Gas around Little Red Dots and High-redshift Active Galactic Nuclei: A Nonstellar Origin of the Balmer Break and Absorption Features}",
      journal = {\apjl},
         year = 2025,
        month = feb,
       volume = {980},
       number = {2},
          eid = {L27},
        pages = {L27},
          doi = {10.3847/2041-8213/adaebd},
archivePrefix = {arXiv},
       eprint = {2409.07805},
 primaryClass = {astro-ph.GA},
       adsurl = {https://ui.adsabs.harvard.edu/abs/2025ApJ...980L..27I}
}





\appendix

\section{Redshift probability distributions for the F466N and F470N excess source sample.}
\label{app:pz_fractions}

In Section \ref{sec:p_z_analysis}, we assess the robustness of the H$\alpha$ classifications by examining the redshift posterior distributions, $P(z)$, and testing the impact on sample selections. The highest redshift emission line galaxy samples ($z_{\rm{1,median}} > 5.5$) include the H$\alpha$ sample at $z \sim 6.1$, but also the \oiiia emitter sample at $z \sim 8.3$. Therefore, we examine the $P(z)$ posteriors for the H$\alpha$ and \oiiia samples to probe for any degeneracies between sample redshift solutions. Table \ref{tab:high_z_pz} shows the $z_{\rm{1,median}} > 5.5$ galaxy sample and considers the integrated $P(z)$ over the broad H$\alpha$ ($5.5 < z < 6.5$) and \oiiia ($7.8 < z < 8.8$) emission line redshift observing windows for the F466N and F470N filters. We conclude that the sum of these two probabilities is always $\sim$100 per cent and so the photo-$z$ fitting is robustly selecting high-redshift emission line galaxies, including the H$\alpha$ sample. In addition, there is not much degeneracy between the H$\alpha$ and \oiiia solutions, except for one source (Source F466N ID 4212). The rest of the high-redshift sample either have $P(5.5 < z < 6.5)$ or $P(7.8 < z < 8.8)$ $>95$ per cent for their primary redshift solutions.

\begin{figure}
\centering
  \includegraphics[width=\linewidth]{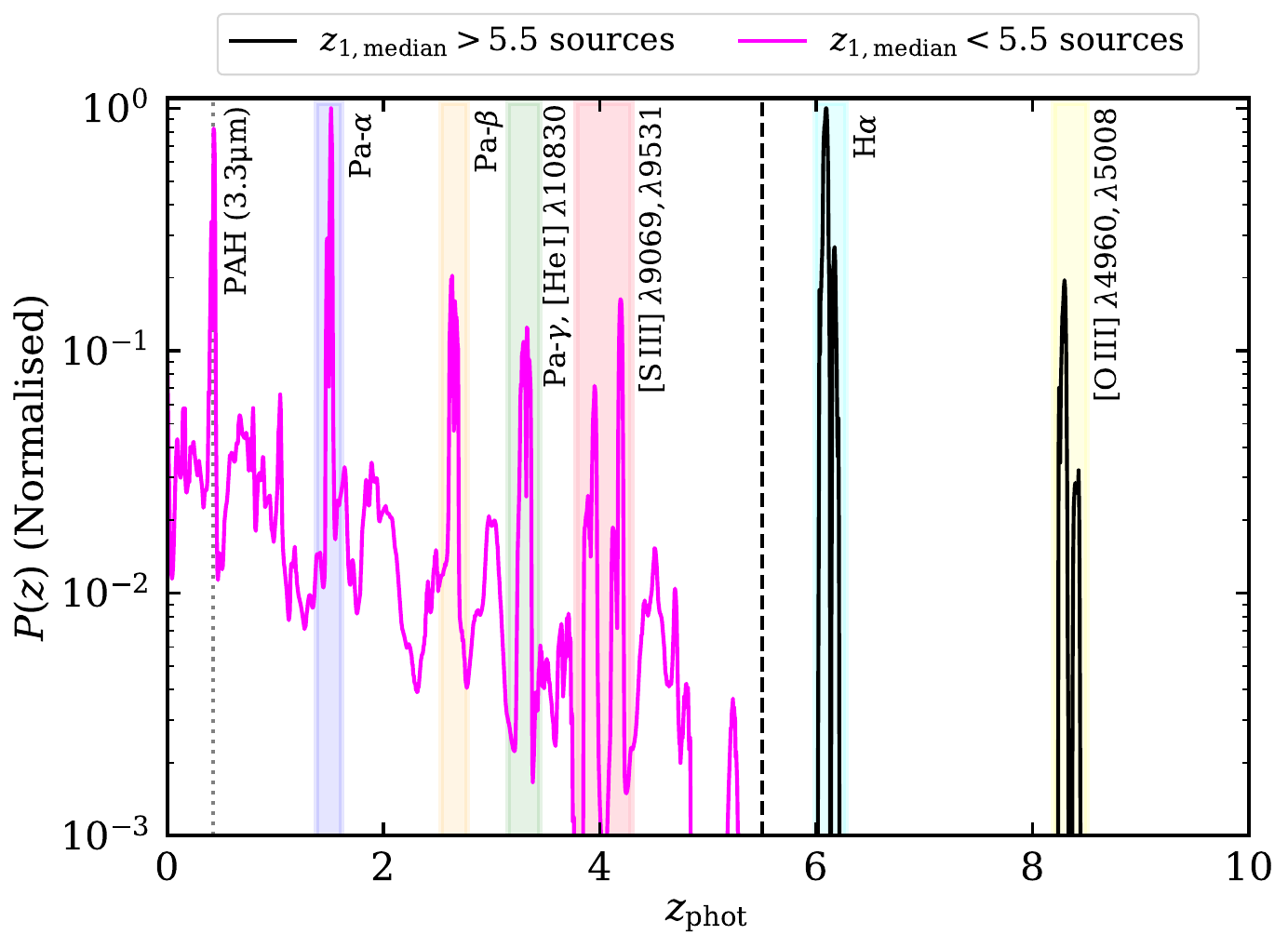}\vspace{-0.4cm}
  
    \caption{The stacked redshift probability distributions (normalised to the largest $P(z)$ peak) for the high-redshift ($z_{\rm{1,median}} > 5.5$) and low-redshift ($z_{\rm{1,median}} < 5.5$) excess source candidates from the F466N and F470N v1.0 catalogues, shown in the black and pink solid lines, respectively. The black dashed line shows $z=5.5$. The shaded regions show the redshift windows in which the spectral feature/emission line can be observed (with the spectral feature shown in text to the right of the relevant shaded region).}

    \label{fig:excess_pz}

\end{figure}

In Fig. \ref{fig:excess_pz}, we show the stacked $P(z)$ (normalised to the largest redshift peak) for the selected high-redshift sample. We find a negligible ($\sim$0.007 per cent) probability that the high-redshift sample is at low-redshift ($z<5.5$) when integrating the stacked $P(z)$. An additional concern in selecting the high-redshift sample is that the criteria enforced might have missed some sources with lower probability solutions, identified with $z_{\rm{1,median}} < 5.5$, but which are actually at high-redshift. We test this by stacking the $P(z)$ for all $z_{\rm{1,median}} < 5.5$ sources (see Fig. \ref{fig:excess_pz}) and find they contributed $<0.1$ per cent probability of being at higher redshift. Given there are a few hundred $z_{\rm{1,median}} < 5.5$ selected emission line galaxies, this corresponds to a probability equivalent to fewer than one source. Therefore, we are confident that the completeness of our H$\alpha$ sample is not impacted by the redshift criteria discussed in Section \ref{sec:halpha_sources}.

\begin{table*}
\centering
\caption{The high-redshift excess source candidates from the v1.0 catalogues (see Section \ref{sec:nb_cat}) where we show the source IDs, sky coordinates (right ascension and declination in degrees), detection filters (F466N or F470N), median redshift of primary redshift peaks ($z_{\rm{1,median}}$), the measured line fluxes and uncertainties ($F_{\rm{line}}$), and the probabilities that the sources lie within the broad redshift ranges for H$\alpha$ ($P(5.5 < z < 6.5)$) and \oiiia ($P(7.8 < z < 8.8)$) emission line galaxy identification. The table is ordered in descending order based on the value of $P(5.5 < z < 6.5)$. For rows where $P(5.5 < z < 6.5)$ values are equivalent, we then ordered by $F_{\rm{line}}$ in ascending order. Source IDs with asterisks have spectroscopic data (see Section \ref{sec:halpha_lum}) confirming their photometric redshifts and hence emission line galaxy classification.}
\label{tab:high_z_pz}
\begin{tabular}{c c c c c c c c}
\hline
Source ID (v1.0) & RA [deg] & Dec [deg] & Detection Filter & $z_{\rm{1,median}}$ & $F_{\rm{line}}$ [$10^{-18}$ erg s$^{-1}$ cm$^{-2}$] & $P(5.5 < z < 6.5)$ & $P(7.8 < z < 8.8)$ \rule[-1.3ex]{0pt}{3.9ex} \\
\hline
1635 & 150.106596 & 2.281332 & F466N & 6.25 & 0.50 $\pm$ 0.33 & 1.0000 & 0.0000 \\
5120 & 150.088831 & 2.339012 & F466N & 6.08 & 0.83 $\pm$ 0.22 & 1.0000 & 0.0000 \\
5446 & 150.098843 & 2.343943 & F466N & 6.08 & 0.97 $\pm$ 0.21 & 1.0000 & 0.0000 \\
5049$^{*}$ & 150.113693 & 2.338062 & F466N & 6.08 & 0.99 $\pm$ 0.21 & 1.0000 & 0.0000 \\
7650 & 150.183716 & 2.372950 & F466N & 6.09 & 1.06 $\pm$ 0.28 & 1.0000 & 0.0000 \\
3297 & 150.101696 & 2.307418 & F466N & 6.12 & 1.09 $\pm$ 0.34 & 1.0000 & 0.0000 \\
4126 & 150.116016 & 2.322179 & F466N & 6.08 & 1.14 $\pm$ 0.24 & 1.0000 & 0.0000 \\
9297 & 150.145339 & 2.403268 & F466N & 6.10 & 1.14 $\pm$ 0.31 & 1.0000 & 0.0000 \\
8557$^{*}$ & 150.138850 & 2.392129 & F466N & 6.05 & 1.15 $\pm$ 0.36 & 1.0000 & 0.0000 \\
4820 & 150.168000 & 2.322630 & F470N & 6.17 & 1.20 $\pm$ 0.20 & 1.0000 & 0.0000 \\
416 & 150.102703 & 2.262541 & F466N & 6.04 & 1.17 $\pm$ 0.39 & 1.0000 & 0.0000 \\
7654 & 150.183744 & 2.373026 & F466N & 6.08 & 1.26 $\pm$ 0.30 & 1.0000 & 0.0000 \\
3535 & 150.142486 & 2.306622 & F470N & 6.17 & 1.40 $\pm$ 0.18 & 1.0000 & 0.0000 \\
7902 & 150.179232 & 2.377068 & F466N & 6.08 & 1.55 $\pm$ 0.29 & 1.0000 & 0.0000 \\
3221 & 150.150189 & 2.306428 & F466N & 6.09 & 1.81 $\pm$ 0.31 & 1.0000 & 0.0000 \\
5849 & 150.081395 & 2.350119 & F466N & 6.08 & 2.01 $\pm$ 0.22 & 1.0000 & 0.0000 \\
9244 & 150.132274 & 2.402047 & F466N & 6.09 & 2.04 $\pm$ 0.33 & 1.0000 & 0.0000 \\
9042$^{*}$ & 150.139768 & 2.398772 & F466N & 6.07 & 2.05 $\pm$ 0.39 & 1.0000 & 0.0000 \\
6680 & 150.165262 & 2.342845 & F470N & 6.16 & 2.24 $\pm$ 0.17 & 1.0000 & 0.0000 \\
6938 & 150.098014 & 2.361185 & F466N & 6.09 & 2.78 $\pm$ 0.30 & 1.0000 & 0.0000 \\
7479 & 150.191302 & 2.370349 & F466N & 6.08 & 3.26 $\pm$ 0.44 & 1.0000 & 0.0000 \\
2144 & 150.155101 & 2.289099 & F466N & 6.08 & 3.27 $\pm$ 0.41 & 1.0000 & 0.0000 \\
4469 & 150.138461 & 2.328439 & F466N & 6.09 & 3.87 $\pm$ 0.29 & 1.0000 & 0.0000 \\
9693 & 150.099447 & 2.413258 & F466N & 6.08 & 3.98 $\pm$ 0.37 & 1.0000 & 0.0000 \\
4970 & 150.156806 & 2.336835 & F466N & 6.09 & 4.25 $\pm$ 0.24 & 1.0000 & 0.0000 \\
2962 & 150.149015 & 2.301786 & F466N & 6.08 & 4.47 $\pm$ 0.31 & 1.0000 & 0.0000 \\
8416 & 150.097882 & 2.388232 & F466N & 6.09 & 5.27 $\pm$ 0.44 & 1.0000 & 0.0000 \\
9300 & 150.145489 & 2.403016 & F466N & 6.08 & 5.38 $\pm$ 0.40 & 1.0000 & 0.0000 \\
11394 & 150.165498 & 2.386310 & F470N & 6.15 & 5.93 $\pm$ 0.28 & 1.0000 & 0.0000 \\
6456 & 150.109598 & 2.356238 & F466N & 6.10 & 5.99 $\pm$ 0.34 & 1.0000 & 0.0000 \\
5887 & 150.077780 & 2.350375 & F466N & 6.07 & 9.31 $\pm$ 0.26 & 1.0000 & 0.0000 \\
9349 & 150.141805 & 2.403731 & F466N & 6.09 & 10.82 $\pm$ 0.45 & 1.0000 & 0.0000 \\
1988 & 150.160247 & 2.286542 & F466N & 6.08 & 13.21 $\pm$ 0.38 & 1.0000 & 0.0000 \\
5364 & 150.090879 & 2.342392 & F466N & 6.09 & 0.68 $\pm$ 0.20 & 0.9998 & 0.0000 \\
3228 & 150.150206 & 2.306384 & F466N & 6.08 & 1.11 $\pm$ 0.36 & 0.9993 & 0.0007 \\
4419 & 150.166023 & 2.317723 & F470N & 6.17 & 1.37 $\pm$ 0.20 & 0.9957 & 0.0043 \\
4873 & 150.168540 & 2.334553 & F466N & 6.08 & 0.73 $\pm$ 0.21 & 0.9931 & 0.0069 \\
6906 & 150.097978 & 2.360992 & F466N & 6.09 & 1.23 $\pm$ 0.26 & 0.9731 & 0.0269 \\
1047 & 150.129161 & 2.275689 & F470N & 6.19 & 0.82 $\pm$ 0.29 & 0.9580 & 0.0000 \\
4212 & 150.072906 & 2.323626 & F466N & 8.25 & 1.46 $\pm$ 0.26 & 0.4120 & 0.5880 \\
9241 & 150.099929 & 2.401969 & F466N & 8.29 & 1.83 $\pm$ 0.33 & 0.0242 & 0.9758 \\
7782$^{*}$ & 150.088773 & 2.375441 & F466N & 8.28 & 1.25 $\pm$ 0.24 & 0.0007 & 0.9993 \\
5312 & 150.136908 & 2.328452 & F470N & 8.41 & 2.19 $\pm$ 0.23 & 0.0000 & 1.0000 \\
5211$^{*}$ & 150.105715 & 2.328452 & F466N & 8.29 & 2.53 $\pm$ 0.23 & 0.0000 & 1.0000 \\
5683 & 150.103274 & 2.347590 & F466N & 8.29 & 5.88 $\pm$ 0.40 & 0.0000 & 1.0000 \\
5685 & 150.103288 & 2.347697 & F466N & 8.29 & 6.10 $\pm$ 0.40 & 0.0000 & 1.0000 \\
\hline
\end{tabular}
\end{table*}

\section{Luminosity dependence of $\kappa_{\rm{H\alpha}}$ using galaxy scaling relations}
\label{app:kappa_halpha_l_dependence}

As discussed in Section \ref{sec:sfrd_v2}, we adopt a luminosity-dependent SFR calibration $\kappa(L)$ to calculate $\rho_{\rm{SFR_{H\alpha}}}$:

\begin{equation}
\label{eq:sfrd_v2}
\rho_{\rm{SFR_{H\alpha}}} = \int_{L_{\rm{lim}}}^{\infty} L \, \kappa(L) \, \Phi(L) \, \rm{d} L
\end{equation}

\noindent We aim to determine the typical $Z_{\star}$ for a given dust-corrected $L_{\rm{H\alpha}}$ to obtain the appropriate $\kappa_{\rm{H\alpha}}$. Given we cannot measure the metallicity directly \citep[e.g.][]{2025ApJ...984..188K,2025arXiv250810099S}, we estimate the typical metallicity as a function of $L_{\rm{H\alpha}}$ by combining the inferred $L_{\rm{H\alpha}}-M_\star$ relation with the mass-metallicity relation (MZR). Specifically, we adopt the star-forming main sequence (SFMS) from \citet{2023MNRAS.519.1526P} at $z=6.1$ and convert the SFR into $L_{\rm{H\alpha}}$ using their chosen $\kappa_{\rm{H\alpha}}$ \citep{2012ARA&A..50..531K}. This gives a $L_{\rm{H\alpha}}-M_{\star}$ relation that covers a wide stellar mass range at this redshift \citep[$8.5 \lesssim \log_{10}(M_{\star}/\rm{M_{\odot}}) \lesssim 11.5$; see Fig. 6 of ][]{2023MNRAS.519.1526P}. Within the overlapping range of stellar masses, the \citet{2023MNRAS.519.1526P} SFMS agrees well with recent constraints on the SFMS using H$\alpha$ as the SFR indicator at $z\sim6$ \citep[e.g.][]{2025arXiv251006681C}, despite using multiple indicators probing a range of star-formation timescales.

We adopt the gas-phase mass-metallicity relation (MZR) from \citet{2023ApJS..269...33N} measured from their $z \sim 4 - 10$ star-forming galaxy sample with a similar stellar mass range to our H$\alpha$ sample ($7.5 < \log_{10}(M_{\star}/\rm{M_{\odot}}) < 9.5$). We note that the MZR is not well constrained at individual redshift epochs or to higher stellar masses ($\log_{10}(M_{\star}/\rm{M_{\odot}}) \gtrsim 9.5$) at $z > 4$ \citep[see also][]{2024A&A...684A..75C}. Therefore, we assume that the MZR does not evolve (or evolves weakly) between $z \sim 4 - 10$ and that extrapolating the \citet{2023ApJS..269...33N} MZR to higher stellar masses is valid. The MZR has been observed to flatten at the highest stellar masses ($\log_{10}(M_{\star}/\rm{M_{\odot}}) \gtrsim 10$) for star-forming galaxies in the local Universe \citep[e.g.][]{2020MNRAS.491..944C}, though studies between $z \sim 2 - 4$ show a reduction in the normalisation of the MZR but no such flattening at high stellar masses \citep{2021ApJ...914...19S,2025arXiv250818369J,2025arXiv251100705S}. Regardless, this will have a negligible impact on the inferred $\rho_{\rm{SFR}}$, given the H$\alpha$ LF drops rapidly at the luminosities corresponding to this stellar mass regime. 

The corresponding stellar metallicities $Z_{\star}$ in high-redshift galaxy samples have been observed to be offset from the gas-phase metallicities $Z_{\rm{g}}$ by $\sim0.35-0.5$ dex across a wide range of stellar masses \citep[e.g.][]{2021MNRAS.505..903C,2022ApJ...925...82K,2024MNRAS.532.3102S,2025ApJ...984..188K}, indicative of $\alpha$-enhancement, which is thought to be ubiquitous at high redshift. We choose to adopt a 0.42 dex decrease in the normalisation of the gas-phase MZR to obtain the stellar MZR, following results from \citet{2024MNRAS.532.3102S} who stacked rest-frame UV and optical spectra for $z=3.5$ star-forming galaxies \citep[see methodologies described in][]{2019MNRAS.487.2038C,2021MNRAS.505..903C}. We then use this MZR to infer the $L_{\rm{H\alpha}}$ -- $Z_{\star}$ relation. From the $L_{\rm{H\alpha}}$ -- $Z_{\star}$ relation, we infer $\kappa_{\rm{H\alpha}}$ as a function of $L_{\rm{H\alpha}}$ using the binary stellar population and spectral synthesis models from BPASSv2.2 \citep{2017PASA...34...58E,2018MNRAS.479...75S}, which provide grids relating $\kappa_{\rm{H\alpha}}$ to $Z_{\star}$ \citep[see Table 4 of ][but adjusted for the v2.2 models and using the correct IMF and upper mass limit]{2017PASA...34...58E}. We finally compute $\kappa(L)$ over a sufficiently large $L_{\rm{H\alpha}}$ range and fold this into the integral in Eq. \ref{eq:sfrd_v2} to compute $\rho_{\rm{SFR_{H\alpha}}}$.


\bsp	
\label{lastpage}
\end{document}